\newcount\mgnf  %ingrandimento
\mgnf=0
 
\ifnum\mgnf=0
\def\openone{\leavevmode\hbox{\ninerm 1\kern-3.3pt\tenrm1}}%
\def\*{\vglue0.3truecm}\fi
\ifnum\mgnf=1
\def\openone{\leavevmode\hbox{\ninerm 1\kern-3.63pt\tenrm1}}%
\def\*{\vglue0.5truecm}\fi

\def\openonesix{\leavevmode\hbox{\sixrm 1\kern-2.6pt\ottorm1}}%
\def\*{\vglue0.3truecm} 

\ifnum\mgnf=0
   \magnification=\magstep0
   \hsize=14truecm\vsize=24.truecm
   \parindent=0.3cm\baselineskip=0.45cm
\font\titolo=cmbx12
\font\titolone=cmbx10 scaled\magstep 2
\font\cs=cmcsc10 %scaled\magstep1%
\font\ottorm=cmr8 %scaled\magstep1%
 %scaled\magstep1%
\font\euftw=eufm10 %scaled\magstep1%
 %scaled\magstep1%
 %scaled\magstep1%
\font\msytw=msbm10 %scaled\magstep1%
 %scaled\magstep1%
 %scaled\magstep1%
 %scaled\magstep1%
\font\indbf=cmbx10 scaled\magstep1%
 %scaled\magstep1%
\font\ottoi=cmmi8\font\ottosy=cmsy8%
\font\ottobf=cmbx8\font\ottott=cmtt8%
\font\ottocss=cmcsc8%
\font\ottosl=cmsl8\font\ottoit=cmti8%

\font\sixrm=cmr6\font\sixbf=cmbx6\font\sixi=cmmi6\font\sixsy=cmsy6%
\font\fiverm=cmr5\font\fivesy=cmsy5
\font\fivei=cmmi5
\font\fivebf=cmbx5
\font\sc=cmcsc10
\font\tenmib=cmmib10
\font\sevenmib=cmmib10 scaled 800
\font\ninerm=cmr9%
\fi
\ifnum\mgnf=1
   \magnification=\magstep1\hoffset=0.truecm
   \hsize=14truecm\vsize=24.truecm
   \baselineskip=18truept plus0.1pt minus0.1pt \parindent=0.9truecm
   \lineskip=0.5truecm\lineskiplimit=0.1pt      \parskip=0.1pt plus1pt
\font\titolo=cmbx12 scaled\magstep 1
\font\titolone=cmbx10 scaled\magstep 3
\font\cs=cmcsc10 scaled\magstep1%
\font\ottorm=cmr8 scaled\magstep1%
\font\euftw=eufm10 scaled\magstep1%
\font\msytw=msbm10 scaled\magstep1%
\font\indbf=cmbx10 scaled\magstep2%

\fi

\def\ottopunti{\def\rm{\fam0\ottorm}%
\textfont0=\ottorm\scriptfont0=\sixrm\scriptscriptfont0=\fiverm%
\textfont1=\ottoi\scriptfont1=\sixi\scriptscriptfont1=\fivei%
\textfont2=\ottosy\scriptfont2=\sixsy\scriptscriptfont2=\fivesy%
\textfont3=\tenex\scriptfont3=\tenex\scriptscriptfont3=\tenex%
\textfont4=\ottocss\scriptfont4=\sc\scriptscriptfont4=\sc%
%\scriptfont4=\ottocss\scriptscriptfont4=\ottocss%
\textfont5=\tenmib\scriptfont5=\sevenmib\scriptscriptfont5=\fivei
\textfont\itfam=\ottoit\def\it{\fam\itfam\ottoit}%
\textfont\slfam=\ottosl\def\sl{\fam\slfam\ottosl}%
\textfont\ttfam=\ottott\def\tt{\fam\ttfam\ottott}%
\textfont\bffam=\ottobf\scriptfont\bffam=\sixbf%
\scriptscriptfont\bffam=\fivebf\def\bf{\fam\bffam\ottobf}%
%\tt\ttglue=.5em plus.25em minus.15em%
\setbox\strutbox=\hbox{\vrule height7pt depth2pt width0pt}%
\normalbaselineskip=9pt\let\sc=\sixrm\normalbaselines\rm}

\global\newcount\numsec\global\newcount\numapp
\global\newcount\numfor\global\newcount\numfig\global\newcount\numsub
\global\newcount\numlemma\global\newcount\numtheorem\global\newcount\numdef
\global\newcount\appflag
\numsec=0\numapp=0\numfig=1
\def\veroparagrafo{\number\numsec}\def\veraformula{\number\numfor}
\def\veraappendice{\number\numapp}\def\verasub{\number\numsub}
\def\verafigura{\number\numfig}
\def\verolemma{\number\numlemma}
\def\verotheorem{\number\numtheorem}
\def\veradef{\number\numdef}
 
\def\section(#1,#2){\advance\numsec by 1\numfor=1\numsub=1%
\numlemma=1\numtheorem=1\numdef=1\appflag=0%
\SIA p,#1,{\veroparagrafo} %
\write15{\string\Fp (#1){\secc(#1)}}%
\write16{ sec. #1 ==> \secc(#1)  }%
\hbox to \hsize{\titolo\hfill \number\numsec. #2\hfill%
\expandafter{\alato(sec. #1)}}\*}
 
\def\appendix(#1,#2){\advance\numapp by 1\numfor=1\numsub=1%
\numlemma=1\numtheorem=1\numdef=1\appflag=1%
\SIA p,#1,{A\veraappendice} %
\write15{\string\Fp (#1){\secc(#1)}}%
\write16{ app. #1 ==> \secc(#1)  }%
\hbox to \hsize{\titolo\hfill Appendix A\number\numapp. #2\hfill%
\expandafter{\alato(app. #1)}}\*}
 
\def\senondefinito#1{\expandafter\ifx\csname#1\endcsname\relax}
 
\def\SIA #1,#2,#3 {\senondefinito{#1#2}%
\expandafter\xdef\csname #1#2\endcsname{#3}\else
\write16{???? ma #1#2 e' gia' stato definito !!!!} \fi}
 
\def \Fe(#1)#2{\SIA fe,#1,#2 }
\def \Fp(#1)#2{\SIA fp,#1,#2 }
\def \Fg(#1)#2{\SIA fg,#1,#2 }
\def \Fl(#1)#2{\SIA fl,#1,#2 }
\def \Ft(#1)#2{\SIA ft,#1,#2 }
\def \Fd(#1)#2{\SIA fd,#1,#2 }
 
\def\etichetta(#1){%
\ifnum\appflag=0(\veroparagrafo.\veraformula)%
\SIA e,#1,(\veroparagrafo.\veraformula) \fi%
\ifnum\appflag=1(A\veraappendice.\veraformula)%
\SIA e,#1,(A\veraappendice.\veraformula) \fi%
\global\advance\numfor by 1%
\write15{\string\Fe (#1){\equ(#1)}}%
\write16{ EQ #1 ==> \equ(#1)  }}

\def\getichetta(#1){Fig. \verafigura%
\SIA g,#1,{\verafigura} %
\global\advance\numfig by 1%
\write15{\string\Fg (#1){\graf(#1)}}%
\write16{ Fig. #1 ==> \graf(#1) }}
 
\def\etichettap(#1){%
\ifnum\appflag=0{\veroparagrafo.\verasub}%
\SIA p,#1,{\veroparagrafo.\verasub} \fi%
\ifnum\appflag=1{A\veraappendice.\verasub}%
\SIA p,#1,{A\veraappendice.\verasub} \fi%
\global\advance\numsub by 1%
\write15{\string\Fp (#1){\secc(#1)}}%
\write16{ par #1 ==> \secc(#1)  }}

\def\etichettal(#1){%
\ifnum\appflag=0{\veroparagrafo.\verolemma}%
\SIA l,#1,{\veroparagrafo.\verolemma} \fi%
\ifnum\appflag=1{A\veraappendice.\verolemma}%
\SIA l,#1,{A\veraappendice.\verolemma} \fi%
\global\advance\numlemma by 1%
\write15{\string\Fl (#1){\lm(#1)}}%
\write16{ lemma #1 ==> \lm(#1)  }}

\def\etichettat(#1){%
\ifnum\appflag=0{\veroparagrafo.\verotheorem}%
\SIA t,#1,{\veroparagrafo.\verotheorem} \fi%
\ifnum\appflag=1{A\veraappendice.\verotheorem}%
\SIA t,#1,{A\veraappendice.\verotheorem} \fi%
\global\advance\numtheorem by 1%
\write15{\string\Ft (#1){\thm(#1)}}%
\write16{ th. #1 ==> \thm(#1)  }}

\def\etichettad(#1){%
\inum\appflag=0{\veroparagrafo.\veradef}%
\SIA d,#1,{\veroparagrafo.\veradef} \fi%
\inum\appflag=1{A\veraappendice.\veradef}%
\SIA d,#1,{A\veraappendice.\veradef} \fi%
\global\advance\numdef by 1%
\write15{\string\Fd (#1){\defz(#1)}}%
\write16{ def. #1 ==> \defz(#1)  }}

\def\Eq(#1){\eqno{\etichetta(#1)\alato(#1)}}
\def\eq(#1){\etichetta(#1)\alato(#1)}
\def\eqg(#1){\getichetta(#1)\alato(fig #1)}
\def\sub(#1){\0\palato(p. #1){\bf \etichettap(#1)\hskip.3truecm}}
\def\lemma(#1){\0\palato(lm #1){\cs Lemma \etichettal(#1)\hskip.3truecm}}
\def\theorem(#1){\0\palato(th #1){\cs Theorem \etichettat(#1)%
\hskip.3truecm}}
\def\definition(#1){\0\palato(df #1){\cs Definition \etichettad(#1)%
\hskip.3truecm}}

\def\equv(#1){\senondefinito{fe#1}$\clubsuit$#1%
\write16{eq. #1 non e' (ancora) definita}%
\else\csname fe#1\endcsname\fi}
\def\grafv(#1){\senondefinito{fg#1}$\clubsuit$#1%
\write16{fig. #1 non e' (ancora) definito}%
\else\csname fg#1\endcsname\fi}

\def\secv(#1){\senondefinito{fp#1}$\clubsuit$#1%
\write16{par. #1 non e' (ancora) definito}%
\else\csname fp#1\endcsname\fi}

\def\lmv(#1){\senondefinito{fl#1}$\clubsuit$#1%
\write16{lemma #1 non e' (ancora) definito}%
\else\csname fl#1\endcsname\fi}
 
\def\thmv(#1){\senondefinito{ft#1}$\clubsuit$#1%
\write16{th. #1 non e' (ancora) definito}%
\else\csname ft#1\endcsname\fi}
 
\def\defzv(#1){\senondefinito{fd#1}$\clubsuit$#1%
\write16{def. #1 non e' (ancora) definito}%
\else\csname fd#1\endcsname\fi}
 
\def\equ(#1){\senondefinito{e#1}\equv(#1)\else\csname e#1\endcsname\fi}
\def\graf(#1){\senondefinito{g#1}\grafv(#1)\else\csname g#1\endcsname\fi}
\def\secc(#1){\senondefinito{p#1}\secv(#1)\else\csname p#1\endcsname\fi}
\def\lm(#1){\senondefinito{l#1}\lmv(#1)\else\csname l#1\endcsname\fi}
\def\thm(#1){\senondefinito{t#1}\thmv(#1)\else\csname t#1\endcsname\fi}
\def\defz(#1){\senondefinito{d#1}\defzv(#1)\else\csname d#1\endcsname\fi}
\def\sec(#1){{\S\secc(#1)}}

\def\BOZZA{
\def\alato(##1){\rlap{\kern-\hsize\kern-1.2truecm{$\scriptstyle##1$}}}
\def\palato(##1){\rlap{\kern-1.2truecm{$\scriptstyle##1$}}}
}
 
\def\alato(#1){}
\def\galato(#1){}
\def\palato(#1){}

{\count255=\time\divide\count255 by 60 \xdef\hourmin{\number\count255}
        \multiply\count255 by-60\advance\count255 by\time
   \xdef\hourmin{\hourmin:\ifnum\count255<10 0\fi\the\count255}}
 
\def\oramin{\hourmin }
 
\def\data{\number\day/\ifcase\month\or gennaio \or febbraio \or marzo \or
aprile \or maggio \or giugno \or luglio \or agosto \or settembre
\or ottobre \or novembre \or dicembre \fi/\number\year;\ \oramin}
\setbox200\hbox{$\scriptscriptstyle \data $}
\footline={\rlap{\hbox{\copy200}}\tenrm\hss \number\pageno\hss}

\let\a=\alpha \let\b=\beta  \let\g=\gamma     \let\d=\delta  \let\e=\varepsilon
\let\z=\zeta  \let\h=\eta    \def\th{\theta}
\let\k=\kappa   \let\l=\lambda
\let\m=\mu                \let\p=\pi      \let\r=\rho
\let\s=\sigma \let\t=\tau        \let\c=\chi
   \let\o=\omega 
 \let\D=\Delta     \let\L=\Lambda  
\let\P=\Pi    \let\Si=\Sigma       
\let\O=\Omega 
 
\def\\{\hfill\break} \let\==\equiv

\let\io=\infty 

\let\0=\noindent

\def\ie{\hbox{\it i.e.\ }}
\def\der{\hbox{\rm d}}
\let\dpr=\partial 
\let\bs=\backslash
 
\def\tende#1{\,\vtop{\ialign{##\crcr\rightarrowfill\crcr
 \noalign{\kern-1pt\nointerlineskip}
 \hskip3.pt${\scriptstyle #1}$\hskip3.pt\crcr}}\,}
\def\otto{\,{\kern-1.truept\leftarrow\kern-5.truept\to\kern-1.truept}\,}
\def\fra#1#2{{#1\over#2}}
 
\def\PP{{\cal P}}\def\EE{{\cal E}}\def\VV{{\cal V}}
\def\WW{{\cal W}}
\def\TT{{\cal T}}
\def\RR{{\cal R}}\def\LL{{\cal L}}
\def\DD{{\cal D}}\def\AA{{\cal A}}\def\SS{{\cal S}}
\def\OO{{\cal O}} 
 
\def\T#1{{#1_{\kern-3pt\lower7pt\hbox{$\widetilde{}$}}\kern3pt}}
\def\VVV#1{{\underline #1}_{\kern-3pt
\lower7pt\hbox{$\widetilde{}$}}\kern3pt\,}
\def\W#1{#1_{\kern-3pt\lower7.5pt\hbox{$\widetilde{}$}}\kern2pt\,}
\def\Re{{\rm Re}\,}\def\Im{{\rm Im}\,}
\def\lis{\overline}

\def\indica{\leaders \hbox to 0.5cm{\hss.\hss}\hfill}
\def\guida{\leaders\hbox to 1em{\hss.\hss}\hfill}
\mathchardef\oo= "0521
 
\def\pp{{\bf p}}\def\xx{{\bf x}}
\def\yy{{\bf y}}\def\kk{{\bf k}}
\def\zz{{\bf z}}\def\uu{{\bf u}}
 \def\bP{{\bf P}}
\def\tt{{\bf t}}\def\bT{{\bf T}}

  \def\grad{{\vec\nabla}} 
\def\vpp{{\vec p}}\def\vxx{{\vec x}}
\def\vkk{{\vec k}}\def\vnn{{\vec n}}

\def\vrr{{\vec r}}\def\vee{{\vec e}}

\def\Halmos{\hfill\vrule height6pt width4pt depth2pt \par\hbox to \hsize{}}
\def\virg{\quad,\quad}
\def\proof{\0{\cs Proof - } }

\def\oo{{\underline \omega}}
 
 \def\vxxx{{\underline\vxx}}
 \def\vkkk{{\underline\vkk}}

\def\qed{\raise1pt\hbox{\vrule height5pt width5pt depth0pt}}

\def\indic{\hbox{\raise-2pt \hbox{\indbf 1}}}

\def\openone{\leavevmode\hbox{\ninerm 1\kern-3.3pt\tenrm1}}%
\def\RRR{\hbox{\msytw R}}

 \def\ZZZ{\hbox{\msytw Z}}
 
\def\TTT{\hbox{\msytw T}}

\def\FFF{\hbox{\euftw F}}

\def\defin{{\buildrel def\over=}}
\def\Tr{{\rm Tr}}
\def\ud{\uparrow\downarrow}

%%% INSERIMENTO FIGURE ( se si usa DVIPS )
%
% Se si vuole utilizzare delle macro postscript personali, contenute
% nel file ini.ps, togliere il commento alla riga seguente
%\special{header=ini.pst}
%
% Il comando seguente inserisce una scatola contenente #3 in modo che
% l'angolo superiore sinistro occupi la posizione (#1,#2)
%
\def\ins#1#2#3{\vbox to0pt{\kern-#2 \hbox{\kern#1 #3}\vss}\nointerlineskip}
%
% Il comando seguente crea una scatola di dimensioni #1x#2 contenente
% il disegno descritto in #4.ps;
% in questo disegno si possono introdurre delle stringhe usando \ins
% e mettendo le istruzioni relative nell'argomento #3.
% Il file #4.ps contiene le istruzioni postscript, che devono essere scritte
% presupponendo che l'origine sia nell'angolo inferiore sinistro della
% scatola, mentre per il resto l'ambiente grafico e' quello standard.
% #5 deve essere della forma \eqg("nome simbolico").
%
% Le istruzioni postscript possono essere inserite nel file che contiene
% l'istruzione \insertplot, racchiudendole fra le istruzioni \initfig{#4}
% e \endfig; inoltre ogni riga deve cominciare con "write13<" e deve finire
% con ">". In questo modo si crea il file #4.ps relativo alla figura.
%
\newdimen\xshift \newdimen\xwidth \newdimen\yshift
 
\def\insertplot#1#2#3#4#5{\par%
\xwidth=#1 \xshift=\hsize \advance\xshift by-\xwidth \divide\xshift by 2%
\yshift=#2 \divide\yshift by 2%
\line{\hskip\xshift \vbox to #2{\vfil%
#3 \includegraphics{#4.pst}}\hfill \raise\yshift\hbox{#5}
}} 
%\raise\yshift\hbox{#5}}}

\openin14=\jobname.aux \ifeof14 \relax \else
\input \jobname.aux \closein14 \fi
\openout15=\jobname.aux

\def\ottopunti{\def\rm{\fam0\ottorm}%
\textfont0=\ottorm\scriptfont0=\sixrm\scriptscriptfont0=\fiverm%
\textfont1=\ottoi\scriptfont1=\sixi\scriptscriptfont1=\fivei%
\textfont2=\ottosy\scriptfont2=\sixsy\scriptscriptfont2=\fivesy%
\textfont3=\tenex\scriptfont3=\tenex\scriptscriptfont3=\tenex%
\textfont4=\ottocss\scriptfont4=\sc\scriptscriptfont4=\sc%
%\scriptfont4=\ottocss\scriptscriptfont4=\ottocss%
\textfont5=\tenmib\scriptfont5=\sevenmib\scriptscriptfont5=\fivei
\textfont\itfam=\ottoit\def\it{\fam\itfam\ottoit}%
\textfont\slfam=\ottosl\def\sl{\fam\slfam\ottosl}%
\textfont\ttfam=\ottott\def\tt{\fam\ttfam\ottott}%
\textfont\bffam=\ottobf\scriptfont\bffam=\sixbf%
\scriptscriptfont\bffam=\fivebf\def\bf{\fam\bffam\ottobf}%
%\tt\ttglue=.5em plus.25em minus.15em%
\setbox\strutbox=\hbox{\vrule height7pt depth2pt width0pt}%
\normalbaselineskip=9pt\let\sc=\sixrm\normalbaselines\rm}

\def\wt{\widetilde}
\def\ul{\underline}

\centerline{\titolone{Fermi liquid behavior
in the 2D Hubbard model}}
\centerline{\titolone{ at low temperatures}}
\vskip.5cm

\centerline{\titolo{G. Benfatto, A. Giuliani, V. Mastropietro}}
\vskip.5cm

\centerline{Dipartimento di Matematica, Universit\`a di Roma ``Tor
Vergata''}

\centerline{Viale della Ricerca Scientifica, 00133, Roma, Italy}
\vskip1cm

\0{\cs Abstract.} {\it We prove that the weak coupling 2D
Hubbard model away from half filling is a Landau Fermi liquid up
to exponentially small temperatures. In particular we show that
the wave function renormalization is an order 1 constant 
and essentially temperature independent in the considered range of 
temperatures and that the interacting Fermi surface is a regular convex curve.
This result is obtained by deriving a convergent expansion
(which is not a power series) for the two point Schwinger function
by Renormalization Group methods and proving at each order
suitable power counting improvements due to the convexity of the
interacting Fermi surface. Convergence follows from determinant
bounds for the fermionic expectations. } \vskip1cm

\section(1, Introduction and Main result)

\sub(1.1) {\it Motivations}

The Hubbard model, according to many (see for instance [L]), has
the same role in the problem of electron correlations as the Ising
model in the problem of spin-spin correlations, that is it is the
simplest possible model displaying many "real world" features. Still, 
we are far from a rigorous understanding of many of its properties,
except in 1D, where Bethe ansatz arguments [LW, G] and
Renormalization Group (RG) analysis [S, M] provide a quite good
understanding of many of its properties up to zero temperature.
The 2D case is actually considered physically the most
interesting one, as it is believed to give some insight on certain
properties of high $T_c$ superconductors. There is no agreement,
even at a qualitative level, on the behavior of the 2D Hubbard
model at zero temperature, where a rather complex behavior is
expected [Me]. On the
other hand, even at finite temperature (above the expected phase
transition) there are several non trivial questions to be
answered, the most important of which is about the Fermi or {\it
non Fermi} liquid behavior of the 2D Hubbard model. The presence
of Fermi liquid behavior is crucially related to the convexity of
the Fermi surface, as it appears by lowest order computations [VR]. 
In particular the 2D Hubbard model is believed to be a
Fermi liquid for densities sufficiently smaller than the half--filling 
density, \ie for 
values of the chemical potential corresponding to a closed and convex 
free Fermi surface. It is well
known however that the interaction produces a {\it deformation} of
the Fermi surface [HM] and it is not guaranteed a priori that the
interacting Fermi surface, if existing at all, is still convex. 
Moreover doubts on the
reliability of lowest order approximations, due to a possible non
convergence of the series expansion, has been raised recently in
the physical literature, see [A], in the debate about the
explanation of high $T_c$ superconductivity.

Our main result can be informally stated in the following way 
(see Theorem \thm(1.1) for a formal statement, the comment 
after the Theorem for the definition of Fermi liquid 
and \equ(1.7) for the definition of Fermi surface at finite temperature).\\
\\
{\it For values of the chemical potential smaller than and not too close 
to $\m=2$ (the half filled band case) the weak coupling 2D Hubbard model 
is a Fermi liquid up to exponentially small temperatures, independently 
on the sign of the interaction. In particular the wave function 
renormalization is essentially temperature independent and the interacting 
Fermi surface is a regular convex surface.}\\
\\
In [R, AMR] it has been proved that in the half-filled
band case the weak coupling 2D Hubbard model is {\it not} a Fermi liquid 
for temperatures larger than an exponentially small one.
Then, combining our results with those in [R], we find
that the 2D Hubbard model for temperatures larger than an exponentially 
small one shows a crossover between a Fermi and a non Fermi liquid
behavior, while varying the chemical potential from $0$ to $2$.

Our main result is proved by showing convergence of a suitable 
resummation of the weak coupling expansion for the interacting two--point
correlation function,
needed to take care of the modification of the Fermi surface due to the 
interaction. In fact the deformation of the Fermi surface has
the effect that the naive power series expansion in terms of the free 
propagator is not convergent at low temperatures. We have then to employ 
RG methods so that the interacting propagator is dynamically changed 
(``dressed'') at
each step of the multiscale analysis: correspondingly the location of the 
singularity of the 
interacting propagator will define an interacting Fermi surface which 
is dynamically modified at each RG step.
Convergence of the resummed series follows from power counting 
improvements at all orders due to the convexity of the
interacting Fermi surface and from determinant
bounds for the fermionic expectations.

\* \sub(1.2) {\it The Hubbard model.}

The Hamiltonian of the 2D Hubbard model is given by
$$\eqalign{&
H=\sum_{\vec x\in \L}\sum_{\s=\uparrow\downarrow} a^+_{\vec
x,\s}\big(-{\D\over 2}-\m\big) a_{\vec x,\s}^- +U\sum_{\vec x\in
\L} a^+_{x,\uparrow} a^-_{\vec x,\uparrow} a^+_{\vec x,\downarrow}
a^-_{\vec x,\downarrow} \cr}\Eq(1.1)$$
where:
\item{1)} $\L\subset\ZZZ^2$ is a square sublattice of
$\ZZZ^2$ with side $L$ (the sites will be labelled by
$(n_1,n_2)\in\ZZZ^2$, $-[L/2]\le n_1,n_2\le [(L-1)/2]$);
\item{2)} $a_{\vec x,\s}^\pm$ are creation or annihilation
fermionic operators with spin index $\s=\uparrow\downarrow$ and
site index $\vec x\in\L$, satisfying periodic boundary conditions
in $\vec x$;
\item{3)} $\D$ is the discrete Laplacean, acting on a function
$f:\ZZZ^2\to\RRR$, periodic of period $L$ in both directions, as:
$\D f(\vec x)=\sum_{j=1,2}f(\vec x+\hat e_j)-2f(\vec x)+f(\vec
x-\hat e_j)$, where $\hat e_j$, $j=1,2$, are the two unit versors
on $\ZZZ^2$;
\item{4)} $\m>0$ is the chemical potential, whose value fixes the
average density of particles; \item{5)} $U$ is the strength of the
on--site density--density interaction; it can be either positive
or negative.
\\

We shall also consider the operators $a^\pm_{\xx,\s}= e^{H
x_0}a^\pm_{\vec x,\s} e^{-H x_0}$ with $\xx=(x_0,\vec x)$ and
$x_0\in[0,\b]$, for some $\b>0$; we shall call $x_0$ the
time variable. The finite temperature Schwinger functions are
defined as
$$S(\xx_1,\s_1,\e_1;\ldots;\xx_n,\s_n,\e_n)={{\rm Tr}\,e^{-\b H}\bT(
a^{\e_1}_{\xx_1,\s_1}\cdots a^{\e_n}_{\xx_n,\s_n}) \over {\rm
Tr}e^{-\b H}}\Eq(1.2)$$
where $\xx_i\in[0,\b]\times\L$, $\s_i=\uparrow\downarrow$,
$\e_i=\pm$ and $\bT$ is the operator of time ordering, acting on a
product of fermionic fields as:
$$\bT(
a^{\e_1}_{\xx_1,\s_1}\cdots a^{\e_n}_{\xx_n,\s_n})= (-1)^\p
a^{\e_{\p(1)}}_{\xx_{\p(1)},\s_{\p(1)}}\cdots a^{\e_{\p(n)}}_{
\xx_{\p(n)},\s_{\p(n)}}\Eq(1.3)$$
where $\p$ is a permutation of $\{1,\ldots,n\}$, chosen in such a
way that $x_{\p(1)0}\ge\cdots\ge x_{\p(n)0}$, and $(-1)^\p$ is its
sign\footnote{${}^1$}{\ottorm If some of the time coordinates are
equal each other, the arbitrariness of the definition is solved by
ordering each set of operators with the same time coordinate so
that creation operators precede the annihilation operators.}.

In the non--interacting case $U=0$ the Schwinger functions of any
order $n$ can be exactly computed as linear combinations of
products of two--point Schwinger functions (via the well--known
{\it Wick rule}). The two--point Schwinger function itself (also
called the {\it free propagator}) is equal to:
$$S_0(\xx-\yy) \= S(\xx,\s,-;\yy,\s,+)\Big|_{U=0}= {1\over \b L^2}
\sum_{\kk\in\DD_{\b,L}} {e^{-i\kk\cdot(\xx-\yy)}\over -i
k_0+\e_0(\vec k)-\m}\Eq(1.4)$$
where:
\item{(a)} $\kk=(k_0,\vec k)$ and $\DD_{\b,L}=\DD_\b\times\DD_L$;
\item{(b)} $\DD_\b=\{k_0={2\pi\over \b}(n_0+{1\over 2})\;:\;n_0\in\ZZZ\}$ and
$\DD_L=\{\vec k={2\pi\over L}(n_1,n_2)\;:\;-[L/2]\le n_1,n_2\le
[(L-1)/2]\}$;
\item{(c)} $\e_0(\vec k)=2-\cos k_1-\cos k_2$ is the {\it
dispersion relation}.
\\

Note that $S_0(\xx)$ is a function of $x_0\in\RRR$ antiperiodic of
period $\b$ and that its Fourier transform $\hat S_0(\kk)$ is
well--defined for any $\kk\in\DD_{\b,L}$, even in the
thermodynamic limit $L\to\io$, since $|k_0|\ge {\pi\over \b}$. We
shall refer to this last property by saying  that the inverse
temperature $\b$ acts as an infrared cutoff for our theory.

In the limit $\b,L\to\io$ the propagator $\hat S_0(\kk)$ becomes
singular on the surface $\{k_0=0\}\times \Sigma_F^{(0)}$, where
$\Sigma_F^{(0)}\= \{\vec k\in [-\p,+\p]\times [-\p,+\p] \;:\;
\e_0(\vec k)-\m=0\}$ is the free {\it Fermi surface}. It is easy
to realize that, if $0<\m<2$, then $\Sigma_F^{(0)}$ is a smooth
convex closed curve, symmetric around the point $\vec k=(0,0)$. It
can be parameterized as $\vec k = \vec p_F^{(0)}(\th)$ in terms of
the polar angle $\th\in[0,2\p]$. We shall also denote $|\vec
p_F^{(0)}(\th)|$ by $u^{(0)}(\th)$.

In order to make apparent the structure of the pole singularity of
$\hat S_0(\kk)$ at $\{0\}\times \Sigma_F^{(0)}$, it is sometimes
convenient to rewrite $\hat S_0(\kk)$ in the form:
$$\hat S_0(\kk)={1\over Z_0}\; {1\over -i k_0+\vec v_F^{(0)}(\th)\cdot\big(
\vec k-\vec p_F^{(0)}(\th)\big)+R(\vec k)}\Eq(1.5)$$
where $\th$ is the polar angle of $\vec k$, $Z_0=1$ is the {\it
free wave function renormalization} and $\vec
v_F^{(0)}(\th)=(\partial\e_0/\dpr\vec k)\big|_{\vec k=\vec
p_F(\th)}$ is the {\it free Fermi velocity}. Moreover, near the
Fermi surface, $|R(\vec k)|\le C \big|\vec k-\vec
p_F^{(0)}(\th)\big|^2$, for some positive constant $C$.\\

\* \sub(1.3) {\it Main results}

In the interacting case $U\neq 0$, the Schwinger functions are not
exactly computable anymore. It is well--known that they can be
written as power series in $U$, convergent for $|U|\le \e_\b$, for
some constant $\e_\b$. The power expansion in $U$ is constructed
in terms of {\it Feynmann diagrams}, using as free propagator the
function $S_0(\xx)$ in
\equ(1.4). The known bounds on the radius of convergence $\e_\b$
of this power series shrink to zero as $\b\to\io$. For example, a
simple scale decomposition can allow to solve the ultraviolet
problem in the $k_0$ variable, by proceeding as in [GLM], and one
can easily get from the explicit structure of the power series the
estimate $\e_\b=\big[C\b^{\a}\big]^{-1}$, with $C,\a$ two positive
constants. The condition $|U|<\e_\b$, guaranteeing convergence of
the power series for the Schwinger functions, can be equivalently
reexpressed as a condition on the inverse temperature $\b$ at
fixed $U$ and it implies that the power series for the Schwinger
functions is well--defined for $\b^{-1}\ge (C|U|)^{1/\a}$, \ie for
temperatures larger than a ``polynomially small'' temperature. It
is not clear at all that one can give sense (even through suitable
{\it resummations}, needed to take care of the modification of the
Fermi surface) to the power series in $U$ up to $\b^{-1}=0$. In
fact the problem of understanding the behavior of the interacting
Schwinger functions of the 2D Hubbard model up to $\b^{-1}=0$ is
one of the major open problems in condensed matter theory and, in
particular, one expects that the nature of the two--point
Schwinger function singularity drastically changes when the temperature
is lowered up to zero temperature, because of a
superconduction instability. An easier but still very non trivial
problem consists in trying to prove that a suitable resummed
version of the expansion (different from the power expansion) is
convergent for $|U|\le \e_\b$ and in trying to optimize $\e_\b$, in order to be
able to describe the thermodynamic behavior of the system at least up to
temperatures exponentially small in the strength of the
interaction, that is for temperatures $\b^{-1}\ge e^{-a/|U|}$:
in fact the onset of superconductivity is expected to be found at such 
temperatures.\\
In the present paper we will show convergence of a resummed power
series for the two--point Schwinger function for temperatures
$\b^{-1}$ larger than an exponentially small one, $\b^{-1}\ge
e^{-a/|U|}$, in the {\it far from half--filling case}, that is for
$0<\m<\m_0={2-\sqrt2\over 2}$ (\ie for values of the average
density sufficiently smaller than $1$). The condition of smallness
of the chemical potential guarantees certain strong convexity
properties of the free Fermi surface we shall need in the proof
below; moreover such condition suppresses a class of possibly
dangerous contributions to the power series (the so--called
``umklapp'' processes with $n\le 4$ quasi--particles, see below
for definitions) that we are not able to control at the present
time. The interacting two--point function
$S(\xx-\yy)=S(\xx,\s,-;\yy,\s,+)$ turns out to have, in the
$L=\io$ limit, the following structure. Let us call $\hat S(\kk)$
the Fourier transform of $S(\xx)$ and $\Sigma(\kk)$ the {\it
self--energy}, defined as usual by the identity
$$\hat S(\kk)={1\over -i k_0+\e_0(\vec k)-\m+\Sigma(\kk)}\Eq(1.6)$$
We introduce the following definitions
\item{a)} The {\it interacting Fermi surface} $\Sigma_F$
is defined as
$$\Sigma_F=\{\vec k\;:\; \e_0(\vec k)+{1\over 2}
{\rm Re}\sum_{j=\pm}\Sigma\big(j\pi\b^{-1},\vec k\big)
=\mu\}\;,\Eq(1.7)$$
see also Remark (3) in Section \secc(1.4).
We shall be able to parameterize $\Sigma_F$ as $\vec k =
\vec p_F(\th)$ in terms of the polar angle $\th\in[0,2\p]$ and we
shall denote $|\vec p_F(\th)|$ by $u(\th)$.
\item{b)} The {\it wave function renormalization} is
$$Z(\th)=1+i\partial_{k_0} \Sigma\Eq(1.8)$$
where $\partial_{k_0} \Sigma= (\b/2\p) [\Sigma ({\pi\over\b},\vec
p_F(\th)) - \Sigma(-{\pi\over\b},\vec p_F(\th))]$
\* \0c) The {\it Fermi velocity} is
$$\vec v_F(\th)={1\over Z(\th)}
{\partial(\e_0+\Sigma)\over\partial \vec k}\Big|_{\kk=(0,\vec
p_F(\th))}\Eq(1.9)$$
Our main result is the following. \* \theorem(1.1) {\it Let us
consider the 2D Hubbard model with $0<\m<\m_0\={2- \sqrt2\over 2}$
and $\b^{-1}\ge e^{-{a\over |U|}}$ where $a>0$ is a suitable
constant. There exists a constant $U_0>0$ such that, if $|U|\le
U_0$, the two point Schwinger function $\hat S(\kk)$ can be
written, in the limit $L=\io$, as
$$\hat S(\kk)=
{1\over Z(\th)}{1\over -i k_0+\vec v_F(\th)\cdot\big(\vec k-\vec
p_F(\th)\big) +R(\kk)}\Eq(1.10)$$
with $Z(\th), \vec v_F(\th)$ and $\vec p_F(\th)$ real and 
$$\eqalign{&Z(\th)=1+ a(\th) U^2+O(U^3) \cr
& \vec v_F(\th)=\vec v_F^{(0)}(\th)+\vec b(\th) U+O(U^2)\cr & \vec
p_F(\th)=\vec p_F^{(0)}(\th)+\vec c(\th) U +O(U^2)\cr}\Eq(1.11)$$
where $a(\th),|\vec b(\th)|,|\vec c(\th)|$ are bounded above and
below by positive $O(1)$ constants in the region $\b^{-1}\ge
e^{-{a\over |U|}}$. Moreover
$$|R(\kk)|\le C \big[|\vec k-\vec p_F(\th)|^2+k_0^2+|\vec k-\vec p_F(\th)|
|k_0|\big]\Eq(1.12)$$
for some constant $C>0$. } \* Let us compare the representation
\equ(1.10) of the interacting two--point Schwinger function with
the free one, given by
\equ(1.5). They are apparently similar but the parameters
$Z(\th)$, $\vec v_F(\th)$ and
$\vec p_F(\th)$, differently from $Z_0$, $\vec v_F^{(0)}(\th)$ and
$\vec p_F^{(0)}(\th)$, are functions of the temperature $\b^{-1}$,
for $\b^{-1}\ge e^{-a/|U|}$. However such dependence can be
stronger or weaker and the different sensitivity to a variation of
the temperature has important physical consequences. In the case of the 2D
Hubbard model \equ(1.1) with $\m<\m_0$ we prove that
$Z(\th)$, $\vec v_F(\th)$ and $\vec p_F(\th)$ are slowly depending
on $\b$ for $\b^{-1}\ge e^{-a/|U|}$, that is they are essentially
constant in $\b$ above an exponentially small temperature. 
This means that, in the considered range of
parameters, the interacting two--point correlation is essentially
identical to the free one, {\it up to a renormalization of the
parameters essentially independent on the temperature}; in this sense
we say that the system shows a {\it Fermi liquid behavior} for temperatures 
larger than an exponentially small one. We think that this notion of 
Fermi liquid 
is the natural mathematical interpretation of the notion of Fermi liquid 
often used in the theoretical physics literature, and it is essentially 
the same as the one adopted for instance in [Sa1, DR].

Of course the property to be a Fermi liquid (in the above sense) is not 
trivial at all
and it is not verified in many cases.
For instance, in the 1D
Hubbard model, the wave function renormalization $Z$ depends logarithmically 
on $\b$, that is $c_1
U^2\log\b\le |Z-1|\le c_2 U^2\log\b$, with $c_1,c_2$ two positive constants,
for temperatures above an exponentially small temperature; so,
with our definition, the 1D Hubbard model is not a Fermi liquid
in such range of temperatures. In the 2D Hubbard model at
half--filling (\ie at $\m=2$) it has been recently proved [R, AMR]
that, for temperatures above an exponentially small temperature,
$c_1 U^2\log^2\b\le |Z-1|\le c_2 U^2\log^2\b$ , so that the system
is not a Fermi liquid at half--filling in that range of
temperatures. On the contrary, an example of Fermi liquid in the
above sense is provided by the continuum approximation of model
\equ(1.1) in $d=2$, the so--called {\it jellium model}, for which
[DR] showed that, in a range of temperatures above an
exponentially small temperature, $c_1 U^2\le |Z-1|\le c_2 U^2$,
and the system is a Fermi liquid. Note that in the jellium model,
due to rotation invariance, the interacting and the free Fermi
surfaces have exactly the same shape, that is a circle, and the
effect of the interaction essentially consists just in changing
its radius.
\* \sub(1.4) {\it Remarks}
\item{1)} The constant $a$ appearing in the bound $\b^{-1}\ge
e^{-{a\over |U|}}$ is a constant depending only on the second
order contributions in the perturbation expansion for $\hat
S(\kk)$ and it can be explicitly computed through our
construction.
\item{2)}
Our result is not uniform in the chemical potential and the proof
we present holds only for $\m<\m_0<2$, where $\m=2$ corresponds to
half--filling. We expect that our condition $\m<\m_0$ is technical
and it is not unlikely that our proof could be extended to any
$\m<2$, under the smallness condition $|U|\le U_0(\m)$, with
$\lim_{\m\to 2^-}U_0(\m) =0$. Our result, combined with the result
of [R, AMR] discussed above, implies that
the 2D Hubbard model shows a transition from Fermi to non--Fermi
liquid behavior, in the above range of temperatures, depending on the choice
of $\m$. It would be
very interesting to explicitly investigate the crossover between
the two regimes.
\item{3)} At non zero
temperature there is an ambiguity in the definition of the Fermi surface
and, in order to resolve this ambiguity, we made the specific choice
\equ(1.7). It will be clear from the proof
below that any ``reasonable'' definition of the Fermi surface will have 
the same regularity properties as those of the one in \equ(1.7).
The claim in Theorem \thm(1.1) implies of course that the interacting 
Fermi surface is a regular and convex curve, uniformly in $\b$ in the allowed
range of temperatures. 

\item{4)} The Theorem above is proved by an iterative resummation
of the original power series expansion for the two--point
correlation function and by a renormalization of the free measure,
which takes into account, in particular, the modification of Fermi
surface. This allows us to reexpress iteratively the original
power series in $U$ as an expansion in an increasing number of
parameters (they are indeed functions), called the ``effective
couplings'' and physically describing the effective interaction at
different momentum scales, denoted by $(\l_1\=U, \l_0,
\l_{-1},\l_{-2},\ldots)$; moreover, the coefficients of the new
series are themselves depending on $U$ trough the renormalized
single scale propagators. The new series will be well--defined
whenever $\bar U=\max_{h\le 1}|\l_{h}|$ will be smaller than
$U_0$, where $U_0$ is a constant {\it independent of the
temperature}. From the physical point of view, this means that the
temperature dependence at all orders in the expansion for $\hat
S(\kk)$ is essentially all included in the effective couplings,
whose size in turn will depend strongly on the temperature. The
possibility of resumming the perturbative expansion in a new
series with the above properties was conjectured in [GS], where
the authors proved the claim for the second order contributions.
We stress that the possibility of resumming the series into a new
series admitting this kind of ``uniform bounds'' is specific of
$d=2$ far from half--filling; for instance in $d=1$ the
coefficient at order $2n\ge 2$ of the resummed expansion for
$Z(\th)$ behaves like $(\bar U/U_0)^{2n}(\log\b)^{n}$, instead of
$(\bar U/U_0)^{2n}$, {\it even assuming that the effective
interactions are bounded}. This is not the case in $d=2$ far from
half--filling; in this case the breaking of Fermi liquid behavior
can be due only to some instability occurring in the effective
interactions.
\item{5)} Our result should be compared to [DR],
in which a proof of Fermi liquid behavior was given for the
jellium model. We have taken from such papers two crucial
technical ingredients: the idea of using anisotropic sectors (and
the relative {\it sector lemma} of [FMRT]) for the bounds and the
idea of further decomposing some sector into isotropic sectors in
order to improve the bounds for the self energy; note however that
the technical implementation of such ideas in the proofs is rather
different with respect to [DR], mainly for the heavy use of
{\it trees} for reorganizing the perturbative series 
and for the fact that we do not need neither a ``1PI analysis'' 
nor an ``arch expansion'' to extract our power counting improvements
needed to prove the Theorem. Moreover, the presence of a non circular Fermi
surface causes many new technical problems with respect to the
case in [DR]. The more important one is that, while in the Jellium
case the interacting Fermi surface is fixed {\it a priori} to be a
{\it circle} as consequence of rotational symmetry, here on the
contrary the shape and the regularity or convexity properties 
of the interacting Fermi surface are completely
unknown: in fact there is no apriori evidence of the fact that the interacting 
Fermi surface is regular and convex uniformly in $\b$ in the considered 
range of temperatures. Hence we cannot in our case {\it fix} the interacting
Fermi surface by properly tuning the chemical potential, as it is
done in [DR]; on the contrary, we proceed as in [BM], by
inserting at each integration step all the quadratic part of the
interaction in the free fermionic measure. In this way to each
fermionic integration at a certain momentum scale corresponds a
different Fermi surface, and one has to check that the geometrical
conditions for defining sectors and to apply the sector lemma are
verified at each scale.
\item{6)} In [BGM] a statement similar to Theorem \thm(1.1) above was proved.
However in [BGM] the interacting Fermi surface was fixed by adding a 
suitable counterterm to the free Hamiltonian. The result of Theorem \thm(1.1)
cannot be recovered simply by the results of [BGM], because the inversion
of the counterterm must be discussed. This is an highly non trivial
problem, that is essentially solved below, by checking that 
the modified Fermi surface satisfies the same convexity properties of the 
free one, at each RG step (the choice of
dynamically changing the Fermi surface instead of inverting the counterterm 
is not substantial and the two approaches are essentially equivalent).
The inversion problem was discussed perturbatively in the series of papers 
[FST][Sa2]. 
In [PS] it was announced that the inversion problem was solved even 
at a non perturbative level; a proof is however not yet published. 
\item{7)} If the temperature is low enough, it is expected that Fermi
liquid behavior breaks down, as a consequence of quantum
instabilities present in the systems; the breakdown of Fermi
liquid behavior is expected both for attractive and repulsive
interaction, as a consequence of the Kohn-Luttinger argument [KL].
It is possible to destroy such instabilities by choosing properly
an highly {\it non symmetric} dispersion relation, for instance by
introducing an external magnetic field; indeed for such a system
in [FKT] a proof of Fermi liquid behavior was indeed given up to
{\it zero temperature} in 2D. We stress that also in [FKT] the
interacting Fermi surface is not fixed by the symmetries and a
counterterm is introduced to fix the interacting Fermi surface;
however, the "inversion problem", which we solve for the Hubbard
problem in the present paper, is still an open problem in that
case.\\

\section(2, Grassman integral and Renormalization Group analysis)

\sub(2.1) {\it Grassmann representation.}

It is well--known that the usual formal power series in $U$ for
the partition function and for the Schwinger functions of model
\equ(1.1) can be equivalently rewritten in terms of Grassmann
functional integrals, defined as follows.

First of all we introduce an ultraviolet cutoff on $k_0$, by
restricting the set $\DD_\b$ of its possible values to the set
$\{k_0={2\pi\over \b}(n_0+{1\over 2}):|n_0|\le M\}$, which we
shall denote with the same symbol, as well as the set
$\DD_{\b,L}$, which is a finite set for each $M$. Given $M$, we
consider the Grassmann algebra generated by the Grassmannian
variables $\{\hat\psi^\pm_{\kk,\s}\}_{ \kk \in
\DD_{\b,L}}^{\s=\uparrow\downarrow}$ and a Grassmann integration
$\int \big[\prod_{\kk\in\DD_{\b,L}}^{\s=\ud} d\hat\psi_{\kk,\s}^+
d\hat\psi_{\kk,\s}^-\big]$ defined as the linear operator on the
Grassmann algebra such that, given a monomial $Q( \hat\psi^-,
\hat\psi^+)$ in the variables $\hat\psi^\pm_{\kk,\s}$, its action
on $Q(\hat\psi^-, \hat\psi^+)$ is $0$ except in the case
$Q(\hat\psi^-, \hat\psi^+)=\prod_{\kk\in\DD_{\b,L}}^{\s=\ud}
\hat\psi^-_{\kk,\s} \hat\psi^+_{\kk,\s}$, up to a permutation of
the variables. In this case the value of the integral is
determined, by using the anticommuting properties of the
variables, by the condition
$$\int \Big[\prod_{\kk\in\DD_{\b,L}}^{
\s=\ud} d\hat\psi_{\kk,\s}^+
d\hat\psi_{\kk,\s}^-\Big]\prod_{\kk\in\DD_{\b,L}}^{\s=\ud}
\hat\psi^-_{\kk,\s} \hat\psi^+_{\kk,\s}=1\Eq(2.1)$$
Defining the free propagator $\hat g_\kk$ as $\hat
g_\kk=\big[-ik_0+ \e_0(\vec k)-\mu\big]^{-1}$ and the ``Gaussian
integration'' $P(d\psi)$ as
$$P(d\psi) = \Big[\prod_{\kk\in\DD_{\b,L}
}^{\s=\ud}(L^2\b\hat g_\kk) d\hat\psi_{\kk,\s}^+
d\hat\psi_{\kk,\s}^-\Big]\cdot\;\exp \Big\{-
\sum_{\kk\in\DD_{\b,L}}^{\s=\ud}\big(L^2\b\hat g_\kk\big)^{-1}
\hat\psi^{+}_{\kk,\s}\hat\psi^{-}_{\kk,\s}\Big\}\;, \Eq(2.2)$$
it holds that
$$\int P(d \psi) \hat \psi^-_{\kk_1,\s_1}\hat \psi^+_{\kk_2,\s_2} =
L^2\b\d_{\s_1,\s_2}\d_{\kk_1,\kk_2} \hat g_{\kk_1}\;,\Eq(2.3)$$
so that
$$\lim_{M\to\io} \fra{1}{L^2\b}\sum_{\kk\in\DD_{\b,L}}
e^{-i\kk(\xx-\yy)}\hat g_{\kk}= \lim_{M\to\io} \int
P(d\psi)\psi^-_{\xx}\psi^+_\yy= S_0(\xx-\yy)\Eq(2.4)$$
where $S_0(\xx-\yy)$ was defined in \equ(1.4) and the Grassmann
fields $\psi_{\xx,\s}^\pm$ are defined by
$$\psi_{\xx,\s}^{\pm}=\fra{1}{L^2\b}\sum_{\kk\in\DD_{\b,L}}
e^{\pm i\kk\xx}\hat\psi^\pm_{\kk,\s}\Eq(2.5)$$

Let us now consider the function on the Grassmann algebra
$$V(\psi)=U\int d\xx \, \psi^+_{\xx,\uparrow}
\psi^-_{\xx,\uparrow} \psi^+_{\xx,\downarrow}
\psi^-_{\xx,\downarrow}\Eq(2.6a)$$
where the symbol $\int d\xx$ must be interpreted as
$$\int d\xx=\int_0^\b dx_0\sum_{\vec x\in\L}\Eq(2.7)$$
and note that the integral $\int P(d\psi)e^{-V(\psi)}$ is well
defined for any $U$; it is indeed a polynomial in $U$, of degree
depending on $M$ and $L$. Standard arguments (see, for example
[NO], where a different cutoff on $k_0$ is used) show that, if
there exists the limit of $\int P(d\psi)e^{-V(\psi)}$ as
$M\to\io$, then the normalized partition function can be written
as
$$e^{-L^2\b F_{L,\b}}\defin\fra{\Tr[e^{-\b H}]}{\Tr[e^{-\b H_0}]}=
\lim_{M\to\io} \int P(d\psi)e^{-V(\psi)}\Eq(2.6)$$
where $H_0$ is equal to \equ(1.1) with $U=0$.

Similarly, the Schwinger functions defined in \equ(1.2) can be
computed as
$$S(\xx_1,\s_1,\e_1;\ldots;\xx_n,\s_n,\e_n)= \lim_{M\to\io} {
\int P(d\psi) e^{-V(\psi)}
\psi^{\e_1}_{\xx_1,\s_1}\cdots\psi^{\e_n}_{\xx_n,\s_n} \over \int
P(d\psi) e^{-V(\psi)}}\;.\Eq(2.8)$$
In the following we shall study the functional integrals by
introducing suitable expansions where the value of $M$ has no role
and we shall indeed be able to control such expansions {\it
uniformly in $M$, if $U$ is small enough}. Our procedure also
implies that we can effectively take the limit $M\to\io$
everywhere, so we shall not stress anymore the dependence on
$M$ and we shall proceed as if $M=\io$.\\

Note that both the Gaussian integration $P(d\psi)$ and the
interaction $V(\psi)$ are invariant under the action of the
following symmetry transformations:
\item{(1)} {\it spin exchange}: $\psi^\e_{\xx,\uparrow}\otto\psi^\e_{\xx,
\downarrow}$;
\item{(2)} {\it global}
$U(1)$: $\psi^\e_{\xx,\s}\to e^{i\e\a_\s}\psi^\e_{\xx,\s}$, with
$\a_\s\in \RRR$ independent of $\xx$;
\item{(3)} {\it global} $SO(2)$:
$\pmatrix{\psi^\e_{\xx,\uparrow}\cr
\psi^\e_{\xx,\downarrow}\cr}\to\pmatrix{\cos\th &\sin\th\cr
-\sin\th &\cos\th\cr}\pmatrix{\psi^\e_{\xx,\uparrow}\cr
\psi^\e_{\xx,\downarrow}\cr}$, with $\th\in\RRR$ independent of
$\xx$;
\item{(4)} {\it parity}: $\psi^\pm_{(x_0,\vxx),\s}\to
\psi^\pm_{(x_0,-\vxx),\s}$;
\item{(5)} {\it complex conjugation}: $\psi^{\pm}_{(x_0,\vxx),\s}
\to \psi^\pm_{(-x_0,\vxx),\s}$, $c\to c^*$, where $c$ is a generic
constant appearing in the formal action.

\* \sub(2.2) {\it The ultraviolet integration.}

It is convenient, for clarity reasons, to start by studying the
free energy $F_{L,\b}$, defined by \equ(2.6). Note that our
lattice model has an intrinsic ultraviolet cut-off in the $\vkk$
variables, while the $k_0$ variable is unbounded. A preliminary
step to our infrared analysis is the integration of the
ultraviolet degrees of freedom corresponding to the large values
of $k_0$. We proceed in the following way. We decompose the free
propagator $\hat g_\kk$ into a sum of two propagators supported in
the regions of $k_0$ ``large'' and ``small'', respectively. The
regions of $k_0$ large and small are defined in terms of a smooth
support function $H_0(t)$, $t\in\RRR$, such that
$$H_0(t) = \cases{
1 & if $t <e_0/\g \;,$ \cr 0 & if $t >e_0\;,$\cr}\Eq(2.9)$$
with $\g=4$ and $e_0$ a parameter to be fixed below. We define
$\c(\kk)= H_0\Big(\sqrt{k_0^2+[\e_0(\vkk)-\m]^2}\Big)$ and
$f_1(\kk)=1-\c(\kk)$, so that we can rewrite $\hat g_\kk$ as:
$$\hat g_\kk=f_1(\kk)\hat g_\kk+\c(\kk)\hat g_\kk\defin
\hat g^{(+1)}(\kk)+\hat g^{(\le 0)}(\kk)\Eq(2.10)$$
We now introduce two independent sets of Grassmann fields
$\{\psi^{(+1)\pm}_{\kk,\s}\}$ and $\{\psi^{(\le
0)\pm}_{\kk,\s}\}$, $\kk\in\DD_{\b,L}$, $\s=\ud$, and the Gaussian
integrations $P(d\psi^{(+1)})$ and $P(d\psi^{(\le 0)})$, defined
by
$$\eqalign{&
\int P(d \psi^{(+1)}) \hat \psi^{(+1)-}_{\kk_1,\s_1}\hat
\psi^{(+1)+}_{ \kk_2,\s_2} = L^2\b\d_{\s_1,\s_2}\d_{\kk_1,\kk_2}
\hat g^{(+1)}(\kk_1)\;,\cr &\int P(d \psi^{(\le 0)}) \hat
\psi^{(\le 0)-}_{\kk_1,\s_1} \hat \psi^{(\le 0)+}_{ \kk_2,\s_2} =
L^2\b\d_{\s_1,\s_2}\d_{\kk_1,\kk_2} \hat g^{(\le 0)}(\kk_1)
\;.\cr}\Eq(2.11)$$
Similarly to $P(d\psi)$, the Gaussian integrations $P(d
\psi^{(+1)})$, $P(d \psi^{(\le 0)})$ also admit an explicit
representation analogous to \equ(2.2), with $\hat g_\kk$ replaced
by $\hat g^{(+1)}(\kk)$ or $\hat g^{(\le 0)}(\kk)$ and the sum
over $\kk$ restricted to the values in the support of $\c(\kk)$ or
$f_1(\kk)$, respectively. The definition of Grassmann integration
implies the following identity (``addition principle''):
$$\int P(d\psi)e^{-V(\psi)}=\int P(d\psi^{(\le 0)})\int P(d\psi^{(+1)})
e^{-V(\psi^{(\le 0)}+\psi^{(+1)})}\Eq(2.12)$$
so that we can rewrite $F_{L,\b}$ as
$$\eqalign{e^{-L^2\b F_{L,\b}}&=
\int P(d\psi^{(\le 0)})\exp\,\big\{\, \sum_{n\ge
1}\fra{1}{n!}\EE^T_1(-V(\psi^{(\le 0)}+\cdot);n)\big\}\=\cr &\=
e^{-L^2\b F_0}\int P(d\psi^{(\le 0)}) e^{-\VV^{(0)}(\psi^{(\le
0)})}\;,\cr}\Eq(2.13)$$
where the {\it truncated expectation} $\EE^T_1$ is defined, given
any polynomial $V_1(\psi^{(+1)})$ with coefficient depending on
$\psi^{(\le 0)}$, as
$$\EE^T_1(V_1(\cdot);n)=\fra{\dpr^n}{\dpr\l^n}
\log\int P(d\psi^{(+1)})e^{\l
V_1(\psi^{(+1)})}\Big|_{\l=0}\Eq(2.14)$$
and $\VV^{(0)}$ is fixed by the condition $\VV^{(0)}(0)=0$. It can
be shown (see Appendix A1) that $\VV^{(0)}$ can be written as
$$\VV^{(0)}(\psi^{(\le 0)})=\sum_{l=1}^\io\sum_{\s_1,\ldots,\s_l=\ud}
\int d\xx_1\cdots d\xx_{2l} \left[\prod_{i=1}^{l} \psi^{(\le
0)+}_{\xx_{2i-1},\s_i}\psi^{(\le
0)-}_{\xx_{2i},\s_i}\right]W^{(0)}_{2l}(\xx_1,\ldots,\xx_{2l})\Eq(2.15)$$
where the integrations $\int d\xx_i$ must be interpreted as in
\equ(2.7). The possibility of representing $\VV^{(0)}$ in the form
\equ(2.15), with the {\it kernels} $W^{(0)}_{2l}$ independent of the spin
indices $\s_i$, follows from the symmetries listed above, after
\equ(2.8). For each fixed $M$, the kernels $W^{(0)}_{2l}$ vanish for $l$
large enough; moreover they are translation invariant and are
given, in the limit $M=+\io$, by power series in $U$, convergent
under the condition $|U|\le U_0$, for $U_0$ small enough; finally,
if $|U|\le U_0$, they satisfy the bounds:
$$\int d\xx_1\cdots d\xx_{2l}\Big[\prod_{1\le i<j\le 2l}|\xx_i-\xx_j|^{m_{ij}}
\Big]\big|W^{(0)}_{2l}(\xx_1,\ldots,
\xx_{2l})\big| \le L^2\b C_m^l|U|^{\max\{1,l/2\}}\;,\Eq(2.16)$$
for some constant $C_m>0$, where $m=\sum_{1\le i<j\le 2l} m_{ij}$. 
The proof of \equ(2.16) is based on a
multiscale analysis of the ultraviolet integration over the time
coordinates (much simpler than the infrared integration we shall
study below) and it is sketched in Appendix A1.

\* \sub(2.3) {\it Renormalization of the free measure.}

The Grassmann integral in the r.h.s. of \equ(2.13) is computed
iteratively in the following way. We step by step decompose the
propagator into a sum of two propagators, the first supported on
momenta $\sim \g^{h}$, $h\le 0$ (here $\g$ is the same {\it
scaling parameter} appearing in \equ(2.9)), the second supported
on momenta smaller than $\g^h$. Correspondingly we rewrite the
Grassmann field as a sum of two independent fields: $\psi^{(\le
h)}=\psi^{(h)}+ \psi^{(\le h-1)}$ and we integrate the field
$\psi^{(h)}$, in analogy with
\equ(2.13). In this way
we inductively prove that, for any $h\le 0$, \equ(2.13) can be
rewritten as
$$e^{- L^2\b F_{L,\b}}=e^{- L^2\b F_h}\int P_{E_h,C_h}
(d\psi^{(\le h)})e^{-\VV^{(h)} (\psi^{\le h})}\;,\Eq(2.17)$$
where: $\VV^{(h)}$ can be represented by an expansion similar to
\equ(2.15), with $\psi^{(\le 0)}$ replaced by $\psi^{(\le h)}$
and the kernels $W^{(0)}_{2l}$ replaced by new kernels
$W^{(h)}_{2l}$; the Grassmann integration $P_{E_h,C_h}(d\psi^{(\le
h)})$ can be represented as
$$\eqalign{&P_{E_h,C_h}(d\psi^{(\le h)}) = \Big[
\prod_{\kk\in\DD_{\b,L}^*}^{\s=\ud}\big(\fra{L^2\b
C_h^{-1}(\kk)}{-i k_0+ E_h(\kk)-\m}\big) d\psi^{(\le
h)+}_{\kk,\s}d\psi^{(\le h)-}_{\kk,\s} \Big]\cdot\cr
&\hskip1.truecm \cdot\exp \left\{-{1\over L^2\b}
\sum_{\kk\in\DD_{\b,L}^*}^{\s=\ud}C_h(\kk) (-i k_0+
E_h(\kk)-\m)\hat\psi^{+(\le h)}_{\kk,\s} \hat\psi^{-(\le
h)}_{\kk,\s}\right\}\;,\cr} \Eq(2.18)$$
where $E_h(\kk)$ is a function to be iteratively constructed
below, with $E_0(k_0,\vkk)\=\e_0(\vkk)$. Moreover $C_h(\kk)^{-1}$
is a compact support function defined as
$$C_h^{-1}(\kk) = H_0
\left[\g^{-h}\big|-ik_0+E_h(\kk)-\m\big|\right]\Eq(2.19)$$
and $\DD_{\b,L}^*$ the restriction of $\DD_{\b,L}$ to the set of
momenta in the support of $C_h^{-1}(\kk)$. Note that, for $h=0$,
the Gaussian integration $P_{E_0,C_0}(d\psi^{(\le 0)})$ coincides
with the integration $P(d\psi^{(\le 0)})$ defined above. So, by
this remark, we see that the representation \equ(2.17)--\equ(2.18)
is true at the first step $h=0$. In order to inductively prove it
for $h<0$ we proceed as follows. We introduce the {\it
localization operator} as a linear operator acting on the kernels
of $\VV^{(h)}$ in the following way:
$$\LL W^{(h)}_{2l}(\xx_1,\ldots,\xx_{2l})=
\cases{W^{(h)}_{2l}(\xx_1,\ldots,\xx_{2l}) \hskip.7truecm {\rm if}
\hskip.7truecm l=1,2;\cr 0 \hskip3.2truecm {\rm if} \hskip.7truecm
l\ge 3\;.} \Eq(2.20)$$
We also define $\RR$ as $\RR=1-\LL$ and rewrite the r.h.s. of
\equ(2.17) as
$$e^{- L^2\b F_h}\int P_{E_h,C_h}(d\psi^{(\le h)})e^{-\LL\VV^{(h)}
(\psi^{(\le h})-\RR\VV^{(h)} (\psi^{(\le h})}\;,\Eq(2.21)$$
where by definition $\LL\VV^{(h)}$ is
$$\eqalign{\LL\VV^{(h)}&=\sum_{\s=\ud}\int d\xx d\yy\, n_h(\xx-\yy)\,
\psi^{(\le h)+}_{\xx,\s} \psi^{(\le h)-}_{\yy,\s}+\cr
&+\sum_{\s,\s'=\ud} \int d\xx_1 d\xx_2 d\xx_3 d\xx_4\,
\l_{h}(\xx_1,\xx_2,\xx_3,\xx_4)\,\psi^{(\le h)+}_{\xx_1,\s}
\psi^{(\le h)-}_{\xx_2,\s}\psi^{(\le h)+}_{\xx_3,\s'}\psi^{(\le
h)-}_{ \xx_4,\s'}\;.\cr}\Eq(2.22)$$
Now, calling $\hat n_h(\kk)$ the Fourier transform of $n_h(\xx)$
and defining
$$E_{h-1}(\kk)=E_h(\kk)+C_h^{-1}(\kk)\hat n_h(\kk)\;,\Eq(2.23)$$
we can rewrite \equ(2.21) as
$$e^{- L^2\b (F_h+t_h)}\int P_{E_{h-1},C_h}(d\psi^{(\le h)})e^{-\LL_4\VV^{(h)}
(\psi^{\le h})-\RR\VV^{(h)} (\psi^{(\le h)})}\;,\Eq(2.24)$$
where $t_h$ is a constant which takes into
account the change in the renormalization factor of the measure, of size 
$|U||h|\g^{2h}$, as it follows from the bound in the first line of \equ(2.36)
below. Moreover
$$\LL_4\VV^{(h)}=
\sum_{\s,\s'=\ud} \int d\xx_1 d\xx_2 d\xx_3 d\xx_4\,
\l_{h}(\xx_1,\xx_2,\xx_3,\xx_4)\,\psi^{(\le h)+}_{\xx_1,\s}
\psi^{(\le h)-}_{\xx_2,\s}\psi^{(\le h)+}_{\xx_3,\s'}\psi^{(\le
h)-}_{ \xx_4,\s'}\;.\Eq(2.25)$$
We now define $\hat\VV^{(h)}\=\LL_4\VV^{(h)}+\RR\VV^{(h)}$ and
again use the addition principle in order to rewrite \equ(2.24) as
$$\eqalign{&e^{- L^2\b (F_h+t_h)}\int P_{E_{h-1},C_{h-1}}(d\psi^{(\le h-1)})
\,\int P_{E_{h-1},f_h^{-1}}(d\psi^{(h)}) e^{-\hat\VV^{(h)}
(\psi^{(\le h-1)}+\psi^{(h)})} \;,\cr}\Eq(2.26)$$
with $P_{E_{h-1},f_h^{-1}}(d\psi^{(h)})$ a Grassmann Gaussian
integration such that
$$\eqalign{&
\int P_{E_{h-1},f_h^{-1}}(d\psi^{(h)}) \hat
\psi^{(h)-}_{\kk_1,\s_1} \hat
\psi^{(h)+}_{\kk_2,\s_2}=L^2\b\d_{\kk_1,\kk_2}\d_{\s_1,\s_2} \hat
g^{(h)}(\kk_1)\;,\cr &\quad{\rm with}\quad \hat
g^{(h)}(\kk)=\fra{f_h(\kk)}{-i k_0+E_{h-1}(\kk)-\m}\cr}\Eq(2.27)$$
and
$$f_h(\kk)=H_0\left[\g^{-h}\big|-ik_0+E_h(\kk)-\m\big|\right]-
H_0\left[\g^{-h+1}\big|-ik_0+E_{h-1}(\kk)-\m\big|\right]\;.\Eq(2.28)$$
If we now define
$$e^{-\VV^{(h-1)}(\psi^{(\le h-1)})- L^2\b \tilde F_h}=
\int P_{E_{h-1},f_h^{-1}}(d\psi^{(h)}) e^{-\hat\VV^{(h)}
(\psi^{(\le h-1)}+\psi^{(h)})}\;,\Eq(2.29)$$
it is easy to see that $\VV^{(h-1)}$ is of the form \equ(2.15),
with $\psi^{(\le 0)}$ replaced by $\psi^{(\le h-1)}$ and the
kernels $W^{(0)}_{2l}$ replaced by new kernels $W^{(h-1)}_{2l}$,
and that
$$F_{h-1}=F_h+t_h+\tilde F_h\;.\Eq(2.30)$$
It is sufficient to use the identity
$$\VV^{(h-1)}(\psi^{(\le h-1)})+L^2\b \tilde F_h=
\sum_{n\ge 1}\fra{1}{n!}(-1)^n\EE^T_h(\hat\VV^{(h)}(\psi^{(\le
h-1)}+\cdot);n) \Eq(2.31)$$
where $\EE^T_h$ is the truncated expectation with respect to the
propagator $\hat g^{(h)}(\kk)$. Moreover the symmetry relations
listed after \equ(2.8) are still satisfied, because the symmetry
properties of the free integration are not modified by the
renormalization procedure, so that the effective potential
$\VV^{(h)}$ on scale $h$ has the same symmetries as $\VV^{(0)}$
and in particular it can be expanded in the form \equ(2.15), with
kernels $W^{(h)}_{2l}$ independent
of the spin labels. \\
\\
We iterate this procedure up to the first scale $h_{\b}$ such that
$$\g^{h_\b-1}e_0< \min \{|k_0-{\rm Im} E_{h_\b}(\kk)|, \kk\in
\DD_{\b,L}, C_{h_\b}^{-1}(\kk)>0\}\;,\Eq(2.31a)$$
where $e_0$ is the same of \equ(2.9). By the properties of
$E_h(\kk)$, that will be described and proved below, it will turn
out that $h_\b$ is finite and actually larger than
$[\log_\g(\p/2e_0\b)]$.

On scale $h_\b$ we define
$$e^{-L^2\b\tilde F_{h_\b}}=\int P_{E_{h_\b},C_{h_\b}}(d\psi^{(\le h_\b)})
e^{-\hat\VV^{(h_\b)}(\psi^{(\le h_\b)})}\;,\Eq(2.32)$$
so that we have
$$F_{L,\b}=F_0+\sum_{h=h_\b}^0(\tilde F_h+t_h)\;.\Eq(2.33)$$
Note that the above procedure allows us to rewrite the effective
coupling $\l_h(\xx_1,\xx_2,\xx_3,\xx_4)$ on scale $h$ and the
renormalized dispersion relation $E_h(\kk)$ as functionals of $U$
and $\l_j,E_j$, with $h< j\le 0$:
$$\eqalign{&E_{h-1}(\kk)=E_h(\kk)+\hat\b^2_h(\kk;E_h,\l_h,\ldots,E_0,\l_0,U)\cr
&\l_{h-1}(\ul\xx)=\l_{h}(\ul\xx)
+\b^4_h(\ul\xx;E_h,\l_h,\ldots,E_0,\l_0,U)\cr}\Eq(2.34)$$
where in the second line we defined
$\ul\xx=\{\xx_1,\xx_2,\xx_3,\xx_4\}$. The functionals $\hat\b^2_h$
and $\b^4_h$ are called the $E$--component and the $\l$--component
of the {\it Beta function}.

The key point of the subsequent discussion will be the fact that
the kernels of the effective potentials $\VV^{(h)}$ will be given
by convergent power series in $U,\l_j$, $h\le j\le 0$, {\it under
some smallness and smoothness conditions of $E_h$ and $\l_j$}.
Once identified the conditions on $E_h$ and $\l_h$ sufficient for
proving convergence of the series, we shall inductively check such
conditions.

The property $\l_h$ has to satisfy in order for the iterative
construction to be well--defined is a smallness property, a bit
stronger than the request:
$$|\hat\l_h(\kk_1,\kk_2,\kk_3,\kk_1-\kk_2+\kk_3)|\le U_0\Eq(2.35)$$
where $\hat\l_h(\kk_1,\kk_2,\kk_3,\kk_1-\kk_2+\kk_3)$ is the
Fourier transform of $\l_h(\xx_1,\xx_2,\xx_3,\xx_4)$ and $U_0$ is
small enough. The precise statement of the smallness property we
need to require on $\l_h$ is a bit technical, in particular it
involves the definitions of ``sectors'' and ``modified running
coupling functions'' and we postpone it to next subsections, see
\equ(2.71a) below.

The smallness and smoothness properties that $E_h(\kk)$ has to
satisfy are the following:
$$\eqalign{&
|E_h(\kk)-E_{h-1}(\kk)|\le C_0 |U||h|\g^{2 h}\;,\cr
&|\dpr_{k_{i_1}}\cdots\dpr_{k_{i_n}}\big(
E_h(\kk)-E_{h-1}(\kk)\big)|\le C_n |U|^2|h|\g^{(2-n)h}\;,\qquad n\ge
1\cr} \Eq(2.36)$$
where $i_j\in\{0,1,2\}$, $j=1,\ldots,n$, and $\dpr_{k_i}$ must be
interpreted as discrete derivatives in the three coordinate
directions\footnote{${}^1$}{\ottopunti In the following we will be
interested in studying the $L\to\io$ limit of the free energy and
two--point correlation function. For this reason, even if,
strictly speaking, the rigorous way to proceed should be
performing bounds at finite $L$, showing uniformity in $L$ and
then performing the limit, in order not to make the notation too
cumbersome and not to hide the ideas behind the proof, we shall
sometimes proceed formally by replacing the discrete derivatives
in spatial direction by their formal $L\to\io$ limit, and the sums
$(2\p/L)^2\sum_{\vkk\in\DD_L}$ by $\int_{-\p}^\p \int_{-\p}^\p
d\vkk$. Being the bounds we perform purely dimensional, it can be
realized that our bounds can be adapted also to the finite $L$
case, and the resulting estimates turn out to be uniform in $L$
for $L$ big enough. A rigorous discussion of the uniformity of the
bounds at finite $L$ for $L$ big enough in a case similar to the
one discussed in this paper can be found in [BM]}, acting on
functions $f:\DD_{\b,L}\to\RRR$ as:
$$\eqalign{&\dpr_{k_0}f(k_0,k_1,k_2)=\fra{\b}{2\p}
\big[f(k_0+\fra{2\p}{\b},k_1,k_2)-f(k_0,k_1,k_2)\big]\;,\cr
&\dpr_{k_1}f(k_0,k_1,k_2)=\fra{L}{2\p}
\big[f(k_0,k_1+\fra{2\p}{L},k_2)-f(k_0,k_1,k_2)\big]\;,\cr
&\dpr_{k_2}f(k_0,k_1,k_2)=\fra{L}{2\p}
\big[f(k_0,k_1,k_2+\fra{2\p}{L})-f(k_0,k_1,k_2)\big]\;.\cr}
\Eq(2.36aa)$$
Moreover it is easy to realize that, by the {\it parity} and {\it
complex conjugation} symmetries listed after \equ(2.8), $E_h(\kk)$
satisfies the symmetry properties
$$\eqalign{
&E_h(k_0,\vkk)=E_h(k_0,-\vkk)\;,\cr
&E_h(k_0,\vkk)=E^*_h(-k_0,\vkk)\;.\cr}\Eq(2.36a)$$
Note that at the first step $h=0$ the symmetry \equ(2.36a) is
true, by the explicit form of $\e_0(\vkk)$ and the definition of
$E_0(\kk)$.
\\

We now proceed as follows. In the following subsections
\secc(2.4)--\secc(2.9) we shall: describe the geometric properties
of the {\it Fermi surface on scale $h$}, which will be crucial for
the subsequent inductive bounds; describe the perturbative
expansion in $(\l_h,\ldots,\l_0,U)$ for the effective potential
$\VV^{(h)}$; resume the inductive procedure described in [BGM]
(adapted to the present case) allowing to prove that, under the
above smallness and smoothness conditions on $E_h$ and $\l_h$, the
expansion for $\VV^{(h)}$ is well--defined; adapt the expansion
for the free energy to the computation of the two--point Schwinger
function and complete the proof of the Main Theorem, under the
above smallness and smoothness conditions on $E_h$ and $\l_h$.
Then in section \sec(3) we shall inductively prove the smallness
and smoothness properties of $E_h$ and $\l_h$. While the smallness
condition on $\l_h$ will be controlled as in [BGM], by imposing a
condition on the temperature, the inductive proof of the smallness
and smoothness conditions on $E_h$ is the main novelty (and the
real new technical difficulty) with respect to [BGM]. This will
conclude the proof of Main Theorem in the introduction.

\* \sub(2.4) {\it The Fermi surface at scale h.}

Given $h_\b\le h\le 0$ and $E_h(\kk)$, we define an effective
dispersion relation on scale $h$ as
$$\e_h(\vec k)\defin \fra{1}{2}
\Big[E_h(\fra{\p}{\b},\vkk)+E_h(-\fra{\p}{\b},\vkk)\Big]\;.
\Eq(2.36c)$$
{\bf Remark.} Note that, thanks to \equ(2.36a), $\e_h(\vec k)$ is
real and $\e_h(\vec k)= \e_h(-\vec k)$.\\
\\
A crucial consequence of the properties
\equ(2.36) and of the explicit form of the unperturbed dispersion
relation $\e_0(\vec k)$ is that at any scale $h\le 0$ we can
define a Fermi surface $\Sigma_F^{(h)}=\{\vkk\;:\;
\e_h(\vkk)-\m=0\}$ with strong convexity properties.

Let us start recalling the properties of the free
dispersion relation $\e_0(\vkk)$.\\

\item{1.} If $\m<\m_0\=\fra{2-\sqrt2}{2}$, there exists $e_0< \m_0-\m$
($e_0$ is the same parameter appearing in \equ(2.9)) such that,
for $|e|\le e_0$, $\e_0(\vec k)-\mu=e$ defines a convex curve
$\Sigma^{(0)}(e)$ encircling the origin, which can be represented
in polar coordinates as $\vpp=u_0(\th,e)\vee_r(\th)$ with
$\vee_r(\th)=(\cos\th,\sin\th)$. Moreover $u_0(\th,e)\ge c^0>0$
and, if $r_0(\th,e)$ is the curvature radius,
$$r_0(\th,e)^{-1}\ge c^0> 0\;.\Eq(2.37)$$
Note that the symmetry property $\e_0(\vkk)=\e_0(-\vkk)$ implies
that the curves $\Si^{(0)}(e)$ are symmetric by reflection with
respect to the origin.
\item{2.} If $|e|\le e_0$ and $\vpp=u_0(\th,e)\vec e_r(\th)$, then
$$0<c_1^0\le \grad\e_0(\vpp)\cdot\vee_r(\th) \le c_2^0\;.\Eq(2.38)$$
\item{3.} Given $e$ as in item 1) and
$\vkk_1,\ldots,\vkk_{2n}\in \Sigma^{(0)}(e)$, $n\le 4$, then
$$\big|\sum_{i=1}^{2n}\vkk_i \big|<2\p\;.\Eq(2.39)$$
We shall refer to this property by saying that
``umklapp processes with $n\le 4$ quasi--particles are not allowed''.\\

\0{\bf{Remark}.} The validity of property (3) can be checked by
noting that: if $\m_1<\m_2$ the surface $\Sigma_{\m_1}^{(0)}(0)$
corresponding to chemical potential $\m_1$ is completely enclosed
into the surface $\Sigma_{\m_2}^{(0)}(0)$ corresponding to
chemical potential $\m_2$; at $\m$ and $e$ fixed and $n\le 4$, the
l.h.s. of \equ(2.39) is maximized by $8\arccos(1-\m-e)$, obtained
in correspondence of the choice $\vkk_i=(\arccos(1-\m-e),0)$,
$\forall i=1,\ldots,2n$; the condition $\m+e_0<(2-\sqrt2)/2$ is
equivalent to $8\arccos(1-\m-e_0)<2\p$.

\* \0 We call $\Si_F^{(0)}\=\Si^{(0)}(0)$ the free {\it Fermi
surface} and we put $u_0(\th,0)\vee_r(\th)={\vec p}_F^{(0)}(\th)$
and $u_0(\th)\=u_0(\th,0)=| {\vec p}^{(0)}_F(\th)|$.

The key remark is that, if \equ(2.36) is true for $|U|\le U_0$
with $U_0$ small enough, for any $h_\b\le h\le 0$, then the same
properties (1)--(3) above still hold (with slightly modified
constants) for the Fermi surfaces corresponding to the dispersion
relations $\e_h(\vkk)$.

\* \lemma(2.1) {\it Let us assume that \equ(2.36) is satisfied for
$h_\b\le h\le 0$ and for $|U|\le U_0$, with $c_0\=|h_\b|U_0$ small
enough. Then there exist constants $\lis e, c, c_1, c_2$ such that
the following properties are true.
\\

\item{1.} If $|e|\le \lis e$, $\e_h(\vec k)-\mu=e$
defines a convex curve $\Sigma^{(h)}(e)$, encircling the origin
and symmetric by reflection with respect to it, which can be
represented in polar coordinates as $\vpp=u_h(\th,e)\vee_r(\th)$.
Moreover $u_h(\th,e)\ge c>0$ and, if $r_h(\th,e)$ is the curvature
radius,
$$r_h(\th,e)^{-1}\ge c> 0\;.\Eq(2.40)$$
\item{2.} If $|e|\le \lis e$ and $\vpp=u_h(\th,e)\vec e_r(\th)$, then
$$0<c_1\le \grad\e_h(\vpp)\cdot\vee_r(\th) \le c_2\;.\Eq(2.41)$$
\item{3.} If $\m<\m_0\=\fra{2-\sqrt2}{2}$ and $|e|\le \lis e$, then
$$n\le 4\;,\qquad \vkk_i\in\Sigma^{(h)}(e),\; i=1,\ldots,2n\;,\qquad
\Rightarrow \qquad \big|\sum_{i=1}^{2n}\vkk_i \big|<2\p\;.
\Eq(2.40a)$$
}

\0{\bf Remark.} Lemma \lm(2.1) says that, under the smallness and
smoothness properties \equ(2.36) and under the symmetry assumption
\equ(2.36a), the qualitative and quantitative properties of the
dispersion relation on scale $h$ are the same as the free one.
From the proof below it will become clear that, if $U_0$ is small
enough, then the constants $\lis e,c,c_1,c_2$ can be chosen equal
to:
$\lis e=e_0/2$, $c={c^0/ 2}$, $c_1={c_1^0/ 2}$ and $c_2=3c_2^0/2$.\\

\proof Given $h_\b\le h\le 0$, we can write $\e_h(\vkk)=\e_0(\vkk)
+\sum_{j=h+1}^0 (\e_{j-1}(\vkk)-\e_j(\vkk))$. From this identity
and the inductive assumption \equ(2.36) we soon find
$$\eqalign{&|\e_h(\vkk)-\e_0(\vkk)|\le C_0|U|\sum_{j=h+1}^0|j|\g^{2j}\le
C_0'|U|\;,\cr &|\grad\e_h(\vkk)-\grad\e_0(\vkk)|\le
2C_1|U|^2\sum_{j=h+1}^0\g^{j}|j|\le C_1'|U|^2\;,\cr &|\dpr^2_{k_i k_r}
\e_h(\vkk)- \dpr^2_{k_i k_r} \e_0(\vkk)|\le C_2 |U|^2\sum_{j=h+1}^0|j| \le C_2
c_0^2\;. \cr}\Eq(2.41a)$$
The first two bounds in \equ(2.41a) show that $\e_h(\vkk)$ and
$\grad\e_h(\vkk)$ are close within $O(U)$ and $O(U^2)$ to $\e_0(\vkk)$ and
$\grad\e_0(\vkk)$ respectively. This properties and the validity
of properties (1)--(3) in the unperturbed case $h=0$ guarantee
that the equation $\e_h(\vkk)-\m=e$ can be inverted for $h<0$ if
$e$ is small enough (as it follows from an application of implicit
function Theorem). Then the set of vectors $\Sigma^{(h)}(e)$
defines a closed curve enclosing the origin (and symmetric around
it, because of the remark after \equ(2.36c)), close to
$\Sigma^{(0)}(e)$ within $O(U)$. Also, the third bound in
\equ(2.41a) implies that the second derivatives of $\e_h(\vkk)$ are
close to the second derivatives of $\e_0(\vkk)$, if $c_0$ small
enough. This means that $\Sigma^{(h)}(e)$ is convex (since so
$\Sigma^{(0)}(e)$ is) and $r_h(\th,e)=r_0(\th,e)+O(c_0^2)$ (because
the curvature radius $r_h(\th,e)$ is computed in terms of the
first two derivatives of $\e_h(\vkk)$) and the Lemma is proved.
\qed

\* \0 In analogy with the definitions corresponding to the
unperturbed dispersion relation $\e_0(\vkk)$, we call
$\Si_F^{(h)}\=\Si^{(h)}(0)$ the {\it Fermi surface on scale $h$}
and we put $u_h(\th,0)\vee_r(\th)={\vec p}_F^{(h)}(\th)$ and
$u_h(\th)\=u_h(\th,0)=| {\vec p}^{(h)}_F(\th)|$.

We conclude this section by listing some more properties of
$E_h(\kk)$ and of $\hat g^{(h)}(\kk)$ following from the inductive
assumptions \equ(2.36). First of all, by proceeding as in the
proof of Lemma \lm(2.1), we see that \equ(2.36) imply that
$$\eqalign{&|E_h(\kk)-E_0(\kk)|\le C_0'|U|\cr
&|\dpr_{k_i} E_h(\kk)-\dpr_{k_i} E_0(\kk)|\le C_1'|U|^2\;,\cr
&|\dpr_{k_{i_1}}\dpr_{k_{i_2}} E_h(\kk)-
\dpr_{k_{i_1}}\dpr_{k_{i_2}} E_0(\kk)|\le C_2'c_0^2\;,\cr
&|\dpr_{k_{i_1}}\cdots\dpr_{k_{i_n}} E_h(\kk)-
\dpr_{k_{i_1}}\cdots\dpr_{k_{i_n}} E_0(\kk)|\le
C_n'|U|^2|h|\g^{(2-n)h}\;,\qquad n\ge 3\cr}\Eq(2.42)$$
with $c_0\=|h_\b|U_0$ small enough.

Finally, let us consider the propagator $\hat g^{(h)}(\kk)$ in
\equ(2.27). By putting $E_{h-1}(\kk)=E_h(\kk)+ (E_{h-1}(\kk)-E_h(\kk))$ and
by expanding $E_h(\kk)$ at first order in $k_0$, we see that the
propagator $\hat g^{(h)}(\kk)$ in
\equ(2.27) can be rewritten as
$$\hat g^{(h)}(\kk)=\fra{f_h(\kk)}{-i k_0\big[1+a_{h}(\vkk)\big]+
\e_{h}(\vkk)+r_{h}(\kk)-\m}\;,\Eq(2.43)$$
where
$$a_{h}(\vec k)=\fra{i}{2}
\Big[\fra{\p}{\b}E_{h}\big(\fra{\p}{\b},\vec k\big)-\fra{\p}{\b}
E_{h}\big(-\fra{\p}{\b},\vec k\big)\Big]=i\dpr_{k_0}E_{h}
(-\fra{\p}{\b},\vkk)\;.\Eq(2.44)$$
By the second of \equ(2.36a), we see that $a_{h}(\vkk)$ is real.
Moreover both $a_{h}$ and the rest $r_{h}$ can be bounded using
\equ(2.36) and \equ(2.42); we find that there exists $C>0$ such that
$|a_{h}(\vec k)|\le C|U|$ and $|r_{h}(\kk)|\le C|U||h|\g^{2h}$ for
any $\kk$ in the support of $f_h(\kk)$. This implies in particular
that, under the inductive assumption \equ(2.36), the scale
$h_{\b}$ defined in \equ(2.31a)) is finite and larger than
$[\log_\g(\p/2e_0\b)]$, if $U$ is small enough. Moreover, by using
the definition of $f_h(\kk)$ in \equ(2.28), one can easily see
that there exists $K>0$ such that
$$f_h(\kk)\neq 0\quad\Rightarrow\quad
K^{-1}e_0\g^h\le |k_0|+|\e_h(\vkk)-\m|\le Ke_0\g^h\;. \Eq(2.42a)$$
In particular, if $f_h(\kk)\neq 0$, we can write
$\vkk=u_h(\th,e)\vec e_r(\th)$ with $e=\e_h(\vkk)-\m$ and $|e|\le
K\g^h e_0$.

\* \sub(2.5) {\it The sector decomposition.}

The smoothness properties of $E_h(\kk)$ and the symmetry and
convexity properties of the Fermi surface described in previous
subsection allow for a further decomposition of the propagator
$\hat g^{(h)}(\kk)$ that is convenient for explicitly performing
the bounds on the kernels of $\VV^{(h)}$: as in [BGM] we
decompose the field $\psi^{(\le h)}$, by slicing the support of
$C_h^{-1}(\kk)$ as follows. As in section 2.3 of [BGM], we
introduce the angles $\th_{h,\o}=\p(\o+\fra{1}{2})\g^{h/2}$, with
$\o$ an integer in the set $O_h=\{0,1,\ldots,\g^{-(h-1)/2}-1\}$
(recall that $\g=4$). Correspondingly we introduce the functions
$\z_{h,\o}(\th)$ with the properties:
$$\eqalign{
&||\th-\th_{h,\o}||<\fra{\p}{4}\g^{h/2}\quad\ \,\Rightarrow\quad
\z_{h,\o}(\th)=1\cr &||\th-\th_{h,\o}||
> \fra{3\p}{4}\g^{h/2}\quad\Rightarrow\quad \z_{h,\o}(\th)=0\cr
&\sum_{\o\in O_h}\z_{h,\o}(\th)=1\;, \qquad \forall\th\in
\TTT^1\cr} \Eq(2.45)$$
where $||\cdot||$ is the usual distance on $\TTT^1$.

We also introduce the support function $F_{h,\o}(\kk)=
f_h(\kk)\z_{h,\o}(\th)$, where, if $\kk=(k_0,\vkk)$, then $\th$ is
the polar angle of $\vkk$. We shall call the functions
$F_{h,\o}(\kk)$ the {\it anisotropic} support functions and the
indices $\o\in O_h$ the {\it anisotropic} sector indices (the name
recalls the fact that the support of $F_{h,\o}(0,\vkk)$ is a
geometric set that is ``wider'' than ``thick''; it is $O(\g^{h/2})$
``wide'' in the direction tangential to the Fermi surface and
$O(\g^h)$ ``thick'' in the normal direction).

Given any $\kk$ belonging to the support of $F_{h,\o}(\kk)$, we
put
$$\vkk=\vpp_F^{(h)}(\th_{h,\o})+k'_1\vnn_h(\th_{h,\o})+k'_2\vec\t_h(\th_{h,\o})
=\vpp_F^{(h)}(\th_{h,\o})+\vkk'\Eq(2.46)$$
where, putting $\vec e_t(\th)=(-\sin\th,\cos\th)$, the vectors
$\vec\t_h(\th)$ and $\vnn_h(\th)$ are defined as
$$\eqalign{&\vec{\t}_h(\th) = {d \vec{p}_F^{(h)}(\th)\over d\th}\left|{d
\vec{p}_F^{(h)}(\th)\over d\th}\right|^{-1}=
{u_h'(\th)\vec{e}_r(\th)+u_h(\th)\vec{e}_t(\th)
\over\sqrt{u_h'(\th)^2+u_h(\th)^2}}\;,\cr &{\vec n}_h(\th) =
{u_h(\th)\vec{e}_r(\th)-u_h'(\th)\vec{e}_t(\th)
\over\sqrt{u_h'(\th)^2+u_h(\th)^2}}\;.\cr}\Eq(2.47)$$
Using \equ(2.42a) and \equ(2.45) it is easy to realize that
$|k_1'|\le C\g^h$ and $|k_2'|\le C\g^{h/2}$ for some constant $C$;
see Lemma 7.3 of [BGM] for details.

Using the decomposition \equ(2.46), we can rewrite
$$\psi_{\xx,\s}^{(h)\pm}\=\sum_{\o\in O_h}
e^{\pm i {\vec p_F^{(h)}}(\th_{h,\o})\vec
x}\psi_{\xx,\s,\o}^{(h)\pm} \virg P(d\psi^{(h)})=\prod_{\o\in O_h}
P(d\psi^{(h)}_{\o})\;,\Eq(2.48)$$
where $P(d\psi^{(h)}_\o)$ is the Grassmanian integration with
propagator
$$g^{(h)}_\o(\xx)=\fra{1}{\b}\sum_{k_0\in\DD_{\b}}\int_{-\p}^\p\int_{-\p}^\p
\fra{d\vkk'}{(2\p)^2}
e^{-i(k_0x_0+\vkk'\vxx)}\fra{F_{h,\o}(\kk'+\pp_F^{(h)}(\th_{h,\o}))}
{-i k_0+E_{h-1}(\kk'+\pp_F^{(h)}(\th_{h,\o}))-\m}\;.\Eq(2.49)$$
where we defined
$\pp_F^{(h)}(\th_{h,\o})\defin(0,\vpp_F^{(h)}(\th_{h,\o}))$. We
insert the decomposition \equ(2.48) into the r.h.s. of \equ(2.29):
$$\int P_{E_{h-1},f_h^{-1}}(d\psi^{(h)})
\exp\Big\{-\hat\VV^{(h)} \big(\psi^{(\le
h-1)\pm}_{\xx,\s}+\sum_{\o\in O_h} e^{\pm i {\vec
p_F^{(h)}}(\th_{h,\o})\vec x}\psi^{(h)\pm}_{\xx,\s,\o}\big)
\Big\}\;,\Eq(2.50)$$
and in this way we induce a decomposition of the kernels of
$\VV^{(h-1)}$ into a sum of contributions labelled by the choices
of the sector labels of the integrated fields
$\psi^{(h)\pm}_{\xx,\s,\o}$, see next section for a description of
this further decomposition of the kernels of $\VV^{(h-1)}$.
\\

The bounds on the (decomposed) kernels of $\VV^{(h-1)}$ are based
on the following key bound on the asymptotic behavior of
$g^{(h)}_\o(\xx)$.

\* \lemma(2.2) {\it Let us assume that the bounds \equ(2.36) are
valid and that $c_0\=|h_\b|U_0$ is small enough. Given $h_\b\le
h\le 0$ and $\o\in O_h$, let us put
$$\vec x=x_1'\vnn_h(\th_{h,\o})+x_2'\vec\t_h(\th_{h,\o})\Eq(2.51)$$
Then, given $N\ge 2$, there exists a constant $C_N$ such that
$$|g^{(h)}_\o(\xx)|\le \fra{C_N\g^{{3\over 2}h}}{1+
(\g^{h}|d_\b(x_0)|+\g^{h}|x_1'|)^N+\g^{-h}(\g^h|x_2'|)^N}\Eq(2.52)$$
where $d_\b(x_0)=\b\p^{-1}\sin(\p\b^{-1}x_0)$.
}\\
\\
{\bf Remark.} The bound \equ(2.52) implies that
$$\eqalign{&\int d\xx\, |\xx|^j\cdot|g^{(h)}_\o(\xx)|\le C_j\g^{-(1+j)h}\;,
\qquad j\ge 0\;,\cr}\Eq(2.52a)$$
that will be widely used in the following.\\

\proof The key remark in order to get the bound \equ(2.52) is the
following. If $\kk=(k_0,\vkk)$ belongs to the support of
$F_{h,\o}(\kk)$, $\o\in O_h$, and we write
$\vkk=\vpp_F^{(h)}(\th_{h,\o})+k_1'\vnn_h(\th_{h,\o})+
k_2'\vec\t_h(\th_{h,\o})$, then
$$\left| {\partial E_{h-1}(\kk)\over\partial k_2'} \right|
\le C \g^{h\over 2}\;,\Eq(2.53)$$
for some constant $C$. In fact, if $\kk=(k_0,\vkk)$ belongs to the
support of $F_{h,\o}(\kk)$, we can write
$E_{h-1}(\kk)=\e_h(\vkk)+(E_{h-1}(\kk)-E_h(\kk))+
(E_h(\kk)-\e_h(\vkk))$ where, by the properties described in
previous section, $\big|\dpr_{k_2'}[(E_{h-1}(\kk)-E_h(\kk))+
(E_h(\kk)-\e_h(\vkk))]\big|\le C\g^h$. Moreover it is easy to
prove that $\dpr_{k_2'}\e_h(\vkk)=O(\g^{h/2})$. In fact we have
$\grad\e_h(\vkk)= |\grad\e_h(\vkk)|\vnn_h(\th,e)$ where
$|\grad\e_h(\vkk)|=O(1)$, $e=\e_h(\vkk)-\m=O(\g^h)$, $\th$ is the
polar angle of $\vkk$ and $\vnn_h(\th,e)$ is the outgoing normal
vector at $\Sigma^{(h)}(e)$ in $\vkk$. Furthermore
$||\th-\th_{h,\o}||=O(\g^{h/2})$ so that
$$\eqalign{\fra{\partial \e_h(\vec k)}{\partial k_2'}&=
|\grad\e_h(\vkk)|\vnn_h(\th,e)\cdot \vec\t_h(\th_{h,\o})=
|\grad\e_h(\vkk)|\vnn_h(\th)\cdot \vec\t_h(\th_{h,\o})+O(\g^h)=\cr
&=|\grad\e_h(\vkk)|\sin(\th-\th_{h,\o})+O(\g^h)=O(\g^{h/2})\cr}\Eq(2.54)$$
and \equ(2.53) follows.

The bound \equ(2.52) is simply obtained by integration by parts
and dimensional bounds on the integrand and on the measure of the
support. First of all, note that by the compact support properties
of $F_{h,\o}(\kk)$ it holds that $|g^{(h)}_\o(\xx)|\le C
\g^{{3\over 2}h}$, for some constant $C$. In order to bound
$|(x_2')^Ng^{(h)}_\o(\xx)|$ with $N\ge 2$, by integrating by parts with respect
to $k_2'$, we rewrite:
$$|(x_2')^Ng^{(h)}_\o(\xx)|=\Big|\fra{1}{\b}\sum_{\kk\in\DD_\b}\int
\fra{d\vkk'}{(2\p)^2}
e^{-i(k_0x_0+\vkk'\vxx)}\dpr_{k_2'}^N\Big[\fra{F_{h,\o}(\kk'
+\pp_F^{(h)}(\th_{h,\o}))} {-i k_0+E_{h-1}(\kk'
+\pp_F^{(h)}(\th_{h,\o}))-\m}\Big]\Big|\;.\Eq(2.55)$$
Using \equ(2.53), \equ(2.42) and the fact that the $n$-th order
derivative of $\z_{h,\o}(\th)$ is of order $\g^{-nh/2}$, it is
easy to see that $\dpr_{k_2'}^N\Big[\fra{F_{h,\o}(\kk'
+\pp_F^{(h)}(\th_{h,\o}))} {-i k_0+E_{h-1}(\kk'
+\pp_F^{(h)}(\th_{h,\o}))-\m}\Big]=O(\g^{-Nh})$ for $N\ge 2$ and any $\kk'$ in
the support of $F_{h,\o}(\kk' +\pp_F^{(h)}(\th_{h,\o}))$; this
implies that $|(x_2')^Ng^{(h)}_\o(\xx)|\le C_N\g^{{3\over
2}h}\g^{-h(N-1)}$ for some constant $C_N$.

Similarly, using the bounds \equ(2.42), we find that, given $N\ge
0$,
$$|[d_\b(x_0)]^Ng^{(h)}_\o(\xx)|\le C_N\g^{{3\over 2}h}\g^{-N h}\;,
\qquad |(x_1')^Ng^{(h)}_\o(\xx)|\le C_N\g^{{3\over 2}h}\g^{-N
h}\;.\Eq(2.56)$$
Combining \equ(2.56) with the analogue bounds on
$|g^{(h)}_\o(\xx)|$ and $|(x_2')^Ng^{(h)}_\o(\xx)|$ we find
\equ(2.52). \qed
\\
\\
Note that $g_\o^{(h)}(\xx)$ is approximately odd in $\xx$: in fact 
$g_\o({\bf 0})$ admits an improved dimensional estimate with respect to 
the bound \equ(2.52), as expressed by the following Lemma.

\*\lemma(2.2a){\it Let us fix $h\le 0$ and $\o\in O_h$. Then
$$|g_\o^{(h)}({\bf 0})|\le C\g^{{5\over 2} h}\Eq(2.56a)$$
}
\\
{\bf Remark.} Note that the bound \equ(2.56a) on the size of 
$g_\o^{(h)}({\bf 0})$ is $\g^h$ smaller than the bound on the size 
of $g_\o^{(h)}({\bf x})$, see \equ(2.52).\\

\proof The propagator $g_\o^{(h)}({\bf 0})$
can be written as
$$g_\o^{(h)}({\bf 0})=\int d\th \z_{h,\o}(\th)\int d k_0 \r d\r  
\fra{f_h(k_0,\r\vec e_r(\th))}{-i k_0[1+a(\th)]+e_\th(\r)+O(c_0\g^{2h})}
\Eq(2.56b)$$
where, referring to \equ(2.43), $a(\th)=a_h(u_h(\th)\vec e_r(\th))$ and 
$e_\th(\r)\defin \e_h(\r\vec e_r(\th))-\e_h(u_{h}(\th) \vec e_r(\th))$ 
(note that $e_\th(u_{h}(\th))=0$ and $e'_\th(u_{h}(\th))=O(1)$ uniformly
in $\th$). The bound on the error term 
in the denominator in \equ(2.56b) comes from
the bounds on $r_h(\kk)$ and $a_h(\vec k)$ discussed after \equ(2.43).
Similarly we find that 
$f_h(k_0,\r\vec e_r(\th))=\tilde f_h\big(\sqrt{k_0^2[1+a(\th)]^2+e_\th(\r)^2}
\,\big)+
f^{R}_h(k_0,\r\vec e_r(\th))$, where
$$\tilde f_h(t)=H_0\big(\g^{-h}t)-H_0\big(\g^{-h+1}t)\Eq(2.56bb)$$
and $f^{R}_h(k_0,\r\vec e_r(\th))$ is an $O(\g^h)$ function, vanishing 
outside a region slightly larger than the support of $f_h(k_0,
\r\vec e_r(\th))$. 
Replacing in \equ(2.56b) $f_h$ with $\tilde f_h+f^{R}_h$ and using that
the error term in the denominator is $O(c_0\g^{2h})$, we find that
$$g_\o^{(h)}({\bf 0})=\int d\th \z_{h,\o}(\th)\int d k_0  \r d\r
\fra{\tilde f_h\big(\sqrt{k_0^2[1+a(\th)]^2+e_\th(\r)^2}\,\big)}
{-i k_0[1+a(\th)]+e_\th(\r)}+R_\o^{(h)}\;,\Eq(2.56d)$$
where the rest $R_\o^{(h)}$ is dimensionally
bounded by $|R_\o^{(h)}|\le c\g^{{5\over 2}h}$.

Now, for any fixed $\th$, we can use Dini's theorem to
invert the relation $e_\th(\r)=e$ into $\r=\r(e)$, and we can rewrite
the first integral in the r.h.s. of 
\equ(2.56d) as
$$\int d\th \z_{h,\o}(\th)\int d k_0 \r(e) d e{1\over 
e'_\th(\r(e))}  
\fra{\tilde f_h\big(\sqrt{k_0^2[1+a(\th)]^2+e^2}\,\big)}
{-i k_0[1+a(\th)]+e}\Eq(2.56c)$$
Note that on the support of $\tilde f_h$ we have $e= O(\g^h)$, so that 
we can rewrite $\r(e)=u_{h}(\th)+O(\g^h)$ and $e'_\th(\r(e))=
e'_\th(u_{h-1}(\th))+O(\g^h)$, where $u_{h-1}(\th)$ and $e'_\th(u_{h-1}(\th))$ 
are bounded below and above by positive $O(1)$ constants, uniformly in $\th$.
Then \equ(2.56c) is equal to 
$$\int d\th \z_{h,\o}(\th)u_{h-1}(\th){1\over 
e'_\th(u_{h-1}(\th))}\int d k_0  d e \fra{\tilde f_h\big(\sqrt{k_0^2
[1+a(\th)]^2+e^2}\,\big)}{-i k_0+e}+
O(\g^{\fra52 h})\Eq(2.56e)$$
and, since the first integral is zero by oddity, the Lemma is proved. \qed\\

In order to perform the inductive bounds, in the following it will
also be convenient to introduce, besides the anisotropic sector
functions, the {\it isotropic} ones, defined as follows. We
introduce the angles
$\lis\th_{h,\bar\o}=\p(\lis\o+\fra{1}{2})\g^{h}$, with $\lis\o$ an
integer in the set $\lis O_h=\{0,1,\ldots,\g^{-(h-1)}-1\}$, and
the functions $\lis\z_{h,\bar\o}(\th)$ with the properties:
$$\eqalign{
&||\th-\lis\th_{h,\bar\o}||<\fra{\p}{4}\g^{h}\quad\
\,\Rightarrow\quad \lis\z_{h,\lis\o}(\th)=1\cr
&||\th-\lis\th_{h,\bar\o}||
> \fra{3\p}{4}\g^{h}\quad\Rightarrow\quad
\lis\z_{h,\lis\o}(\th)=0\cr &\sum_{\bar\o\in\bar
O_h}\lis\z_{h,\bar\o}(\th)=1\;, \qquad \forall\th\in \TTT^1\;.\cr}
\Eq(2.57)$$
We also introduce the support functions $\lis
F_{h,\bar\o}(\kk)=f_h(\kk)\lis\z_{h,\bar\o}(\th)$, to be called
the {\it isotropic} support
functions. From now on we shall use the convention that \\
\\
{\it we shall denote all the quantities associated with the
isotropic sectors by symbols obtained by overlining the symbols
for the corresponding quantities associated
with the anisotropic sectors.}\\

Given any $\kk$ belonging to the support of $\lis
F_{h,\bar\o}(\kk)$, we put
$$\vkk=\vpp_F^{(h)}(\th_{h,\o})+\vkk'\Eq(2.58)$$
where $\vkk'=O(\g^h)$. Correspondingly we can decompose
$g^{(h)}(\kk)$ into a sum of isotropic propagators:
$$\eqalign{&
g^{(h)}(\xx)=\sum_{\bar\o\in\bar
O_h}e^{i\vpp_F^{(h)}(\bar\th_{h,\bar\o}) \vxx}\lis g_{\bar
\o}^{(h)}(\xx)\;,\cr &\lis g^{(h)}_{\bar\o}(\xx)=
\fra{1}{\b}\sum_{k_0\in\DD_{\b}}\int_{-\p}^\p\int_{-\p}^\p
\fra{d\vkk'}{(2\p)^2} e^{-i(k_0x_0+\vkk'\vxx)}\fra{\lis F_{h,\bar
\o}(\kk' +\lis\pp_F^{(h)}(\lis\th_{h,\bar\o}))} {-i
k_0+E_{h-1}(\kk'
+\lis\pp_F^{(h)}(\lis\th_{h,\bar\o}))-\m}\;.\cr}\Eq(2.59)$$
with $\lis g^{(h)}_{\bar\o}(\xx)$ satisfying the following
analogue of Lemma \lm(2.2) and of Lemma \lm(2.2a)
(to be proven via a repetition of the proof of Lemma \lm(2.2) and Lemma 
\lm(2.2a)).

\* \lemma(2.3) {\it Let us assume that the bounds \equ(2.36) are
valid and that $c_0\=|h_\b|U_0$ is small enough. Given $h_\b\le
h\le 0$ and $\lis\o\in\lis O_h$ and $N\ge 0$, there exists a
constant $C_N$ such that
$$|\lis g^{(h)}_{\bar\o}(\xx)|\le \fra{C_N\g^{2h}}{1+
(\g^{h}|d_\b(x_0)|+\g^{h}|\vxx|)^N}\;.\Eq(2.60)$$
}
\\
{\bf Remark.} \equ(2.60) implies the analogue of \equ(2.52a):
$$\int d\xx\, |\xx|^j\cdot|\lis g^{(h)}_{\bar\o}(\xx)|\le C_j\g^{-(1+j)h}\;,
\qquad j\ge 0\;.\Eq(2.52b)$$
Note that the dimensional bound on the integral of an isotropic propagator 
is the same as the bound on the integral of an anisotropic one.\\

\lemma(2.2b) {\it Let us fix $h\le 0$ and $\lis\o\in \lis O_h$. Then 
$$ |\lis g_{\bar\o}^{(h)}({\bf 0})|\le C\g^{3 h}\Eq(2.60a)$$}
\\
{\bf Remark.} Note that the bound \equ(2.60a) on the size of 
$\lis g_{\bar\o}^{(h)}({\bf 0})$ is $\g^h$ smaller than the bound on the size 
of $\lis g_{\bar\o}^{(h)}({\bf x})$, see \equ(2.60).\\

\* \sub(2.6) {\it The tree expansion.}

In this section and in the two following ones we shall describe
the expansion for the effective potential $\VV^{(h)}$ and the
inductive bounds we use to prove convergence of the expansion for
the free energy, {\it under the smoothness and smallness
assumption \equ(2.36) and under a smallness assumption on
$\l_h(\xx)$} to be stated in a precise form below. In particular
we shall summarize the bounds described in [BGM] in a form
adapted to the present case and suitable for proving the improved
dimensional bounds on $E_h(\kk)$ (\ie allowing for an inductive
proof of \equ(2.36), see next Chapter).\\

Our expansion of $\VV^{(h)}$, $0\ge h\ge h_\b$, is obtained by
integrating iteratively the field variables of scale $j\ge h+1$
and sector index $\o\in O_h$ (for the moment we do not consider
isotropic sectors; they will be introduced in next Chapter in
order to optimize the bounds) and by applying at each step the
{\it localization procedure} described above. The result can be
expressed in terms of a {\it tree expansion}, very similar to the
one described in [BGM]. For completeness we list here the definition of trees.
See Fig.1 for an example of a possible tree appearing in 
the expansion for the effective potentials.\\
\\
\\

\insertplot{300pt}{150pt}%
{\ins{30pt}{85pt}{$r$}\ins{50pt}{85pt}{$v_0$}\ins{130pt}{100pt}{$v$}%
\ins{35pt}{-2pt}{$h$}\ins{55pt}{-2pt}{$h+1$}\ins{135pt}{-2pt}{$h_v$}%
\ins{215pt}{-2pt}{$0$}\ins{235pt}{-2pt}{$+1$}\ins{255pt}{-2pt}{$+2$}}%
{fig51}{\eqg(fig51)}
\\
\\

\0 1) Let us consider the family of all trees which can be
constructed by joining a point $r$, the {\it root}, with an
ordered set of $n\ge 1$ points, the {\it endpoints} of the {\it
unlabeled tree} (see Fig. 1), so that $r$ is not a branching
point. $n$ will be called the {\it order} of the unlabeled tree
and the branching points will be called the {\it non trivial
vertices}. The unlabeled trees are partially ordered from the root
to the endpoints in the natural way; we shall use the symbol $<$
to denote the partial order.

Two unlabelled trees are identified if they can be superposed by a
suitable continuous deformation, so that the endpoints with the
same index coincide. It is then easy to see that the number of
unlabeled trees with $n$ end-points is bounded by $4^n$.

We shall consider also the {\it labelled trees} (to be called
simply trees in the following); they are defined by associating
some labels with the unlabeled trees, as explained in the
following items.

\0 2) We associate a label $h\le -1$ with the root and we denote
by $\TT_{h,n}$ the corresponding set of labelled trees with $n$
endpoints. Moreover, we introduce a family of vertical lines,
labelled by an integer taking values in $[h,1]$, and we represent
any tree $\t\in\TT_{h,n}$ so that, if $v$ is an endpoint or a non
trivial vertex, it is contained in a vertical line with index
$h_v>h$, to be called the {\it scale} of $v$, while the root is on
the line with index $h$. There is the constraint that, if $v$ is
an endpoint, $h_v>h+1$.

The tree will intersect in general the vertical lines in set of
points different from the root, the endpoints and the non trivial
vertices; these points will be called {\it trivial vertices}. The
set of the {\it vertices} of $\t$ will be the union of the
endpoints, the trivial vertices and the non trivial vertices. Note
that, if $v_1$ and $v_2$ are two vertices and $v_1<v_2$, then
$h_{v_1}<h_{v_2}$.

Moreover, there is only one vertex immediately following the root,
which will be denoted $v_0$ and can not be an endpoint (see
above); its scale is $h+1$.

Finally, if there is only one endpoint, its scale must be equal to
$h+2$.

\0 3) With each endpoint $v$ of scale $h_v=+1$ we associate one of the
monomials in \equ(2.15) contributing to $\VV^{(0)}$ and a set
$\xx_v$ of space-time points (the corresponding integration
variables); with each
endpoint of scale $h_v\le 0$ we associate a contribution {\it of
type $\l$}, that is a contribution of the form \equ(2.25), with
$h=h_v-1$, and the corresponding set $\xx_v$ of space-time points.
We impose the constraint that, if $v$ is an endpoint,
$h_v=h_{v'}+1$, if $v'$ is the non trivial vertex immediately
preceding $v$.

Given a vertex $v$, which is not an endpoint, $\xx_v$ will denote
the family of all space-time points associated with one of the
endpoints following $v$.

\0 4) If $v$ is not an endpoint, the {\it cluster } $L_v$ with
scale $h_v$ is the set of endpoints following the vertex $v$; if
$v$ is an endpoint, it is itself a ({\it trivial}) cluster. The
tree provides an organization of endpoints into a hierarchy of
clusters.

\0 5) The trees containing only the root and an endpoint of scale
$h+1$ will be called the {\it trivial trees}.

\0 6) We introduce a {\it field label} $f$ to distinguish the
field variables appearing in the terms associated with the
endpoints as in item 3); the set of field labels associated with
the endpoint $v$ will be called $I_v$. Analogously, if $v$ is not
an endpoint, we shall call $I_v$ the set of field labels
associated with the endpoints following the vertex $v$; $\xx(f)$,
$\e(f)=\pm$ and $\s(f)=\ud$ will denote the space-time point, 
the creation/annihilation index and the spin index,
respectively, of the field variable with label $f$.

\\
In terms of these trees, the effective potential $\VV^{(h)}$ can
be written as
$$\VV^{(h)}(\psi^{(\le h)}) + L\b \tilde F_{h+1}=
\sum_{n=1}^\io\sum_{\t\in\TT_{h,n}} \VV^{(h)}(\t,\psi^{(\le
h)})\;,\Eq(2.61)$$
where, if $v_0$ is the first vertex of $\t$ and $\t_1,..,\t_s$
($s=s_{v_0}$) are the subtrees of $\t$ with root $v_0$,
$\VV^{(h)}(\t,\psi^{(\le h)})$ is defined inductively by the
relation
$$ \VV^{(h)}(\t,\psi^{(\le h)})= {(-1)^{s+1}\over s!}
\EE^T_{h+1}[ \bar\VV^{(h+1)}(\t_1,\psi^{(\le h+1)});\ldots;
\bar\VV^{(h+1)}(\t_{s},\psi^{(\le h+1)})]\;,\Eq(2.62)$$
and $\bar\VV^{(h+1)}(\t_i,\psi^{(\le h+1)})$

\0 a) is equal to $\RR\VV^{(h+1)}(\t_i,\psi^{(\le h+1)})$ if the
subtree $\t_i$ is not trivial;

\0 b) if $\t_i$ is trivial and $h< -1$, it is equal to
$\LL_4\VV^{(h+1)}$ or, if $h=-1$, to one of the monomials
contributing to $\VV^{(0)}(\psi^{\le 0})$.

\0 $\EE^T_{h+1}$ denotes the truncated expectation with respect to
the measure $\prod_{\o} P(d \psi_\o^{(h+1)})$, that is
$$\eqalign{
&\EE^T_{h+1}(X_1;\ldots;X_p) \={\dpr^p\over\dpr\l_1\ldots\dpr\l_p}
\left.\log\int\prod_{\o} P(d \psi_\o^{(h+1)}) e^{\l_1 X_1+
\cdots\l_pX_p}\right|_{\l_i=0}.\cr}\Eq(2.63)$$
This means, in particular, that, in \equ(2.62), one has to use for
the field variables the sector decomposition \equ(2.48). The
sector decomposition induces a further decomposition of the
functions $\VV^{(h)}(\t,\psi^{(\le h)})$ in the r.h.s. of
\equ(2.61) and in order to describe it we need some more definitions.

We associate with any vertex $v$ of the tree a subset $P_v$ of
$I_v$, the {\it external fields} of $v$. These subsets must
satisfy various constraints. First of all, if $v$ is not an
endpoint and $v_1,\ldots,v_{s_v}$ are the vertices immediately
following it, then $P_v \subset \cup_i P_{v_i}$; if $v$ is an
endpoint, $P_v=I_v$. Given a vertex $v$, $|P_v|=2$ is not allowed 
and vertices $v$ with $|P_v|=4$ are necessarily endpoints.
We shall denote $Q_{v_i}$ the intersection of
$P_v$ and $P_{v_i}$; this definition implies that $P_v=\cup_i
Q_{v_i}$. The subsets $P_{v_i}\bs Q_{v_i}$, whose union ${\cal
I}_v$ will be made, by definition, of the {\it internal fields} of
$v$, have to be non empty, if $s_v>1$. Given $\t\in\TT_{h,n}$, 
there are many possible choices of the
subsets $P_v$, $v\in\t$, compatible with all the constraints. We
shall denote $\PP_\t$ the family of all these choices and $\bP$
the elements of $\PP_\t$.

Moreover, we associate with any $f\in {\cal I}_v$ a scale label
$h(f)=h_v$ and, if $h(f)\le 0$, an index $\o(f)\in O_{h(f)}$,
while, if $h(f)=+1$, we put $\o(f)=0$. Note that, if $h(f)\le 0$,
$h(f)$ and $\o(f)$ single out a sector of scale $h(f)$ and sector
index $\o(f)$ associated with the field variable of index $f$. In
this way we assign $h(f)$ and $\o(f)$ to each field label $f$,
except those which correspond to the set $P_{v_0}$; we associate
with any $f\in P_{v_0}$ the scale label $h(f)=h$ and a sector
index $\o(f)\in O_h$. We shall also put, for any $v\in\t$, $\O_v=
\{\o(f),f\in P_v\}$. We shall call ${\cal O}_\t$
the family of possible values of $\O=\{\o(f), f\in \cup_v I_v\}$.

With these definitions, we can rewrite $\VV^{(h)}(\t,\psi^{(\le h)})$ 
in the r.h.s. of \equ(2.61) as:
$$\eqalign{&\VV^{(h)}(\t,\psi^{(\le
h)})=\sum_{\bP\in\PP_\t,\O\in {\cal O}_\t}
\VV^{(h)}(\t,\bP,\O)\;,\cr &\VV^{(h)}(\t,\bP,\O)=\int d\xx_{v_0}
\tilde\psi^{(\le h)}_{\O_{v_0}} (P_{v_0})
K_{\t,\bP,\O}^{(h+1)}(\xx_{v_0})\;,\cr}\Eq(2.64)$$
where 
$$\tilde\psi^{(\le h)}_{\O_{v}}
(P_{v})=\prod_{f\in
P_v}e^{i\e(f)\vpp_F^{(h)}(\th_{h,\o(f)})\vxx(f)}\psi^{ (\le
h)\e(f)}_{\xx(f),\s(f),\o(f)}\Eq(2.65)$$
and $K_{\t,\bP,\O}^{(h+1)}(\xx_{v_0})$ is defined inductively by
the equation, valid for any $v\in\t$ which is not an endpoint,
$$K_{\t,\bP,\O}^{(h_v)}(\xx_v)={1\over s_v !}
\prod_{i=1}^{s_v} [K^{(h_v+1)}_{v_i}(\xx_{v_i})]\; \;\EE^T_{h_v}[
\tilde\psi^{(h_v)}_{\O_1}(P_{v_1}\bs Q_{v_1}),\ldots,
\tilde\psi^{(h_v)}_{\O_{s_v}}(P_{v_{s_v}}\bs
Q_{v_{s_v}})]\;,\Eq(2.65)$$
where $\O_i=\{\o(f), f\in P_{v_i}\bs Q_{v_i}\}$ and
$\tilde\psi^{(h_v)}_{\O_i}(P_{v_i}\bs Q_{v_i})$ has a definition
similar to \equ(2.65). Moreover, if $v$ is an endpoint and $h_v\le
0$, $K^{(h_v)}_v(\xx_v)= \l_{h_v-1}(\xx_v)$, while if $h_v=+1$
$K^{(1)}_v$ is equal to one of the kernels of the monomials in
\equ(2.15).

\equ(2.61)--\equ(2.64) is not the final form of our expansion;
we further decompose $\VV^{(h)}(\t,\bP,\O)$, by using the
following representation of the truncated expectation in the
r.h.s. of
\equ(2.65). Let us put $s=s_v$, $P_i\=P_{v_i}\bs Q_{v_i}$;
moreover we order in an arbitrary way the sets $P_i^\pm\=\{f\in
P_i,\e(f)=\pm\}$, we call $f_{ij}^\pm$ their elements and we
define $\xx^{(i)}=\cup_{f\in P_i^-}\xx(f)$, $\yy^{(i)}=\cup_{f\in
P_i^+}\xx(f)$, $\xx_{ij}=\xx(f^-_{i,j})$,
$\yy_{ij}=\xx(f^+_{i,j})$. Note that $\sum_{i=1}^s
|P_i^-|=\sum_{i=1}^s |P_i^+|\=n$, otherwise the truncated
expectation vanishes. A couple
$l\=(f^-_{ij},f^+_{i'j'})\=(f^-_l,f^+_l)$ will be called a line
joining the fields with labels $f^-_{ij},f^+_{i'j'}$, sector
indices $\o^-_l=\o(f^-_l)$, $\o^+_l=\o(f^+_l)$ and spin indices
$\s^-_l=\s(f^-_l)$, $\s^+_l=\s(f^+_l)$, connecting the points
$\xx_l\=\xx_{i,j}$ and $\yy_l\=\yy_{i'j'}$, the {\it endpoints} of
$l$. Moreover, if $\o^-_l=\o^+_l$, we shall put
$\o_l\=\o^-_l=\o^+_l$. Then, it is well known that, up to a sign,
if $s>1$,
$$\EE^T_{h}(\tilde\psi^{(h)}_{\O_1}(P_1),...,
\tilde\psi^{(h)}_{\O_s}(P_s))=\sum_{T}\prod_{l\in T} \tilde
g^{(h)}_{\o_l}(\xx_l-\yy_l) \d_{\o^-_l,\o^+_l} \d_{\s^-_l,\s^+_l}
\int dP_{T}(\tt) \det G^{h,T}(\tt)\;,\Eq(2.66)$$
where
$$ \tilde g^{(h)}_\o(\xx) = e^{-i\vpp_F(\th_{h,\o})\vxx}
g^{(h)}_\o(\xx)\Eq(2.67)$$
and $T$ is a set of lines forming an {\it anchored tree graph} between
the clusters of points $\xx^{(i)}\cup\yy^{(i)}$, that is $T$ is a
set of lines, which becomes a tree graph if one identifies all the
points in the same cluster. Moreover $\tt=\{t_{i,i'}\in [0,1],
1\le i,i' \le s\}$, $dP_{T}(\tt)$ is a probability measure with
support on a set of $\tt$ such that $t_{i,i'}=\uu_i\cdot\uu_{i'}$
for some family of vectors $\uu_i\in \RRR^s$ of unit norm. Finally
$G^{h,T}(\tt)$ is a $(n-s+1)\times (n-s+1)$ matrix, whose elements
are given by
$$G^{h,T}_{ij,i'j'}=t_{i,i'} \tilde
g^{(h)}_{\o_l}(\xx_{ij}-\yy_{i'j'})\d_{\o^-_l,\o^+_l}\Eq(2.67a)$$
with $(f^-_{ij}, f^+_{i'j'})$ not belonging to $T$.

In the following we shall use \equ(2.66) even for $s=1$, when $T$
is empty, by interpreting the r.h.s. as equal to $1$, if
$|P_1|=0$, otherwise as equal to $\det
G^{h}=\EE^T_{h}(\tilde\psi^{(h)}(P_1))$. \*

If we apply the expansion \equ(2.66) in each non trivial vertex of
$\t$, we get an expression of the form
$$ \VV^{(h)}(\t,\bP,\O) = \sum_{T\in {\bf T}} \int d\xx_{v_0}
\tilde\psi^{(\le h)}_{\O_{v_0}}(P_{v_0}) W_{\t,\bP,\O\bs
\O_{v_0},T}^{(h)}(\xx_{v_0}) \= \sum_{T\in {\bf T}}
\VV^{(h)}(\t,\bP,\O,T)\;,\Eq(2.68)$$
where ${\bf T}$ is a special family of graphs on the set of points
$\xx_{v_0}$, obtained by putting together an anchored tree graph
$T_v$ for each non trivial vertex $v$. Note that any graph $T\in
{\bf T}$ becomes a tree graph on $\xx_{v_0}$, if one identifies
all the points in the sets $x_v$, for any vertex $v$ which is also
an endpoint. Given $\t\in\TT_{h,n}$ and the labels $\bP,\O,T$,
calling $v_i^*,\ldots,v_n^*$ the endpoints of $\t$ and putting
$h_i=h_{v_i^*}$, the explicit representation of $W_{\t,\bP,\O\bs
\O_{v_0},T}^{(h)}(\xx_{v_0})$ in \equ(2.68) is
$$\eqalign{&
W_{\t,\bP,\O\bs \O_{v_0}, T}(\xx_{v_0}) =\left[\prod_{i=1}^n
K_{v_i^*}^{h_i} (\xx_{v_i^*})\right] \;\cdot\cr &\cdot\;
\Bigg\{\prod_{v\,\atop\hbox{\ottorm not e.p.}}{1\over s_v!} \int
dP_{T_v}(\tt_v)\det G^{h_v,T_v}(\tt_v)\Big[\prod_{l\in
T_v}\d_{\o_l^+,\o_l^-} \d_{\s^-_l,\s^+_l} \tilde
g^{(h_v)}_{\o_l}(\xx_l-\yy_l)]\Big]\Bigg\}\;,\cr}\Eq(2.68a)$$

\* \sub(2.7) {\it Modification of the running coupling functions.}

Let us consider the expansion described in previous section and
let us remark that, thanks to momentum conservation and compact
support properties of propagator Fourier transforms,
$\VV^{(h)}(\t,\bP,\O)$ vanishes for some choices of $\O$. In the
following bounds it will be crucial to take into account this
constraint, and for this reason we introduce a different
representation of the running coupling functions $\l_h$, in order
to include in the new definitions the momentum constraints on the
external lines of the corresponding vertices.

We define, for any $h\le 0$ and $\o\in O_h$, the {\it s-sector}
$S_{h,\o}$ as
$$ S_{h,\o} = \{\vkk=\r\vee_r(\th) \in\RRR^2: |\e_h(\vkk)-\m|\le \g^h e_0,\;
\z_{h,\o}(\th)\not=0\}\Eq(2.69)$$
and note that the  definition of s-sector has the property, to be
used extensively in the following, that the s-sector $S_{h+1,\o}$
of scale $h+1$ contains the union of two s-sectors of scale $h$:
$S_{h+1,\o}\supseteq \left\{ S_{h,2\o}\cup S_{h,2\o+1}\right\}$,
as follows from the definition of $\z_{h,\o}$. By construction and 
by the inductive assumption on $E_h(\kk)$, see \equ(2.36), we see that 
for any $h\le 0$ and any $\o\in O_{h+1}$ the s-sectors $S_{h,2\o}$ and 
$S_{h,2\o+1}$ are the two only sectors on scale $h$ strictly contained into 
$S_{h+1,\o}$; here we used that $\g=4$ (so that $\g^{1/2}=2$).

We now observe that the field variables
$\hat\psi_{\kk(f),\o(f),\s(f)}^{\le h_{v_0},\e(f)}$ have the same
supports as the functions $C^{-1}_{h_{v_0}}(\kk(f))$
$\z_{h_{v_0},\o(f)}(\th(f))$ and $h(f)\le h_i-1$, $\forall f\in
P_{v_i^*}$; hence in the expression \equ(2.68a), we can freely
multiply $\hat K_{v_i^*}^{h_i}(\kk_{v_i^*})$ by $\prod_{f\in
P_{v^*_i}} \tilde F_{h_i-1,\tilde\o(f)}(\vkk)$, where $\tilde
F_{h,\o}(\vkk)$ is a smooth function equal to $1$ on $S_{h,\o}$
and with a support slightly greater than $S_{h,\o}$, while
$\tilde\o(f)\in O_{h_i-1}$ is the unique sector index such that
$S_{h(f),\o(f)} \subseteq S_{h_i-1,\tilde\o(f)}$. In order to
formalize this statement, it is useful to introduce the following
definition.

Let $G(\vxxx)$ be a function of $2p$ variables $\vxxx=(\vxx_1,
\ldots, \vxx_{2p})$ with Fourier transform $\hat G(\vkkk)$,
defined so that $G(\vxxx)=\int d\vkkk (2\p)^{-4p}
\exp(-i\sum_{l=1}^{2p} \e_i\vkk_i\vxx_i) \hat G(\vkkk)$, where
$\e_1, \ldots, \e_p=-\e_{p+1}= \ldots = -\e_{2p}=+1$. Then, we
define, given $h\le 0$ and a family $\oo=\{\o_i\in O_h,
i=1,\ldots, 2p \}$ of sector indices,
$$ (\FFF_{2p,h,\oo} * G)(\vxxx) = \int {d\vkkk\over(2\p)^{4p}}
e^{-i\sum_{l=1}^{2p} \e_i\vkk_i\vxx_i} \left[ \prod_{i=1}^{2p}
\tilde F_{h,\o_i}(\vkk_i) \right] \hat G(\vkkk)\;.\Eq(2.70)$$
In order to extend this definition to the case $h=1$, when the
sector index can take only the value $0$, we define $\tilde
F_{1,0}(\vkk)$ as a smooth function of compact support, equal to
$1$ on the support of $\bar C_0^{-1}(\vkk)$, defined above.

Hence, if we put $p_i=|P_{v_i^*}|$, $\O_{v_i^*} = \{\o(f), f\in
P_{v_i^*}\}$ and we define
$$ \tilde K_{v_i^*, \O_{v_i^*}}^{h_i}(\xx_{v_i^*}) =
\left( \FFF_{p_i,h_i-1,\O_{v_i^*}} * K_{v_i^*}^{h_i}
\right)(\xx_{v_i^*})\;,\Eq(2.71)$$
we can substitute in \equ(2.68a) each
$K_{v_i^*}^{h_i}(\xx_{v_i^*})$ with $\tilde K^{h_i}_{v_i^*,
\O_{v_i^*}}(\xx_{v_i^*})$, to be called the {\it modified coupling
functions}. In particular, if $v_i^*$ is of type $\l$, we shall
denote the corresponding modified coupling functions by the symbol
$\tilde\l_{h_i-1,\O_{v_i^*}} (\xx_{v_i^*})$. The {\it smallness
condition on $\l$} can be now stated in terms of
$\tilde\l_{h_i-1,\O_{v_i^*}} (\xx_{v_i^*})$ as follows:
$$\fra{1}{L^2\b}\int d\xx_{v_i^*}
|\tilde\l_{h_i-1,\O_{v_i^*}} (\xx_{v_i^*})|\le C|U|\;,\Eq(2.71a)$$
for some constant $C>0$. \\

We shall call $W^{(mod)}_{\t,\bP,\O,T}(\xx_{v_0})$ the expression
we get from $W_{\t,\bP,\O\bs \O_{v_0},T}(\xx_{v_0})$ by the
substitution of the kernels $K_{v_i^*}^{h_i}(\xx_{v_i^*})$ with
the modified ones. Note that $W^{(mod)}_{\t,\bP,\O,T}(\xx_{v_0})$
is not independent of $\O_{v_0}$, unlike $W_{\t,\bP,\O\bs
\O_{v_0},T}(\xx_{v_0})$, and that
$W^{(mod)}_{\t,\bP,\O,T}(\xx_{v_0})$ is equal to $W_{\t,\bP,\O\bs
\O_{v_0},T}(\xx_{v_0})$, only if $|P_{v_0}|=0$; however, the
previous considerations imply that, if $p_0=|P_{v_0}|>0$,
$$ \left( \FFF_{p_0,h,\O_{v_0}} *
W^{(mod)}_{\t,\bP,\O,T} \right)(\xx_{v_0})= \left(
\FFF_{p_0,h,\O_{v_0}}
* W_{\t,\bP,\O\bs \O_{v_0},T} \right)(\xx_{v_0})\;,\Eq(2.72)$$
a trivial remark which will be important in the discussion of the
running coupling functions flow in next Chapter.

\* \sub(2.8) {\it Bounds for the effective potentials and the free
energy.}

In order to explicitly keep track of the constraints satisfied by
the sector indices $\O$, which is crucial for performing the
dimensional bounds on the free energy, we introduce the following
constraint functions. Given a tree $\t\in\TT_{h,n}$ with all its labels, 
a vertex $v\in\t$ and the set of anisotropic sector indices 
$\O_{v}=\{\o(f)\in O_{h(f)}, f\in P_v\}$ labelled by $P_v$, we define 
$$\c(\O_v) = \cases {\openone
\Big( \forall f\in P_v, \exists \vkk(f)\in S_{h(f),\o(f)}:
\sum_{f\in P_v}\e(f)\vkk(f)=0\Big)\;,\qquad {\rm if}\qquad |P_v|\le 8\cr
\ \cr
1\;,\hskip8.7truecm {\rm if}\qquad |P_v|\ge 10\cr}\Eq(2.73)$$
where the function $\openone(condition)$ is the 
function $=1$ if $condition$ is satisfied and $=0$ otherwise.

The previous considerations imply that
$$F_{L,\b} \le \sum_{h=h_\b}^1 \sum_{n=1}^\io
\sum_{\t\in\TT_{h,n}} \sum_{\bP\in\PP_\t\atop |P_{v_0}|=0}
\sum_{T\in {\bf T}} J^{(0)}_{h,n}(\t,\bP,T)\;,\Eq(2.75)$$
with
$$\eqalign{
J_{h,n}^{(F)}(\t,\bP,T)&= \sum_{\O\setminus\O_{ext}^{(F)}}
\left[\prod_{v}\c(\O_v) \right]\int d(\xx_{v_0}\bs \xx^*)
\left| W^{(mod)}_{\t,\bP,\O,T}(\xx_{v_0})
\right|\;,\cr}\Eq(2.76)$$
where $\xx^*$ is an arbitrary point in $\xx_{v_0}$, $\O\in\OO_\t$ and, if
$2l_0=|P_{v_0}|>0$ and $0<F\le 2l_0$, $\O_{ext}^{(F)}\subset \O_{v_0}$
is an arbitrary subset of the sector indices in $\O_{v_0}$
of cardinality $F$, $|\O_{ext}^{(F)}|=F$: in particular, if $l_0=0$ or $F=0$,
$\sum_{\O\setminus\O_{ext}^{(F)}\in\OO_\t}$ coincides with 
$\sum_{\O\in\OO_\t}$. Note
that we could freely insert $\prod_{v}\c(\O_v)$ in
\equ(2.76), because of the constraints following from momentum
conservation and the compact support properties of propagator
Fourier transforms: here we used that umklapp processes are impossible if 
$|P_v|\le 8$ (because of the condition on $\m$ chosen in Theorem \thm(1.1));
for vertices with $|P_v|\ge 10$ we discarded any possible constraint coming 
from momentum conservation (modulo $2\p\ZZZ^2$).

The following Theorem will be the starting point of our analysis.

\* \theorem(2.1) { \it Given $h_\b\le h\le 0$, $\t\in\TT_{h,n}$,
$\bP\in \PP_\t$, $T\in\bT$, if
$E_j(\kk)$ satisfies \equ(2.36) for any $j\ge h$, $\tilde
\l_{h_{v^*}-1,\O_{v^*}}$ satisfies \equ(2.71a) for any endpoint $v^*\in\t$
and $U_0|h_\b|=c_0$ is small enough, then
$$\eqalign{
& J_{h,n}^{(0)}(\t,\bP,T) \le (c|U|)^n
\g^{h(2-\fra{3}{4}|P_{v_0}|)} \prod_{v\ {\rm not}\ {\rm e. p.
}}\fra{1}{s_v!}\g^{\d(|P_v|)}\;,\cr & J_{h,n}^{(1)}(\t,\bP,T) \le
(c|U|)^n \g^{h(\fra{5}{2}- \fra{3}{4}|P_{v_0}|)}\prod_{v\ {\rm
not}\ {\rm e. p. }} \fra{1}{s_v!}\g^{\d(|P_v|)}\;,\cr}\Eq(2.77)$$
where
$$\d(p) = 1-\fra{p}{4}+\openone(p\ge 10)\;.\Eq(2.78)$$
}\\

{\bf Remark.} By construction, clusters with $|P_v|=2$ are
not allowed, and clusters with $|P_v|=4$ are necessarily
endpoints. Then the exponent $\d(|P_v|)$ appearing in \equ(2.77)
is always $\le -1/2$ and, as a consequence, it can be easily
proved that the sums over $\bP$ and $\t$ in \equ(2.75) converge
exponentially, see [BGM], so that the sum in the r.h.s. of
\equ(2.75) is absolutely convergent and, under the assumptions of
the Theorem, $\lim_{L\to\io}F_{L,\b}$ does exist and is $O(U)$.\\

In the remaining part of this section we shall sketch the proof of
Theorem \thm(2.1), already presented in section 2.7 of [BGM] (see
in particular eq. 2.101) in a slightly different case and with a
smallness condition on the $\tilde \l_h$ different from
\equ(2.71a) (but dimensionally equivalent). Moreover,
the proof will be presented here in a form convenient for
performing the improved dimensional bounds on $E_h(\kk)$ in next Chapter.\\

\proof We shall describe the proof in the case that all the
endpoints are of type $\l$. A posteriori, it will be clear that
the possible presence of endpoints of scale $+1$ with $p\ge 6$
external legs (produced by the ultraviolet integration, see
Appendix A1) does not qualitatively change the argument.

We begin with considering the case $F=1$. 
By using the definition of $W^{(mod)}_{\t,\bP,\O,T}(\xx_{v_0})$
and the expression \equ(2.68a), we can bound the r.h.s. of
\equ(2.76) as
$$\eqalign{&\sum_{\O\bs\O^{(1)}_{ext}}
\left[\prod_{v}\c(\O_v) \right]\int\prod_{l\in T^*}
d(\xx_l-\yy_l) \left[\prod_{i=1}^n |\wt\l_{h_i-1,\O_{v_i^*}}
(\xx_{v_i^*})|\right]\cdot\cr
&\qquad\qquad\cdot\Bigg\{\prod_{v\,\atop\hbox{\ottorm not
e.p.}}{1\over s_v!} \max_{\tt_v}|\det
G^{h_v,T_v}(\tt_v)|\prod_{l\in T_v}|
g^{(h_v)}_{\o_l}(\xx_l-\yy_l)|\Bigg\}\;,\cr}\Eq(2.79)$$
where $T^*$ is a tree graph obtained from $T=\cup_vT_v$, by adding
in a suitable (obvious) way, for each endpoint $v_i^*$,
$i=1,\ldots,n$, one or more lines connecting the space-time points
belonging to $\xx_{v_i^*}$.

A standard application of Gram--Hadamard inequality, combined with
the dimensional bound on $g^{(h)}_\o$ given by Lemma \lm(2.2),
implies (see [BGM]) that
$$|\det G_\a^{h_v,T_v}(\tt_v)| \le
c^{\sum_{i=1}^{s_v}|P_{v_i}|-|P_v|-2(s_v-1)}\cdot\;
\g^{{h_v}{3\over
4}\left(\sum_{i=1}^{s_v}|P_{v_i}|-|P_v|-2(s_v-1)\right)}\;.
\Eq(2.80)$$
By the decay properties of $g^{(h)}_\o$ given by Lemma \lm(2.2) it
also follows that
$$\prod_{v\,\hbox{\ottorm not e.p.}}
{1\over s_v!}\int \prod_{l\in T_v} d(\xx_l-\yy_l)
|g^{(h_v)}_{\o_l}(\xx_l-\yy_l)|\le c^n \prod_{v\,\hbox{\ottorm not
e.p.}} \fra{1}{s_v!} \g^{-h_v(s_v-1)}\;.\Eq(2.81)$$
The smallness assumption \equ(2.71a) on the size of
$\tilde\l_{h_v-1,\O_v}$ implies that
$$\int\prod_{l\in T^*\setminus\cup_v T_v}d(\xx_l-\yy_l)
\prod_{i=1}^n |\wt\l_{h_i-1,\O_{v_i^*}} (\xx_{v_i^*})|\le
(C|U|)^n\;. \Eq(2.82)$$
Finally, as we shall explain below, by suitably taking into
account the constraint functions $\c(\SS(P_v))$, the sum over
the choices of the sector indices gives
$$\eqalign{&\sum_{\O\setminus\O_{ext}^{(1)}} \prod_{v\in\t}
\left(\c(\O_v)\prod_{l\in T_v} \d_{\o_l^+,\o_l^-}\right)\le
\cr & \le c^n \g^{-{1\over 2}h m_4(v_0)} \prod_{v {\rm \ not\ e.\
p.}}\g^{\left[ -{1\over 2} m_4(v)+{1\over 2}(|P_v|-3)\openonesix (4\leq
|P_v|\leq 8)+ {1\over 2}(|P_v|-1)\openonesix(|P_v|\geq
10)\right]}\;,\cr}\Eq(2.83)$$
where $m_4(v)$ denotes the number of endpoints of type $\l$
following $v$ on $\t$. 
It is straightforward to check that Theorem \thm(2.1) in the case $F=1$ 
follows by
combining the bounds \equ(2.80), \equ(2.81), \equ(2.82) and \equ(2.83).

If $F=0$ the Theorem simply follows by the remark that the r.h.s. of 
\equ(2.76), that is of the form $\sum_\O[\cdots]$ can be rewritten as 
$\sum_{\o_1\in O_h}\sum_{\O\bs\o_1}[\cdots]$, where $\o_1$ is an arbitrary
sector index in $\O_{v_0}$. Now, the sum $\sum_{\O\bs\o_1}[\cdots]$ can be 
bounded as in the second line of \equ(2.77), the first sum $\sum_{\o_1\in O_h}$
gives a contribution $O(\g^{-h/2})$ and the first line of \equ(2.77) follows.\\

Let us now give a sketch of the proof of \equ(2.83), that will be
the starting point of the proof of the improved dimensional bounds
on $E_h(\kk)$, to be described in next Chapter.

Let us first note that, by the definition of $s$--sector
$S_{h,\o}$ and by the properties of $E_h(\kk)$, the following crucial property
is true:\\
\\
{\it given $\o\in O_h$ and the $s$--sector $S_{h,\o}$, for any
$j>h$ there is a unique $\o'(j;\o)\in O_j$ such that
$S_{j,\o'(j;\o)} \supset S_{h,\o}$.}\\

This property allows us to give a meaning to the following
definition. Given $\t\in\TT_{h,n}$, $\bP\in\PP_\t$ and $v\in\t$,
we introduce the symbol $\O_v^{(j)}$ to denote the set
$$\eqalign{&\O_v^{(j)}=\{\o(f), f\in P_v\ :\ h(f)\ge j\}
\cup \{\o'(j;\o(f)), f\in P_v\ :\ j>h(f)\}\=\cr
&\hskip.63truecm\=\{\o^{(h)}_f\in O_{j^{(h)}_f}, f\in P_v\}
\;,\cr}\Eq(2.84)$$
where the last identity {\it defines} the scales $j_f^{(h)}$ and the sector 
indices $\o_f^{(h)}$, $f\in P_v$.
The definition \equ(2.84) implies in particular that $\O_v=\O_v^{(h)}$.
With reference to definition \equ(2.84), 
in the following we shall also denote:
$$\c(\O_v^{(j)})=\openone\Big( \forall f\in P_f\,, \exists \vkk(f)
\in S_{j_f^{(h)},\o_f^{(h)}}\, :\, \sum_{f\in P_f}\e(f)\vkk(f)={\vec 0}
\Big)\;,\Eq(2.84a)$$
Note in particular that:
$$\c(\O_v^{(j)})\le \c(\O_v^{(k)})\;, \qquad{\rm if}\qquad
j\le k\;.\Eq(2.85)$$
Given $\t\in\TT_{h,n}$, we define the set $V_c(\t)$ 
of $c$--vertices of $\t$ as the set of vertices $v$ of $\t$ which either 
are endpoints or have the property that their  
set ${\cal{I}}_v$ of internal lines is non empty\footnote{${}^1$}{\ottopunti
With reference to the definition of $\c$--vertices introduced in Sect.3.1 of 
[BGM], a vertex is called a $c$--vertex if it is either an endpoint
or a $\c$--vertex. The prefix $c$-- recalls that a constraint function $\c$
is associated to any $c$--vertex.}; in the following we shall 
often drop the dependence on $\t$ (when it is clear from the context).
Note that by definition, if $\t\in\TT_{h,n}$, then $|V_c(\t)|=O(n)$ 
and it holds $\prod_{v\in\t} \c(\O_v) = \prod_{v\in V_c}
\c(\O_v)$. Moreover, using that $\O_v=\O_v^{(h)}$,
we see that we can replace the product
$\prod_{v\in\t}\c(\O_v)$ in the l.h.s. of \equ(2.83) by $\prod_{v\in
V_c}\c(\O_v^{(h)})$.

We now begin to inductively bound the l.h.s. of
\equ(2.83); first of all we shall bound the sum corresponding to the
first $c$--vertex following the root; then we will iteratively
enter its structure. After each step we will be left with a
product of sector sums of the same form of the initial one, but on
larger scales.

We call $\tilde v_0$ the first $c$--vertex following the root and
$h_0\=h_{\tilde v_0}-1$ the maximal possible value for the scale
of its external legs; using \equ(2.85), we find
$$\prod_{v\in V_c}
\c(\O_v^{(h)})\le \c
(\O_{\tilde v_0}^{(h)})\prod_{v>\tilde v_0,v\in V_c}
\c(\O_v^{(h_{0})})\;.\Eq(2.87)$$
Substituting \equ(2.87) into the l.h.s. of \equ(2.83), we find
$$\eqalign{&\sum_{\O\setminus\O_{ext}^{(1)}} \prod_{v\in V_c}
\left(\c(\O_v^{(h)})\prod_{l\in T_v}
\d_{\o_l^+,\o_l^-}\right)\le \cr &\le\sum^*_{\O_{\tilde
v_0}^{(h_{0})}}\Big[ \sum^*_{\O_{\tilde v_0}^{(h)}\prec \O_{\tilde
v_0}^{(h_{0})}}\c(\O_{\tilde
v_0}^{(h)})\Big]\sum_{\O\setminus\O_{\tilde v_0}} \prod_{v>\tilde
v_0, v\in V_c} \c\big(\O_v^{(h_{0})}\big)\prod_{l\in T}
\d_{\o_l^+,\o_l^-}\;,\cr}\Eq(2.88)$$
where $\O_{\tilde v_0}^{(h)}\prec \O_{\tilde v_0}^{(h_{0})}$ means
that the indices in $\O_{\tilde v_0}^{(h)}$ satisfy the following constraint:
given $f\in P_v$ and the corresponding index 
$\o_f^{(h_0)}\in O_{j_f^{(h_0)}}$ 
of $\O_{\tilde v_0}^{(h_{0})}$, 
then the index $\o_f^{(h)}\in O_{j_f^{(h)}}$ 
of $\O_{\tilde v_0}^{(h)}$ is such that 
$S_{j_f^{(h)},\o_f^{(h)}}\subset S_{j_f^{(h_0)},\o_f^{(h_0)}}$. 
The symbol $*$ on the sums in the second line means 
that the sector index in $\O_{ext}^{(1)}$, associated to one of the
fields in $P_{\tilde v_0}$ (say to the field $f_0\in
P_{\tilde v_0}$), is not summed over.

By the {\it sector counting Lemma} (see Appendix A3), the sum in
square brackets can be bounded, {\it uniformly in $\O_{\tilde
v_0}^{(h_{0})}$}, as
$$\sum^*_{\O_{\tilde v_0}^{(h)}\prec
\O_{\tilde v_0}^{(h_0)}}\c
(\O_{\tilde v_0}^{(h)})\le c \g^{(h_{0}-h) \left[ {1\over
2}(|P_{\tilde v_0}|-3 ) \openonesix (4\leq |P_{\tilde v_0}|\leq 8)+{1\over
2}(|P_{\tilde v_0}|-1)\openonesix(|P_{\tilde v_0}|\ge
10)\right]}\Eq(2.89)$$
so that the r.h.s. of \equ(2.88) can be bounded by the r.h.s. of
\equ(2.89) times
$$\sum^*_{\cup_v \O_v^{(h_{0})}} \Big[
\prod_{v>\tilde v_0,v\in V_c}\ \c
(\O_v^{(h_{0})})\Big]\Big[\prod_{l\in T}\d_{\o_l^+,\o_l^-}
\Big]\;,\Eq(2.90)$$
where the $*$ on the sum recalls again that we are not summing
over the sector index of $f_0\in P_{\tilde v_0}$.

We will now prove that \equ(2.90) can be reduced to a product of
contributions analogue to the l.h.s. of \equ(2.88), with $h_0$
replacing $h$. In fact, calling $\ul{\tilde
v}_0=\{v_1,\ldots,v_{s_{\tilde v_0}}\}$ the set of $c$--vertices
immediately following $\tilde v_0$ on $\t$ and $\O_{\ul{\tilde
v}_0}^{(h_0)}= \cup_{v\in\ul{\tilde v}_0} \O_{v}^{(h_{0})}$, we
can rewrite \equ(2.90) as
$$\eqalignno{&\sum_{\O_{\ul{\tilde v}_0}^{(h_0)}}^*
\prod_{v\in\ul{\tilde v}_0}\left[\ \sum_{\cup_{w> v}
\O_w^{(h_{0})}\setminus \O_{v}^{ (h_{0})}}\ \prod_{w\ge v, w\in
V_c} \c\big(\O_w^{(h_{0})}\big)
\prod_{l\in\cup_{w\ge v}T_w} \d_{\o_l^+,\o_l^-}\right]\prod_{l\in
T_{\tilde v_0}}\d_{\o_l^+,\o_l^-}\=&\cr &\=\sum_{\O_{\ul{\tilde
v}_0}^{(h_0)}}^* \prod_{v\in\ul{\tilde v}_0}
F_{v}(\O_{v}^{(h_{0})}) \prod_{l\in T_{\tilde
v_0}}\d_{\o_l^+,\o_l^-}  \;.&\eq(2.91) \cr}$$
The function $F_{v}(\O_{v}^{(h_{0})})$, defined by \equ(2.91), is
the sum over the ``internal sector indices'' of the product of the
constraint functions corresponding to the vertex $v$. Note that
also the l.h.s. of
\equ(2.88) could have been written in terms of one of this
functions; the l.h.s. of
\equ(2.88) is in fact equal to $\sum^*_{\O_{\tilde v_0}^{(h)}}
F_{\tilde v_0}(\O_{\tilde v_0}^{(h)})$.

We now choose as the root of $T_{\tilde v_0}$ the vertex $v_i\in
\ul{\tilde v}_0$ such that $f_0\in P_{v_i}$; then we select a leaf
$v^*$ of $T_{\tilde v_0}$ and we call $l^*$ the branch of
$T_{\tilde v_0}$ anchored to $v^*$. Calling $\O_{\ul{\tilde
v}_0\setminus v^* }^{(h_0)}=\cup_{v\in\ul{\tilde v}_0\setminus
v^*} \O_{v}^{(h_{0})}$, we denote by ${\cal F}_{\ul{\tilde
v}_0\setminus v^*}(\O_{\ul{\tilde v}_0 \setminus v^*})$ the
product of the constraint functions corresponding to the set of
vertices $\ul{\tilde v}_0\setminus v^*$:
$${\cal F}_{\ul{\tilde v}_0\setminus v^*}(\O^{(h_0)}_{\ul{\tilde v}_0
\setminus v^*})\defin \prod_{v\in\ul{\tilde v}_0\setminus v^*}
F_{v}(\O_{v}^{(h_{0})}) \prod_{l\in T_{\tilde v_0}\setminus
l^*}\d_{\o_l^+,\o_l^-}\Eq(2.92)$$
so that we can rewrite \equ(2.91) as
$$
\sum_{\o_{l^*}^+,\o_{l^*}^-}\d_{\o_{l^*}^+,\o_{l^*}^-} 
\sum_{\O_{\ul{\tilde v}_0\setminus v^*}}^{**}\
{\cal F}_{\ul{\tilde v}_0\setminus v^*}\big(\O^{(h_0)}_{\ul{\tilde v}_0
\setminus v^*}\big)
\sum_{\O_{v^*}^{(h_0)}}^*F_{v^*}\big(\O_{v^*}^{(h_{0})}\big)
\;,\Eq(2.93)$$
where the $**$ on the second sum means that we are not summing
neither on $\o_{f_0}$ nor on $\o_{l^*}$ and the $*$ on the third
sum recalls that we are not summing over $\o_{l^*}$. Bounding the
last sum by $\sup_{\o_{l^*}}\sum_{\O_{v^*}^{(h_0)}}^*F_{v^*}
(\O_{v^*}^{(h_{0})})$, we see that the last sum can be factorized
out:
$$\equ(2.93)\le \sum_{\O_{\ul{\tilde v}_0\setminus v^*}}^*
{\cal F}_{\ul{\tilde v}_0\setminus v^*}(\O^{(h_0)}_{\ul{\tilde v}_0
\setminus v^*}) \cdot\sup_{\o_{l^*}}
\Big[\sum_{\O_{v^*}^{(h_0)}}^*F_{v^*}
(\O_{v^*}^{(h_{0})})\Big]\;.\Eq(2.94)$$
It is now clear that we can iterate the same procedure choosing
another leaf of $T_{\tilde v_0}\setminus l^*$ and by factorizing
out the corresponding contribution. At the end of the procedure we
reach the root of $T_{\tilde v_0}$ and we finally find that
\equ(2.90) can be bounded by
$$\prod_{v\in \ul{\tilde v}_0}\sup_{\o_v^*}\sum_{\O_v^{(h_0)}}^*F_{v}
(\O_v^{(h_0)})\Eq(2.95)$$
where $\o_v^*$ is the sector index corresponding to the line of
$T_{\tilde v_0}$ exiting from $v\in\ul{\tilde v}_0$ and the $*$ on
the sum $\sum_{\O_v^{(h_0)}}^*$ means that we are not summing over
$\o_v^*$.

Now, if $v\in\ul{\tilde v}_0$ is an endpoint, then the
corresponding contribution in \equ(2.95) can be easily bounded by
$$\sum_{\O_v^{(h_0)}}^*F_{v}
(\O_v^{(h_0)})\le c\g^{-\fra{h_0}{2}}\;,\qquad {\rm if} \ v\ {\rm
is}\ {\rm an}\ {\rm endpoint}\;,\Eq(2.96)$$
where we used the sector counting Lemma proved in Appendix A3.

If $v\in\ul{\tilde v}_0$ is not an endpoint, the corresponding
factor in \equ(2.95) has exactly the same form as the l.h.s. of
\equ(2.88), and we can bound it by repeating the same procedure
described above: then, proceeding by induction, we find
\equ(2.83). \qed
\\

Before concluding this section, let us state a generalization of
the bound \equ(2.83), proved in Appendix \secc(A4), 
that will be useful in the following.

\*\lemma(2.4) {\it Given $h_\b\le h\le 0$ and a tree $\t\in\TT_{h,n}$
with all its labels, let us consider the sum
$$\sum_{\O\setminus\O_{ext}^{(F)}}\prod_{v\in\t}\Big(
\c(\O_v)\prod_{l\in T_v}\d_{\o_l^+,\o_l^-}\Big)\Eq(2.97)$$
with $F$ compatible with $\bP$, \ie $F\le |P_{v_0}|$.
If $F=3$, \equ(2.97) can be bounded by the r.h.s. of \equ(2.83)
times $\g^{\fra{h}2}$ and, if $F=5$, 
\equ(2.97) can be bounded by the r.h.s. of \equ(2.83)
times $\g^{{h}}$.}
\\

Lemma \lm(2.4) allows to get the following generalization of 
Theorem \thm(2.1).

\*\lemma(2.5) { \it Given $h_\b\le h\le 0$, $\t\in\TT_{h,n}$,
$\bP\in \PP_\t$, $T\in\bT$, if
$E_k(\kk)$ satisfies \equ(2.36) for any $k\ge h$, if $\tilde
\l_{h_{v^*}-1,\O_{v^*}}$ satisfies
\equ(2.71a) for any endpoint $v^*\in\t$ and if $U_0|h_\b|=c_0$ is small enough,
then
$$\eqalign{&
J_{h,n}^{(3)}(\t,\bP,T) \le (c|U|)^n \g^{h(3-\fra{3}{4}|P_{v_0}|)}
\prod_{v\ {\rm not}\ {\rm e. p.
}}\fra{1}{s_v!}\g^{\d(|P_v|)}\;,\cr &J_{h,n}^{(5)}(\t,\bP,T) \le
(c|U|)^n \g^{h(\fra{7}{2}- \fra{3}{4}|P_{v_0}|)}\prod_{v\ {\rm
not}\ {\rm e. p. }} \fra{1}{s_v!}\g^{\d(|P_v|)}\;,\cr}\Eq(2.98)$$
}
\\

The proof of Lemma \lm(2.5) consists in a repetition of the proof
of Theorem \thm(2.1), unless for the fact that one has to use the 
estimates described in Lemma \lm(2.4) in order to bound 
the sum over the sector indices. Lemma \lm(2.4) guarantees
exactly that the analogue of the sum in \equ(2.83) with 3 
external fixed sectors is bounded by the
r.h.s. of \equ(2.83) times a gain $\g^{h/2}$ and the analogue of 
the sum in \equ(2.83)
with 5 external fixed sectors is bounded by the r.h.s. of
\equ(2.83) times $\g^{h}$, and \equ(2.98) follows.\\

\* \sub(2.9) {\it The two point Schwinger function.}

In this section we want to describe how to modify the expansion for the 
free energy described in previous section in order to compute the two point
Schwinger function. 

The Schwinger functions \equ(2.8) can be derived from the 
{\it generating function} defined as
$$\WW(\phi)=\log\int P(d\psi)e^{-\VV(\psi)+\int d\xx \left[\phi^+_{\xx,\s}
\psi^-_{\xx,\s}+\psi^+_{\xx,\s}
\phi^-_{\xx,\s}\right]}\Eq(4.1)$$
where the variables $\phi^\e_{\xx,\s}$ are Grassmann variables, anticommuting 
among themselves and with the variables $\psi^\e_{\xx,\s}$. 
In particular the two--point Schwinger function 
$S(\xx-\yy)\=S(\xx,\s,-;$ $\yy,\s,+)$, see \equ(2.8), is given by
$$S(\xx-\yy)={\partial^2\over \phi^+_{\xx,\s} 
\phi^-_{\yy,\s}}\WW(\phi)\Big|_{\phi=0}\Eq(4.2)$$
We start by studying the generating function \equ(4.1) and, in analogy with 
the procedure described in Section \secc(2.2), we begin 
by decomposing the field $\psi$ in an ultraviollet and an infrared component:
$\psi=\psi^{(1)}+\psi^{(\le 0)}$, see \equ(2.10). Proceeding through the 
analogues of \equ(2.12) and \equ(2.13), after the integration of the 
$\psi^{(1)}$ variables, we can rewrite:
$$e^{\WW(\phi)}=e^{-L^2\b F_0+S^{(\ge 0)}(\phi)}\int P(d\psi^{(\le 0)})
e^{-\VV^{(0)}(\psi^{(\le 0)})-B^{(0)}(\psi^{(\le 0)},\phi)
+\int d\xx \left[\phi^+_{\xx,\s}\psi^-_{\xx,\s}+\psi^+_{\xx,\s}
\phi^-_{\xx,\s}\right]}\Eq(4.3)$$
where $S^{(\ge 0)}(\phi)$ collects the terms depending on $\phi$ but not 
on $\psi^{(\le 0)}$ and $B^{(0)}(\psi^{(\le 0)},\phi)$ the terms depending
both on $\phi$ and $\psi^{(\le 0)}$, at least quadratic in $\phi$.

Proceeding as in Section \secc(2.3), we can show inductively that for any 
$h_\b\le h< 0$, $e^{\WW(\phi)}$ can be rewritten in a way similar to the 
r.h.s. of \equ(4.3) (and analogous to \equ(2.17)):
$$\eqalign{&e^{\WW(\phi)}=e^{-L^2\b F_h+S^{(\ge h)}(\phi)}\cdot\cr
&\cdot
\int P_{E_h,C_h}
(d\psi^{(\le h)})e^{-\VV^{(h)}(\psi^{(\le h)})-B^{(h)}(\psi^{(\le h)},\phi)
+\int d\kk \left[\hat\phi^+_{\kk,\s}\hat Q^{(h+1)}_\kk\hat\psi^-_{
\kk,\s}+\hat\psi^+_{\kk,\s}\hat Q^{(h+1)}_\kk\hat\phi^-_{\kk,\s}\right]}\;,
\cr}\Eq(4.4)$$
where $\int d\kk$ must be interpreted as equal to 
$\fra1{\b L^2}\sum_{\kk\in\DD_{\b,L}}$ and
$B^{(h)}(\psi^{(\le h)},\phi)$ can be written as 
$B^{(h)}_\phi(\psi^{(\le h)})+\lis W_R^{(h)}$, with $\lis W_R^{(h)}$ 
containing three or more $\phi$ fields, and $B^{(h)}_\phi(\psi^{(\le h)})$ 
of the form
$$\eqalign{&\int d\xx\Big[
\phi^+_{\cdot,\s}*\,G^{(h+1)}*\,{\partial\over 
\partial\psi^{(\le h)+}_{\cdot,\s}}\VV^{(h)}(\psi^{(\le h)})+{\partial\over 
\partial\psi^{(\le h)-}_{\cdot,\s}}\VV^{(h)}(\psi^{(\le h)})\,* 
G^{(h+1)}*\phi^-_{\cdot,\s}\Big](\xx)+\cr
+& \int d\xx\Big[\phi^+_{\cdot,\s}*G^{(h+1)}*
\fra{\dpr^2}{\dpr\psi^{(\le h)+}_{\cdot,\s}
\dpr\psi^{(\le h)-}_{\cdot,\s'}}\hat
\VV^{(h)}(\psi^{(\le h)})*G^{(h+1)}*\phi^-_{\cdot,\s'}\Big](\xx)+\lis W_R^{(h)}
\;,\cr}\Eq(4.6)$$ 
where $\lis W_R^{(h)}$ contains the terms of third or higher order in $\phi$,
$G^{(h+1)}$ is defined as:
$$G^{(h+1)}(\xx)=\sum_{j\ge h+1} g^{(k)}*Q^{(k)}(\xx)\Eq(4.7)$$
and $\hat Q^{(h)}_\kk$ is defined inductively by the relations
$$\hat Q^{(h)}_\kk=\hat Q^{(h+1)}_\kk-\hat n_h(\kk)\hat G^{(h+1)}(\kk)\;,
\quad\quad Q^{(1)}_\kk\=1\;,\Eq(4.8)$$
where $\hat n_h(\kk)$ was defined in \equ(2.22)--\equ(2.23).
Note that, by the compact support properties 
of $\hat g^{(h)}(\kk)$, it holds that, if $\hat g^{(h)}(\kk)\not=0$
$$\hat Q^{(h)}_\kk=1-\hat n_h(\kk) \hat g^{(h+1)}(\kk)\hat Q^{(h+1)}_\kk
\;,\Eq(4.8a)$$
so that, proceeding by induction, we see that on the support of 
$\hat g^{(h)}(\kk)$ we have
$$|\hat Q^{(h)}_\kk-1|\le C|U||h|\g^{h}\;,\quad\quad
|\dpr_\kk^n\hat Q^{(h)}_\kk|\le C_n|U||h|\g^{(1-n)h}\;,\Eq(4.8a)$$
uniformly in $h$. In order to derive \equ(4.8a), we used the inductive 
assumption \equ(2.36). 

Using \equ(4.8a) and the definition \equ(4.7), 
repeating the proof of \equ(2.52a), we find that 
$$\int d\xx\,|\xx|^j\,G^{(h)}(\xx)\le C_j\g^{-(1+j)h}\;.\Eq(4.8b)$$
For $h=0$ the assumption \equ(4.4) is clearly true 
(it coincides with \equ(4.3)). 
Assuming inductively that \equ(4.4) is true up to a certain value of $h<0$,
we can show that the same representation is valid for $h-1$. In fact 
we can rewrite the term $\VV^{(h)}$ in the exponent of \equ(4.4) as 
$\VV^{(h)}=\LL\VV^{(h)}+\RR\VV^{(h)}$, 
using \equ(2.20), and we ``absorb'' the quadratic term in $\LL\VV^{(h)}$ 
(the one in the first line of \equ(2.22))
in the fermionic integration, as explained in \S 2.3. Similarly we rewrite
$$\fra\dpr{\dpr\psi^{(\le h)\pm}_{\xx,\s}}\VV^{(h)}(\psi^{(\le h)})=
\int d\yy\, n_{h}(\xx-\yy)\psi^{(\le h)\mp}_{\yy,\s}+
\fra\dpr{\dpr\psi^{(\le h)\pm}_{\xx,\s}}\hat \VV^{(h)}(\psi^{(\le h)})\,,
\Eq(4.9)$$
where $\hat \VV^{(h)}$ was defined after \equ(2.25). This rewriting
induces a decomposition of the first line of \equ(4.6) into two pieces, 
the first proportional to $n_h$, the second 
identical to the first line of \equ(4.6) itself, with $\VV^{(h)}$ replaced by
$\hat\VV^{(h)}$. We choose to ``absorb'' the term proportional to 
$n_h$ into the definition of $Q^{(h)}$, and this gives the recursion 
relation \equ(4.8). 

After these splittings and redefinitions, we integrate the field $\psi^{(h)}$,
as in Section \secc(2.3), and we end up with an expression given by the r.h.s.
of \equ(4.4), with $h$ replaced by $h-1$ and the inductive assumption \equ(4.4)
is proved. 

For $h=h_\b$ we define 
$$\eqalign{&e^{-L^2\b\tilde F_{h_\b}+S^{(<h_\b)}(\phi)}
=\int P_{E_{h_\b},C_{h_\b}}(d\psi^{(\le h_\b)})\,\cdot\cr
&\cdot\Big\{
e^{-\hat\VV^{(h_\b)}(\psi^{(\le h_\b)})-B^{(h_\b)}(\psi^{(\le h_\b)},\phi)
+\int d\kk\left[\hat\phi^+_{\kk,s}\hat Q^{(h_\b+1)}_\kk\hat\psi^-_{
\kk,\s}+\hat\psi^+_{\kk,\s}\hat Q^{(h_\b+1)}_\kk\hat\phi^-_{\kk,\s}
\right]}\Big\}\;,\cr}\Eq(4.10)$$
where 
$$S^{(<h_\b)}(\phi)=\sum_{\s=\ud}\int d\xx
\,\Big( \phi^+_{\s}*Q^{(h_\b)}*g^{(h_\b)}*Q^{(h_\b)}*\phi^-_{\s}-
\phi^+_{\s}*G^{(h_\b)}*n_{h_\b-1}*G^{(h_\b)}*\phi^-_{\s}\Big)%(\xx)
\Eq(4.10aa)$$ 
and $g^{(h_\b)}$ is the propagator 
associated to the integration $P_{E_{h_\b},C_{h_\b}}(d\psi^{(\le h_\b)})$,
while $n_{h_\b-1}$ is a function satisfying the 
same bounds as $n_h$, with $h_\b-1$ replacing $h$. 

From the definitions and the construction above, we get
$$\eqalign{S(\xx-\yy)&={\partial^2\over \phi^+_{\xx,\s} 
\phi^-_{\yy,\s}}\Big[S^{(<h_\b)}(\phi)+S^{(\ge h_\b)}(\phi)\Big]
\Big|_{\phi=0}=\cr
&=\sum_{h=h_\b}^1\Big[
\big(Q^{(h)}*g^{(h)}*Q^{(h)}\big)(\xx-\yy)-
\big(G^{(h)}*n_{h-1}*G^{(h)}\big)(\xx-\yy)\Big]\;.\cr}\Eq(4.10a)$$
Taking the Fourier transform and defining
$h_\kk=\min\{ h: \hat g^{(h)}(\kk)\not=0\}$, we get 
$$\eqalign{&\hat S(\kk)=g^{(h_\kk)}(\kk)\big(Q^{(h_\kk)}_\kk\big)^2+
g^{(h_\kk+1)}(\kk)\big(Q^{(h_\kk+1)}_\kk\big)^2-\cr
&-\big[g^{(h_\kk)}(\kk)Q^{(h_\kk)}_\kk+g^{(h_\kk+1)}(\kk)Q^{(h_\kk+1)}_\kk
\big]^2\hat n_{h_\kk-1}(\kk)-
\big[g^{(h_\kk+1)}(\kk)Q^{(h_\kk+1)}_\kk
\big]^2\hat n_{h_\kk}(\kk)\cr}\Eq(4.11)$$
Denoting by $D_h(\kk)=-i k_0+E_{h-1}(\kk)-\m$, we can rewrite \equ(4.11)
as
$$\hat S(\kk)=\fra1{D_{h_\kk}(\kk)}\big(1+W(\kk)\big)\;,\Eq(4.12)$$
where we defined 
$$\eqalign{W(\kk)&=f_{h_\kk}(\kk)\Big[\big(Q^{(h_\kk)}_\kk\big)^2-1\Big]+
f_{h_\kk+1}(\kk)\Big[\fra{D_{h_\kk}(\kk)}{D_{h_\kk+1}(\kk)}
\big(Q^{(h_\kk+1)}_\kk\big)^2-1\Big]-\cr
&-\fra{n_{h_\kk-1}(\kk)}{D_{h_\kk}(\kk)}\Big(f_{h_\kk}(\kk)
Q^{(h_\kk)}_\kk+f_{h_\kk+1}(\kk)Q^{(h_\kk+1)}_\kk
\fra{D_{h_\kk}(\kk)}{D_{h_\kk+1}(\kk)}\Big)^2-\cr
&-
\fra{n_{h_\kk}(\kk)}{D_{h_\kk}(\kk)}\Big(f_{h_\kk+1}(\kk)Q^{(h_\kk+1)}_\kk
\fra{D_{h_\kk}(\kk)}{D_{h_\kk+1}(\kk)}\Big)^2\;.\cr}
\Eq(4.13)$$
From the estimates \equ(2.36) and \equ(4.8a) it follows that 
$$|W(\kk)|\le C|U||h|\g^h\;,\qquad |\dpr^n_\kk W(\kk)|\le C_n|U||h|\g^{(1-n)h}
\;.\Eq(4.14)$$
Substituting \equ(4.12) into the definition of $\Sigma(\kk)$ we find:
$$\Sigma(\kk)=E_{h_{\kk}-1}(\kk)-\e_0(\vec k)+\Big[\fra1{1+W(\kk)}-1\Big]
D_{h_\kk}(\kk)\;,\Eq(4.15)$$
so that, using \equ(4.14), we find
$$|\Sigma(\kk)|\le C|U|\;,\qquad |\dpr_\kk\Sigma(\kk)|\le C|U|^2\;,\qquad
|\dpr_\kk^2\Sigma(\kk)|\le Cc_0\;.\Eq(4.16)$$
Moreover, by the same symmetries implying \equ(2.36a), we have that 
$\sum_{j=\pm}\Im\Sigma(j\p\b^{-1},\vkk)=0$ and 
$\sum_{j=\pm}j\Re\Sigma(j\p\b^{-1},\vkk)=0$.
Combining this bound with an explicit computation of the lowest order 
contributions to $\Sigma(\kk)$ and its first derivatives with respect 
to $k_0$ and $\vec k$ (showing that the functions $a(\th),|\vec b(\th)|,
|\vec c(\th)|$ in the statement of Theorem \thm(1.1) are bounded above 
and below by positive $O(1)$ constants), the proof of Theorem \thm(1.1)
under the assumption \equ(2.36) and the assumption that \equ(2.71a) is true
for any $h_i-1\ge h_\b$ is complete. \\

In next Chapter we shall inductively check that \equ(2.36) and 
the assumption that \equ(2.71a) is true for any $h_i-1\ge h_\b$ are indeed 
true.\\

\* \section(3, Smoothness of the effective dispersion relation)

\sub(3.1) In this section we first want to prove \equ(2.36), under the
assumption that the $\l$ smallness condition \equ(2.71a) is
verified for any $h_i>h$; we shall then prove the smallness
assumption on $\l_h$ by an iterative argument.

We actually want to prove a statement slightly stronger than
\equ(2.36) and, in order to do this, we introduce some definitions.
We denote by $\b^2_h(\xx)$ the Fourier transform of
$\hat\b^2_h(\kk)\=E_{h-1}(\kk)-E_h(\kk)$, see
\equ(2.34), and, given an isotropic sector index $\lis\o\in\lis
O_h$, we define
$$\b^2_{h,\bar\o}(\xx)\defin (\lis{\FFF}_{2,h,(\bar\o,\bar\o)}*\b^2_h)(\xx)\;,
\Eq(3.1)$$
where $\lis{\FFF}_{2,h,(\bar\o,\bar\o)}$ is the ``isotropic
analogue'' of the operator in \equ(2.70) (\ie the operator
obtained by replacing $\tilde F_{h,\o_i}(\vkk)$ in the r.h.s. of
\equ(2.70) with $\tilde{\lis F}_{h,\bar\o}(\vkk)$, where
$\tilde{\lis F}_{h, \bar\o}(\vkk)$ is a smooth function $=1$ on
$\lis S_{h,\bar\o}$ and with a support slightly larger than $\lis
S_{h,\bar\o}$). We want to prove the following.

\*

\theorem(3.1) {\it Given $h_\b\le h\le 0$, let us assume that
$E_j(\kk)$ satisfies \equ(2.36) for any $j\ge h$ and that, given
any tree $\t\in\TT_{h,n}$ (with all its labels), $\tilde
\l_{h_{v^*}-1,\O_{v^*}}$ satisfies \equ(2.71a) for all the
endpoints $v^*\in\t$. Then, if $U_0|h_\b|=c_0$ is small enough,
$$\eqalign{&\int d\xx |\b^2_{h,\bar\o}(\xx)|\le C_0 |U||h|\g^{2h}\;,\cr
& \int d\xx |\xx|^n |\b^2_{h,\bar\o}(\xx)|\le C_n
|U|^2|h|\g^{(2-n)h}\;,\qquad n\ge 1\;. \cr}\Eq(3.2)$$
}

Note that the bound \equ(2.36) easily follows from Theorem
\thm(3.1). In fact, given $\kk$ in the support of $C_h^{-1}(\kk)$,
let us define $\lis\o(\vec k)\in\lis O_h$ as the isotropic sector
index such that $\tilde{\lis F}_{h,\bar\o(\vkk)}( \vkk)=1$. With
this definition, we can rewrite $E_{h-1}(\kk)-E_h(\kk)=\tilde{\lis
F}_{h,\bar\o(\vkk)}( \vkk)\hat\b^2_h(\kk)$, so that
$E_{h-1}(\kk)-E_h(\kk)$ and its derivatives can be rewritten as
$$\eqalign{&
E_{h-1}(\kk)-E_h(\kk)=\int d\xx\,
e^{i\kk\xx}\b^2_{h,\bar\o}(\xx)\;,\cr &
\dpr_{k_{i_1}}\cdots\dpr_{k_{i_n}}
\big(E_{h-1}(\kk)-E_h(\kk)\big)=i^n \int d\xx\ x_{i_1}\cdots
x_{i_n} e^{i\kk\xx}\b^2_{h,\bar\o}(\xx)\;,\cr}\Eq(3.3)$$
hence they can be bounded by the l.h.s. of \equ(3.2) and
\equ(2.36) follows.

So now we shall focus on the proof of Theorem \thm(3.1).\\

\proof The same iterative construction leading to the tree
expansion for $\VV^{(h)}$ allows us to represent
$\b^2_{h,\bar\o}(\xx)$ as a sum over trees. Let $\xx, \yy$ be the
two points where the two external fields (the fields in $P_{v_0}$)
are hooked on; then, by \equ(2.22), \equ(2.23) and \equ(2.68),
$$\eqalign{
&\b^2_{h,\bar\o}(\xx -\yy) =
\sum_{n=1}^\io\sum_{\t\in\TT_{h,n}}\sum_{\bP\in\PP_\t \atop
|P_{v_0}|=2}\sum_{T\in\bT}\sum_{\O\setminus\O_{v_0}}
\Big[\prod_{v\in V_c}\c(\O_v)\Big]\cdot\cr
&\cdot\lis{\FFF}_{2,h,(\bar\o,\bar\o)}* \int d(\xx_{v_0}\setminus
\{\xx,\yy\}) W^{(mod)}_{\t,\bP,\O,T}(\xx_{v_0}) \;,\cr} \Eq(3.4)$$
where $W^{(mod)}_{\t,\bP,\O,T}(\xx_{v_0})$ can be represented as
in \equ(2.68a), with the kernels $K^{h_i}_{v_i^*}$ replaced by the
modified ones. Again, for simplicity, we shall explicitly study
only contributions coming from trees $\t$ such that all the
endpoints are of type $\l$. Given $\t\in\TT_{h,n}$, $\bP\in\PP_\t$
with $|P_{v_0}|=2$, $T\in\bT$, in analogy with definition
\equ(2.76), we define $\lis J^{(2)}_{h,n,\bar\o}(\t,\bP,T;j)$ as
$$\eqalign{&\lis J_{h,n,\bar\o}^{(2)}(\t,\bP,T;j)= \sum_{\O\setminus\O_{v_0}}
\left[\prod_{v}\c(\O_v) \right]\int d(\xx-\yy) \, |\xx
-\yy|^j\cdot\cr &\cdot \left|
\lis{\FFF}_{2,h,(\bar\o,\bar\o)}*\int d(\xx_{v_0}\setminus
\{\xx,\yy\}) W^{(mod)}_{\t,\bP,\O,T}(\xx_{v_0}) \right|
\;.\cr}\Eq(3.5)$$
We want to prove an improved version of the bound \equ(2.77),
valid if $|P_{v_0}|=2$:
$$\sup_{\bar\o\in\bar O_h}
\lis J_{h,n,\bar\o}^{(2)}(\t,\bP,T;j) \le (c_j|U|)^n |h|\g^{(2-j)h}
\prod_{v\in V_c} \fra{1}{s_v!}|P_v|^5\g^{\d(|P_v|)(h_v-h_{v'})}
\;, \qquad n\ge 2\;,\Eq(3.6)$$
where $v'$ is the $c$--vertex immediately preceding $v$ on $\t$.
Note that the proof of Theorem \thm(2.1), adapted to the present
case, would easily imply a bound like \equ(3.6) with $|h|\g^{2h}$
replaced by $\g^h$ and $|P_v|^5$ replaced by $1$. The bound
\equ(3.6) is valid for $n\ge 2$; if $n=1$ there is only one
possible choice of $\t\in\TT_{h,1}$ and, calling $v$ the only
endpoint in $\t$ and $\{f_2,f_3\}=P_v\setminus P_{v_0}$, we shall
prove that
$$\eqalign{&
\lis J_{h,1,\bar\o}^{(2)}(\t,\bP,T;j)=\sum_{\o(f_2),\o(f_3)\in
O_h}\d_{ \o(f_2),\o(f_3)}\int d(\xx_1-\xx_4) \,
|\xx_1-\xx_4|^j\cdot\cr &\cdot
\Big|\lis{\FFF}_{2,h,(\bar\o,\bar\o)}* \int d\xx_2 d\xx_3\;
g^{(h)}_{\o(f_3)}(\xx_2-\xx_3)
\tilde\l_{h,\O_v}(\xx_1,\xx_2,\xx_3,\xx_4)\Big|\le
c_j|U|^{2-\d_{j,0}}|h|^{\d_{j,0}} \g^{(2-j)h}\;.\cr}\Eq(3.7)$$
It is easy to realize that the bounds \equ(3.6) and \equ(3.7) and
the representation \equ(3.4) allow us to get the bound \equ(3.2)
on $\b^2_{h,\bar\o}$. So we will now prove \equ(3.6) and
\equ(3.7). We shall proceed as follows; we shall first prove
\equ(3.6) under the assumption that, given $h'$ and $k$ with $h\le h' \le
k$, $\lis \o_1, \lis \o_4\in \lis O_k$, with $|\lis \o_1 -
\lis\o_4|\le 1$, and $P=P^i\cup P^a$, with $P^i=\{f_1,f_4\}$,
$P^a=\{f_2,f_3\}$,
$$\eqalign{&
\sum_{\o\in O_{h'}}\int d(\xx_1-\xx_4) \, |\xx_1-\xx_4|^j \left|
\int d\xx_2 d\xx_3\; g^{(h')}_{\o}(\xx_2-\xx_3)
\tilde\l_{k,\tilde\O_4}(\xx_1,\xx_2,\xx_3,\xx_4) \right| \le\cr
&\le c_j|U|^{2-\d_{j,0}}(1+|k|^{\d_{j,0}}) \g^{2h'-jk} \;,\cr}\Eq(3.8)$$
where $\tilde\O_4=\{\lis\o_f\in \lis O_k, f\in
P^i\}\cup\{\o(f)=\o, f\in P^a\}$ and, with a small abuse of
notation,
$$\eqalign{&\tilde\l_{k,\tilde\O_4}(\xx_1,\xx_2,\xx_3,\xx_4)=\fra1{\b^4}
\sum_{\ul k_{0}}\int
\fra{d\ul\vkk}{(2\p)^8}e^{-i\sum_{i=1}^4\kk_i\xx_i}
\cdot\cr&\qquad\cdot \tilde{\lis F}_{k,\bar\o_{f_1}}(\vkk_1)
\tilde F_{h',\o}(\vkk_2) \tilde F_{h',\o}(\vkk_3) \tilde{\lis
F}_{k,\bar\o_{f_4}}(\vkk_4)
\,\hat\l_k(\kk_1,\kk_2,\kk_3,\kk_4)\;.\cr}\Eq(3.8a)$$
After having proved \equ(3.6) under the assumption
\equ(3.8), we shall prove \equ(3.8) inductively in $h'$, so that,
in particular, \equ(3.7) will follow.\\

In order to prove \equ(3.6), we will first further expand
$W^{(mod)}_{\t,\bP,\O,T}(\xx)$ in \equ(3.5) by extracting some
``loop propagators'' from the Gram determinant in the second line
of \equ(2.79) (via an interpolation technique described below and
in Appendix \secc(A2)). Such extraction of loop propagators will
imply some new constraints on the sector indices appearing in the
sum $\sum_{\O\setminus\O_{v_0}}$ in \equ(3.5). Then, depending on
the explicit structure of the terms obtained by this further
expansion, we shall decide whether rewriting the extracted loop
propagators as sums of isotropic propagators or not. In both cases
we shall describe a new inductive procedure to bound the sum over
the sector indices, in order to take into account the new
constraints coming from the extraction of the loop propagators. We
shall compare the new inductive procedure to bound the sector sums
with the one described in \sec(2.8) and by comparison we will show
how to get the desired dimensional gains.
\\

We can now turn to the proof of \equ(3.6) and we shall first 
consider the case $j=0$. The proof is
done by distinguishing among several different cases. First of
all, we need to introduce some definitions. We call $\ul
v_0=\{v_1,\ldots,v_{s_{v_0}}\}$ the set of $c$--vertices
immediately following $v_0$ on $\t$ (note that, for the trees
contributing to $\b^2_{h,\bar\o}(\xx -\yy)$, $v_0$ is necessarily
a $c$--vertex). Given $T_{v_0}$, we identify the two (possibly
coinciding) clusters $v_\xx$ and $v_\yy$ in which the two fields
in $P_{v_0}$ (call them $f_\xx,f_\yy$) enter. Correspondingly, we
call $l_{\xx,\yy}$ the (possibly trivial) path on $T_{v_0}$
connecting $v_\xx$ and $v_\yy$. Moreover we identify $v_\xx$ with
the root of $T_{v_0}$: in this way the concept of leaf of
$T_{v_0}$ is well defined (note that, if $v_\xx\neq v_\yy$,
$v_\yy$ is a leaf of $T_{v_0}$).\\

\0(A) Let us assume that $s_{v_0}\ge 2$. Let us further distinguish the case
$T_{v_0}\neq l_{\xx,\yy}$ and $T_{v_0}\= l_{\xx,\yy}$.\\

\0(A1) $s_{v_0}\ge 2$ and $T_{v_0}\neq l_{\xx,\yy}$. In this case
there must be at least one leaf of $T_{v_0}$ different from
$v_{\yy}$; let us choose one such leaf and call it $v^*$; see Fig.
\graf(A1), where the solid lines represent the tree propagators in
$T_{v_0}$, the wiggling lines represent the loop fields and the
broken lines represent the external fields.

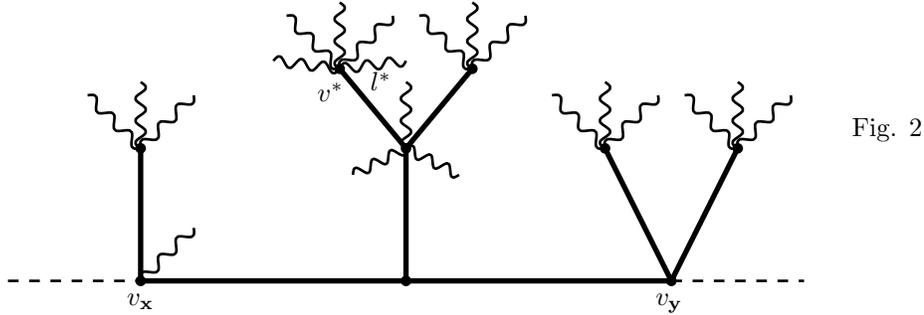
\insertplot{300pt}{130pt}{\ins{45pt}{5pt}{$v_\xx$}
\ins{245pt}{5pt}{$v_\yy$}
\ins{117pt}{85pt}{$v^*$}
\ins{137pt}{89pt}{$l^*$}
}%
{A1}{\eqg(A1)}

\*

By construction, $v^*$ has one exiting line belonging to the
spanning tree $T_{v_0}$ (call it $l^*$ and $f^*$ its field label)
and $|P_{v^*}|-1\ge 3$ loop lines. Let $\tilde P_{\ul v_0}\subset
\cup_{v\in\ul v_0}P_v$ be the set of fields contracted into the
Gram determinant $\det G^{h_{v_0},T_{v_0}}$ appearing in
\equ(2.68a). If, given a set ${\cal I}$ of field labels, we define
${\cal I}^\pm=\{f\in {\cal I}\;:\;\e(f)=\pm\}$, we can think
$G^{h_{v_0},T_{v_0}}$ as a matrix whose rows are associated to the
fields $f\in\tilde P_{\ul v_0}^-$ and whose columns are associated
to the fields $f\in\tilde P_{\ul v_0}^+$. Let us distinguish the
field labels belonging to $\tilde P_{v^*}\= P_{v^*}\setminus f^*$
from those belonging to $\tilde P_{\ul v_0} \setminus\tilde
P_{v^*}$ and let us correspondingly write $G^{h,T_{v_0}}$ in
blocks, as follows:
$$\det G^{h,T_{v_0}}=\det\pmatrix{
A & B \cr C & D\cr}\Eq(3.9)$$
where: $A$ is the block with both row and column indices belonging
to $\tilde P_{v^*}$; $B$ is the block with row indices in $\tilde
P_{v^*}$ and column indices in $\tilde P_{\ul v_0}\setminus\tilde
P_{v^*}$; $C$ is the block with row indices in $\tilde P_{\ul
v_0}\setminus\tilde P_{v^*}$ and column indices in $\tilde
P_{v^*}$; $D$ is the block with both row and column indices
belonging to $\tilde P_{\ul v_0}\setminus\tilde P_{v^*}$. Note
that by construction neither $A$ nor $D$ are square matrices and,
because of this, $\det\pmatrix{ A & 0 \cr 0 & D\cr}=0$, so that we
can rewrite the r.h.s. of \equ(3.9) as:
$$\eqalign{&\det\pmatrix{ A & B \cr C & D\cr}-\det\pmatrix{ A & 0 \cr 0 & D\cr}
=\int_0^1 ds {\der\over \der s}\det\pmatrix{ A & sB \cr sC &
D\cr}=\cr
&\qquad\qquad=\int_0^1 ds \sum_{i\in P^-_{v^*}\ ,\ j\in \tilde
P^+_{\ul v_0}\setminus \tilde P^+_{v^*}}(-1)^{i+j}B_{ij}\det
m_{ij}\big(G(s)\big)+\cr
&\qquad\qquad+\int_0^1 ds \sum_{i\in \tilde P^-_{\ul v_0}
\setminus \tilde P^-_{v^*}\ ,\ j\in \tilde
P^+_{v^*}}(-1)^{i+j}C_{ij}\det m_{ij}\big(G(s)\big)\;,\cr}
\Eq(3.10)$$
where $G(s)=\pmatrix{ A & sB \cr sC & D\cr}$ and, given a matrix
$M$, $m_{ij}(M)$ is the minor of $M$ corresponding to the entry
$i,j$. The identity \equ(3.10) is essentially a first order
expansion of $\det\pmatrix{ A & B \cr C & D\cr}$ around the point
$\pmatrix{ A & 0 \cr 0 & D\cr}$ and it corresponds to the
operation of ``extracting one loop'' connecting $\tilde P_{v^*}$
with $\tilde P_{\ul v_0} \setminus\tilde P_{v^*}$. The key remark
is that the r.h.s. of \equ(3.10) is a sum of less than
$\prod_{v\in\ul v_0}|P_v|$ terms, each of them equal to the
product of a propagator $B_{ij}$ or $C_{ij}$ times a determinant
{\it that is still a Gram determinant}, then it can still be
bounded via the Gram--Hadamard inequality, the result being the
following dimensional estimate, analogue to the r.h.s. of
\equ(2.80):
$$|\det m_{ji}\big(G(s)\big)|\le c^{\sum_{i=1}^{s_{v_0}}|P_{v_i}|-|P_v|-2
s_{v_0}}\g^{\fra34
h(\sum_{i=1}^{s_{v_0}}|P_{v_i}|-|P_v|-2s_{v_0})}\Eq(3.11)$$
A procedure analogue to the first order expansion leading to
\equ(3.10) allows us to expand $\pmatrix{ A & B \cr C & D\cr}$ up
to fifth order around the point $\pmatrix{ A & 0 \cr 0 & D\cr}$
and to rewrite $\det G^{h,T_{v_0}}$ as
$$\eqalign{&\det G^{h,T_{v_0}}(\tt_{v_0})=
\sum^*_{\{i_1j_1,i'_1j'_1\}}(-1)^{\e_1}
G^{h,T_{v_0}}_{i_1j_1,i'_1j'_1}(\tt_{v_0}) \,\det \wt
G^{h,T_{v_0}}\big(\{i_1j_1,i'_1j'_1\};0,\tt_{v_0} \big)+\cr
&+\fra1{3!}\sum^*_{\{i_qj_q,i'_qj'_q\}_{q=1}^3}(-1)^{\e_1+\e_2+\e_3}
\Big[\prod_{q=1}^3 G^{h,T_{v_0}}_{i_qj_q,i'_qj'_q}(\tt_{v_0})
\Big] \,\det \wt G^{h,T_{v_0}}
\big(\{i_qj_q,i'_qj'_q\}_{q=1}^3;0,\tt_{v_0} \big)+\cr
&+\fra1{5!}\sum^*_{\{i_qj_q,i'_qj'_q\}_{q=1}^5}(-1)^{\e_1+\cdots+\e_5}
\Big[\prod_{q=1}^5 G^{h,T_{v_0}}_{i_qj_q,i'_qj'_q}(\tt_{v_0})\Big]
\, \det \wt G^{h,T_{v_0}}\big(\{i_qj_q,i'_qj'_q\}_{q=1}^5;\bar
s,\tt_{v_0} \big)\cr}\Eq(3.12)$$
where: the $*$'s on the three sums constraint the indices
$\{i_qj_q,i'_qj'_q\}$ to run over choices such that
$f^-_{i_qj_q}\in\tilde P_{v^*}$ and $f^+_{i'_qj'_q}\in \tilde
P_{\ul{v}_0}\setminus\tilde P_{v^*}$ or viceversa; $\e_q=\pm$ is a
sign, depending on the parity of the element index
$\{i_qj_q,i_q'j_q'\}$, $q=1,\ldots,5$; the matrices $\wt
G^{h,T_{v_0}}\big(\{i_qj_q,i'_qj'_q\}_{q=1}^L;s, \tt_{v_0})$,
$L=1,3,5$ are the minors of $G^{h,T_{v_0}}(s,\tt_{v_0})$
complementary to the elements $(i_qj_q,i'_qj'_q)$, $q=1,\ldots,L$;
the parameter $0\le \bar s\le 1$ appearing in the argument of $\wt
G^{h,T_{v_0}}$ in the third line is an interpolation parameter
corresponding to the integration variable $s$ appearing in
\equ(3.10). Note that the determinants on the first two lines are
computed at $s=0$, and this implies that they are the product of
two determinants, the first involving only fields in $\tilde
P_{v^*}$, the second involving only fields in $\tilde P_{\ul
v_0}\setminus\tilde P_{v^*}$. The three determinants in
\equ(3.12) are Gram determinants which can be bounded using
Gram--Hadamard inequality: the one on the first line can be
bounded exactly as in \equ(3.11), while the other two can be
bounded by the r.h.s. of \equ(3.11) times a factor $\g^{-3h}$ or
$\g^{-6h}$ respectively.

The splitting \equ(3.12) induces a similar splitting in $\lis
J^{(2)}_{ h,n,\bar\o}(\t,\bP,T;0)$ with $|P_{v_0}|=2$:
$$\lis J^{(2)}_{h,n,\bar\o}(\t,\bP,T;0)\le\sum_{L=1,3,5}
\lis J^{(2)}_{h,n,\bar\o}(\t,\bP,T;0,L)\Eq(3.13)$$
and, calling $s_L=\d_{L,5}\bar s$,
$$\eqalign{&\lis J^{(2)}_{h,n,\bar\o}(\t,\bP,T;0,L)=\sum_{\O\setminus
\O_{v_0}} \Big[\prod_{v} \c(\tilde\O_{v,\bar\o})
\Big]\Biggl|\int\prod_{l\in T^*} d(\xx_l-\yy_l) \Big[\prod_{v\
{\rm e. p.}} \tilde\l_{h_v-1,\tilde\O_{v,\bar\o}}(\xx_{v}) \Big]
\cdot\cr & \cdot\Big\{\int d P_{T_{v_0}}(\tt_{v_0})
\fra1{s_{v_0}!}\fra1{L!}\sum^*_{\{i_qj_q,i'_qj'_q\}_{q=1}^L}
\Big[\prod_{q=1}^L G^{h,T_{v_0}}_{i_qj_q,i'_qj'_q}\Big]\cdot\cr
&\cdot \det \wt
G^{h,T_{v_0}}\big(\{i_qj_q,i'_qj'_q\}_{q=1}^L;s_L,\tt_{v_0} \big)
\prod_{l\in T_{v_0}}
g^{(h)}_{\o_l}(\xx_l-\yy_l)\d_{\o_l^+,\o_l^-}\Big\}\cdot\cr
&\cdot\Big\{\int dP_{T_v}(\tt_v)\prod_{v>v_0\, \atop\hbox{\ottorm
not e.p.}}{1\over s_v!} \det G^{h_v,T_v}(\tt_v)\prod_{l\in T_v}
g^{(h_v)}_{\o_l}(\xx_l-\yy_l)\d_{\o_l^+,\o_l^-}\Big\}\Biggr|\;,\cr}\Eq(3.14)$$
where
$$\tilde\O_{v,\bar\o}=
\big\{\o(f)\in O_{h(f)},\ f\in P_v\setminus\{f_\xx,f_\yy\}
\big\}\cup\big\{\lis\o_f\=\lis \o, \ f\in
P_v\cap\{f_\xx,f_\yy\}\big\} \Eq(3.15)$$
and, by writing $\c(\tilde\O_{v,\bar\o})$, we implicitly
introduced the notion of a $\c$ function depending on a set of
sector indices which contains both isotropic and anisotropic
sector indices: in general, if $P_v^a\subset P_v$,
$P_v^i=P_v\setminus P_v^a$ and $\O_v'$ is the set of sector
indices $\O_v'=\{\o(f)\in O_{h(f)}, f\in
P_v^a\}\cup\{\lis\o_f\in\lis O_{j_f}, f\in P_v^i\}$ labelled by
$P_v^a$ and $P_v^i$, we define
$$\c(\O_v') =
\c\left( \forall f\in P_v\,,\ \exists \vkk(f)\in\cases{
S_{h(f),\o(f)},\ {\rm if}\ f\in P_v^a\cr \lis
S_{j_f,\bar\o_f},\hskip.6truecm {\rm if}\ f\in P_v^i\cr}\ :\
\sum_{f\in P_v}\e(f)\vkk(f)=0\right)\;.\Eq(3.16)$$
The modified coupling function
$\tilde\l_{h_v-1,\tilde\O_{v,\bar\o}}(\xx_{v})$ is defined in a
way similar to \equ(3.8a). The presence of the functions
$\c(\tilde\O_{v,\bar\o})$ and of
$\tilde\l_{h_v-1,\tilde\O_{v,\bar\o}}(\xx_{v})$ is due to the
convolution operator $\lis{\FFF}_{2,h,(\bar\o,\bar\o)}$ appearing
in the definition \equ(3.5).\\

We now want to show how to get the desired dimensional gain for
the three contributions $\lis J^{(2)}_{h,n,\bar\o}(\t,\bP,T;0,L)$
in \equ(3.14), $L=1,3,5$. In order to do that, we shall suppose
that the modified coupling functions satisfy the following bound,
essentially equivalent to the smallness condition \equ(2.71a):
$$\fra{1}{L^2\b}\int d\xx_v
|\tilde\l_{h_v-1,\tilde\O_{v,\bar\o}}(\xx_{v})|\le
C|U|\;.\Eq(2.71b)$$
At the end, by an iterative argument, we shall prove this bound,
together with the bound \equ(3.8).
\\
\\
(A1.1) $L=3$: there are exactly three loop lines connecting
$\tilde P_{v^*}$ with $\tilde P_{v_0}\setminus \tilde P_{v^*}$,
call them $l_1,l_2,l_3$ and $f_1,f_2,f_3$ their field labels, see
Fig. \graf(A11).

\insertplot{300pt}{150pt}{\ins{45pt}{5pt}{$v_\xx$}
\ins{245pt}{5pt}{$v_\yy$}
\ins{118pt}{100pt}{$v^*$}
\ins{135pt}{90pt}{$f^*$}
\ins{80pt}{68pt}{$l_1$}
\ins{125pt}{60pt}{$l_2$}
\ins{240pt}{135pt}{$l_3$}
}%
{A11}{\eqg(A11)}

\*

Let us consider the product $\prod_{q=1}^3
G^{h,T_{v_0}}_{i_qj_q,i'_qj'_q}= \prod_{q=1}^3 t_{i_q,i_q'}
g_{\o(f_q)}^{(h)}(\xx_{l_q}-\yy_{l_q}) \d_{\o_{l_q}^+,\o_{l_q}^-}$
in \equ(3.14) and let us substitute both the extracted loop
propagators $g_{\o(f_q)}^{(h)}$ with $q=1,2,3$ and the propagator
of the spanning tree $g_{\o(f^*)}^{(h)}$ (we recall that $f^*$ is
the field label of the line $l^*$ of $T_{v_0}$ exiting from $v^*$)
with the sums of isotropic propagators $\lis g_{\bar\o_q}^{(h)}$,
$q=1,2,3$, and $\lis g_{\bar\o^*}^{(h)}$, such that $\lis
S_{h,\bar \o_q}\subset S_{h,\o(f_q)}$ and $\lis S_{h,\bar
\o^*}\subset S_{h,\o(f^*)}$. This operation is allowed since
$\sum_{\o\in O_h} g^{(h)}_\o = \sum_{\bar\o\in \bar O_h} \lis
g^{(h)}_{\bar \o}$ and our definitions are such that any isotropic
sector is strictly contained in one anisotropic sector.

For any fixed configuration of internal anisotropic sectors, the
new sums on the isotropic sectors can be written as:
$$\sum_{\bar\o_q\prec \o(f_q)\atop q=1,2,3}\ \sum_{\bar\o^*\prec\o(f^*)}\;,
\Eq(3.17)$$
where the meaning of the symbol $\prec$ is the same as in
\equ(2.88). Note also that, by construction, the summand in the
r.h.s. of \equ(3.14) is $0$ unless
$\c(\{\lis\o_1,\lis\o_2,\lis\o_3,\lis\o^*\})=1$, so that we can
freely multiply the summand in the r.h.s. of \equ(3.14) by
$\c(\{\lis\o_1,\lis\o_2,\lis\o_3,\lis\o^*\})$.

Having done this, we can bound the resulting expression by
replacing all the modified coupling functions, all the propagators
and all the Gram determinants with their absolute values. Now, we
can bound the integral over the absolute values of the modified
coupling functions as in \equ(2.82) (by using \equ(2.71b) in place
of \equ(2.71a)), the integral over the absolute values of the
propagators of the spanning tree as in
\equ(2.81) (note that the integral $\int d(\xx_{l^*}-\yy_{l^*})
|\lis g_{\bar\o^*}^{(h)}(\xx_{l^*}-\yy_{l^*})|$ admits the same
dimensional bound as the integral of an anisotropic propagator,
see Lemma \lm(2.3)). Moreover, the product of the determinants
times $\prod_{q=1}^3|\lis g_{\bar
\o_q}^{(h)}(\xx_{l_q}-\yy_{l_q})|$ can be bounded by the product
over $v$ of the r.h.s. of \equ(2.80) {\it times a dimensional
gain} $\g^{3h/2}$ coming from the fact that the bound on the size
of $|\lis g_{\bar \o_q}^{(h)}(\xx_{l_q}-\yy_{l_q})|$ is $\g^{h/2}$
smaller than the bound on the size of an anisotropic propagator of
scale $h$, see Lemma \lm(2.2) and Lemma \lm(2.3). After these
bounds, the sum $\sum^*_{\{i_qj_q,i'_qj'_q\}_{q=1}^3}$ can be
bounded by $\prod_{v\in\ul v_0}|P_v|^3\le \prod_{v\in
V_c}|P_v|^5$, where the latter is the same product appearing in
the r.h.s. of
\equ(3.6).

We are still left with the sum over the sector indices:
$$\sum_{\cup_v \O_v\setminus\O_{v_0}}
\sum_{\bar\o_q\prec \o(f_q)\atop
\bar\o^*\prec\o(f^*)}\Big[\prod_{v>v_0\atop v\in V_c}\ \c
(\tilde\O_{v,\bar\o})\Big]\c(\{\lis\o_1,\lis\o_2,\lis\o_3,\lis\o^*\})
\Big[\prod_{l\in T}\d_{\o_l^+,\o_l^-}\Big]
\prod_{q=1}^3\d_{\o_{l_q}^+,\o_{l_q}^-}\;.\Eq(3.19)$$
where the product $\c(\{\lis\o_1,\lis\o_2,\lis\o_3,\lis\o^*\})
\prod_{q=1}^3\d_{\o_{l_q}^+,\o_{l_q}^-}$ takes into account the
new constraints coming from the extraction of the three loop
propagators.

We want to prove that \equ(3.19) can be bounded by the r.h.s. of
\equ(2.83) times a dimensional factor $\g^{-h/2}|h|$. Combining this loss
with the gain $\g^{3h/2}$ coming from the bound on the size of the
loop propagators and discussed above, we see that globally the
contribution under analysis has a dimensional gain $\g^h|h|$ with
respect to the bounds described in \sec(2.8), that is the desired
gain.

We proceed keeping in mind the procedure followed in the proof of
\equ(2.83) and by comparison we shall show how to get the
dimensional gain. Using the definitions introduced in \sec(2.8)
after \equ(2.90) we can rewrite \equ(3.19) as
$$\eqalign{&\sum_{\O_{\ul{v}_0}\setminus\O_{v_0}}
\sum_{\bar\o_q\prec \o(f_q)\atop \bar\o^*\prec\o(f^*)}\Biggl\{
\prod_{v\in\ul{v}_0}\Big[\ \sum_{\cup_{w> v} \O_w\setminus
\O_{v}}\ \prod_{w\ge v, w\in V_\c} \c\big(\tilde\O_{w,\bar\o}\big)
\prod_{l\in\cup_{w\ge v}T_w} \d_{\o_l^+,\o_l^-}\Big]\cdot\cr
&\qquad\qquad\qquad\cdot\c(\{\lis\o_1,\lis\o_2,\lis\o_3,\lis\o^*\})\prod_{l\in
T_{v_0}}\d_{\o_l^+,\o_l^-}
\prod_{q=1}^3\d_{\o_{l_q}^+,\o_{l_q}^-}\Biggr\}\=\cr &\=
\sum_{\O_{\ul{v}_0}\setminus\O_{v_0}} \sum_{\bar\o_q\prec
\o(f_q)\atop \bar\o^*\prec\o(f^*)}\Big[ \prod_{v\in\ul v_0}
F_{v}(\tilde\O_{v,\bar\o})\Big]\cdot
\c(\{\lis\o_1,\lis\o_2,\lis\o_3,\lis\o^*\})\prod_{l\in
T_{v_0}}\d_{\o_l^+,\o_l^-} \prod_{q=1}^3\d_{\o_{l_q}^+,\o_{l_q}^-}
\;.\cr}\Eq(3.20) $$
Defining
$${\cal F}_{\ul{v}_0\setminus v^*}(\tilde\O_{\ul{v}_0
\setminus v^*})\defin \prod_{v\in\ul{v}_0\setminus v^*}
F_{v}(\tilde\O_{v,\bar\o}) \prod_{l\in T_{v_0}\setminus
l^*}\d_{\o_l^+,\o_l^-}\;,\Eq(3.21)$$
and calling $\O^{(4)}=\{\o(f_1),\o(f_2),\o(f_3),\o(f^*)\}$, we can
bound
\equ(3.20) as
$$\eqalign{&\equ(3.20)\le
\sup_{\O^{(4)},\O_{v_0}}\Big[\sum_{\tilde\O_{\ul{v}_0\setminus
v^*}}^{**} \ {\cal F}_{\ul{v}_0\setminus
v^*}\big(\tilde\O_{\ul{v}_0 \setminus v^*}\big)\Big]\,\cdot\,
\sup_{\O^{(4)}}\Big[ \sum_{\tilde\O_{v^*,\bar\o}}^*
F_{v^*}\big(\tilde\O_{v^*,\bar\o}\big)\Big]\,\cdot\cr
&\qquad\cdot\sum_{\bar\o_1,\bar\o_2,\bar\o_3,\bar\o^*}
\c(\{\lis\o_1,\lis\o_2,\lis\o_3,\lis\o^*\}) \;,\cr}\Eq(3.22)$$
where the $**$ on the first sum means that we are not summing over
the indices in $\O_{v_0}\cup\O^{(4)}$, while the $*$ on the second
sum means that we are not summing over the indices in $\O^{(4)}$.
Now, the first two sums can be bounded using Lemma \lm(2.4) and,
with respect to the cases in which the two sums have just one
fixed external sector (as it is the case in the proof of
\equ(2.83), see \equ(2.94)), they have a dimensional gain $\g^h$
and $\g^{h/2}$ respectively. The last sum can be bounded using the
following Lemma, see Appendix \secc(A2) for a proof.

\*\lemma(3.1) {\it Given $h_\b\le h\le 0$, the following bound
holds:
$$\sup_{\bar\o_1\in\bar O_h}
\sum_{\bar\o_q\in\bar O_h\atop q=2,3,4}
\c\big(\{\lis\o_1,\lis\o_2,\lis\o_3,\lis\o_4\}\big) \le c
\g^{-h}|h|\;. \Eq(3.23)$$ }
Using Lemma \lm(3.1) we see that
$$\sum_{\bar\o_1,\bar\o_2,\bar\o_3,\bar\o^*}
\c(\{\lis\o_1,\lis\o_2,\lis\o_3,\lis\o^*\})\le
c\g^{-2h}|h|\;.\Eq(3.24)$$
Combining all the gains and losses described above, we see that
\equ(3.19) can be bounded by the r.h.s. of \equ(2.83) times a dimensional
factor $\g^{-h/2}|h|$. As discussed after \equ(3.19) this implies
that $\lis J^{(2)}_{h,n,\bar\o}(\t,\bP,T;0,L=3)$ with
$|P_{v_0}|=2$ admits a dimensional bound given by the r.h.s. of
\equ(3.6) with $j=0$.
\\
\\
(A1.2) L=5: there are five extracted loop lines connecting $\tilde
P_{v^*}$ with $\tilde P_{v_0}\setminus \tilde P_{v^*}$, call them
$l_1,l_2,l_3,l_4,l_5$ and $f_1,f_2,f_3,f_4,f_5$ their field
labels, see Fig. \graf(A12); the other lines exiting from $\tilde
P_{v^*}$ can be contracted with any other loop line.

\insertplot{300pt}{150pt}{\ins{45pt}{5pt}{$v_\xx$}
\ins{245pt}{5pt}{$v_\yy$}
\ins{114pt}{100pt}{$v^*$}
\ins{135pt}{90pt}{$f^*$}
\ins{80pt}{68pt}{$l_1$}
\ins{125pt}{60pt}{$l_3$}
\ins{240pt}{135pt}{$l_5$}
\ins{80pt}{35pt}{$l_2$}
\ins{155pt}{102pt}{$l_4$}
}%
{A12}{\eqg(A12)}

\*

We replace all the modified coupling functions, all the
propagators and all the Gram determinants with their absolute
values. Having done this, we bound the integral over the modified
coupling functions as in \equ(2.82), the integral over the
propagators of the spanning tree as in \equ(2.81) and we bound the
product of the determinants times
$\prod_{q=1}^5|g_{\o(f_q)}^{(h)}(\xx_{l_q}-\yy_{l_q})|$ by the
product over $v$ of the r.h.s. of \equ(2.80). After these bounds,
the sum $\sum^*_{\{i_qj_q,i'_qj'_q\}_{q=1}^5}$ can be bounded by
$\prod_{v\in\ul v_0}|P_v|^5\le \prod_{v\in V_c}|P_v|^5$, where the
latter is the same product appearing in the r.h.s. of
\equ(3.6).

We are still left with the sum over the sector indices:
$$\sum_{\cup_v\O_v\setminus\O_{v_0}}\Big[
\prod_{v>v_0\atop v\in V_c}\c(\tilde\O_{v,\bar\o})\Big]
\Big[\prod_{l\in T}\d_{\o_l^+,\o_l^-}\Big]
\prod_{q=1}^5\d_{\o_{l_q}^+,\o_{l_q}^-}\;,\Eq(3.25)$$
where the product $\prod_{q=1}^5\d_{\o_{l_q}^+,\o_{l_q}^-}$ takes
into account the new constraints coming from the extraction of the
three loop propagators.

We want to prove that \equ(3.25) can be bounded by the r.h.s. of
\equ(2.83) times a dimensional factor $\g^{h}$, that
is even more than the desired dimensional gain. With the same
notations as above, we rewrite \equ(3.25) in the form
$$\sum_{\O_{\ul{v}_0}\setminus\O_{v_0}}\Big[
\prod_{v\in\ul{\tilde v}_0}F_{v}(\tilde\O_{v,\bar\o})\Big]\cdot
\prod_{l\in T_{v_0}}\d_{\o_l^+,\o_l^-}
\prod_{q=1}^5\d_{\o_{l_q}^+,\o_{l_q}^-} \;.\Eq(3.26)$$
Defining ${\cal F}_{\ul{v}_0\setminus v^*}(\tilde\O_{\ul{v}_0
\setminus v^*})$ as in \equ(3.21)
and calling $\O^{(6)}=\{\o(f_q)\}_{q=1}^5 \cup\{\o(f^*)\}$, we can
bound
\equ(3.26) by
$$
\sup_{\O_{v_0}}\Big[\sum_{\tilde\O_{\ul{v}_0\setminus v^*} }^*
{\cal F}_{\ul{v}_0\setminus v^*}\big(\tilde\O_{\ul{v}_0 \setminus
v^*}\big)\Big]\,\cdot\, \sup_{\O^{(6)}}\Big[
\sum_{\tilde\O_{v^*,\bar\o}\setminus\O^{(6)}}
F_{v^*}\big(\tilde\O_{v^*}\big)\Big]\;.\Eq(3.27)$$
where the $*$ on the first sum means that we are not summing over
the sectors in $\O_{v_0}$. Now, the first sum can be bounded
exactly as in the proof of \equ(2.38), while the second one, using
Lemma \lm(2.4), has a dimensional gain $\g^h$, and this concludes
the proof in the present case.
\\
\\
(A1.3) L=1: there is exactly one loop line connecting $\tilde
P_{v^*}$ with $\tilde P_{v_0}\setminus \tilde P_{v^*}$, call it
$l_1$ and $f_1$ its field label, see Fig. \graf(A13).

\insertplot{300pt}{150pt}{\ins{45pt}{5pt}{$v_\xx$}
\ins{245pt}{5pt}{$v_\yy$}
\ins{118pt}{100pt}{$v^*$}
\ins{135pt}{90pt}{$f^*$}
\ins{240pt}{135pt}{$l_1$}
}%
{A13}{\eqg(A13)}

\*

We consider the extracted loop propagator
$G^{h,T_{v_0}}_{i_1j_1,i'_1j'_1}=t_{i_1,i_1'}
g_{\o(f_1)}^{(h)}(\xx_{l_1}-\yy_{l_1}) \d_{\o_{l_1}^+,\o_{l_1}^-}$
and we rewrite both $g_{\o(f_1)}^{(h)}$ and the propagator of the
spanning tree $g_{\o(f^*)}^{(h)}$ as sums of isotropic
propagators, to be denoted by $\lis g_{\bar\o_1}^{(h)}$ and $\lis
g_{\bar\o^*}^{(h)}$. For any fixed configuration of internal
anisotropic sectors, the new sums on the isotropic sectors can be
written as:
$$\sum_{\bar\o_1\prec \o(f_1)\atop\bar\o^*\prec\o(f^*)}\,\openone
(|\bar\o_1-\bar\o^*|\le 1) \;,\Eq(3.28)$$
where the characteristic function is due to the remark that the
value of the integrand in the r.h.s. of \equ(3.14) is $0$ unless
$|\bar\o_1-\bar\o^*|\le 1$, because of momentum conservation.

We now rewrite the determinant $\det \wt
G^{h_{v_0},T_{v_0}}(\{i_1j_1, i_1'j_1'\};0,\tt_{v_0})$ as a
product of two determinants, the first involving only fields in
$\tilde P_{v^*}\setminus f_1$, the second involving only fields in
$\tilde P_{v_0}\setminus (\tilde P_{v^*}\cup f_1)$:
$$\det \wt G^{h,T_{v_0}}(\{i_1j_1,
i_1'j_1'\};0,\tt_{v_0})=\det\wt G^{h}_{v^*}(\tt_{v^*})
\cdot\det\wt G^{h,T_{v_0}}_{\ul v_0\setminus v^*}
(\tt_{v_0})\;.\Eq(3.29)$$

The splitting \equ(3.29) allows us to bound the resulting
expression by
$$\eqalign{&\sum^*_{\{i_1j_1,i'_1j'_1\}}
\sum_{\bar\o_1,\bar\o^*} \openone(|\bar\o_1-\bar\o^*|\le 1)\;
[v^*] \cdot [\ul v_0 \setminus v^*] \cdot \sup_\xx |\lis
g_{\bar\o_1}^{(h)}(\xx)| \int d\xx |\lis g^{(h)}_{\bar\o^*}
(\xx)|\le\cr
& \le c \g^{2h} \g^{-h} \sum^*_{\{i_1j_1,i'_1j'_1\}}
\sum_{\bar\o_1,\bar\o^*} \openone(|\bar\o_1-\bar\o^*|\le 1)\;
[v^*] \cdot [\ul v_0 \setminus v^*]\;,}\Eq(3.30)$$
where $[v^*]$ collects all the spanning tree propagators, the
determinants and the endpoints associated with the vertices $v\ge
v^*$, except the propagators of $l_1$ and $l^*$, while $[\ul v_0
\setminus v^*]$ collects all the other terms, again except the
propagators of $l_1$ and $l^*$.

Note that $[v^*]$ has the structure of a contribution to the
effective potential with two external fields whose {\it isotropic}
sectors are fixed. If we denote by $\t'$ the subtree of $\t$
rooted in $v_0$ and containing $v^*$ and by $\bP',\O',T'$ the
subsets of $\bP,\O,T$ corresponding to the subtree $\t'$ of $\t$,
we can write
$$\eqalign{ & [v^*] = \sum_{\O'\setminus(\bar\o_1\cup\bar\o^*)}
\Big[\prod_{v\in \t'}
\c(\tilde\O_{v,\bar\o_1,\bar\o^*})\Big]\;\cdot\cr
& \cdot\; \Bigg|\int d(\xx_{v^*}\setminus\xx^*){\lis{\FFF}}_{
2,h,(\bar\o_1,\bar\o^*)}*\big[\det\wt G^{h}_{v^*}(\tt_{v^*})\cdot
W^{(mod)}_{
\t',\bP',\O',T'}(\xx_{v^*})\big]\Bigg|\;,\cr}\Eq(3.30a)$$
where $\xx^*$ is the space time point of $\xx_{v^*}$ where the
line $l^*$ is hooked on and, in analogy to \equ(3.15), we defined
$$\tilde\O_{v,\bar\o_1,\bar\o^*} =
\big\{\o(f)\in O_{h(f)},\ f\in P_v\setminus\{f_1,f^*\} \big\}
\cup\big\{\lis\o_{f_1} = \lis \o_1, \lis\o_{f^*} = \lis \o^*
\big\}\;. \Eq(3.30b)$$

Note that, in the case that $v^*$ is an endpoint, the r.h.s. of
\equ(3.30a) can be bounded by \equ(3.8), with
$k=h'=h$ and $j=0$, which has a dimensional factor $\g^h|h|$ more
than the dimensional estimate one would get simply using
\equ(2.71a) and Lemma \lm(2.2). If $v^*$ is not an endpoint, we
will prove below that
$$[v^*] \le (c|U|)^{m_4(v^*)} \g^{2h}|h| \prod_{v\in V_c, v> v^*}{1\over
s_v!}|P_v|^5\g^{\d(|P_v|)(h_v-h_{v'})} \;,\Eq(3.31)$$
where $m_4(v^*)$ is the number of endpoints of $\t'$ and $v'$ is
the $c$--vertex immediately preceding $v$ on $\t'$. The proof of
\equ(3.31) will be postponed to Sect. \secc(3.2) below. Note that
the bound in \equ(3.31) has a $\g^h|h|$ more than the dimensional
bound one would get by repeating the proof of Theorem \thm(2.1).

Let us now consider the factor $[\ul v_0 \setminus v^*]$. It can
be written in the following way:
$$[\ul v_0 \setminus v^*] = \int d\xx_1 d\xx_2 d\yy
\; |G^{(4)}_{\bar\o, \bar\o_1, \bar\o^*}(\xx, \yy, \xx_1, \xx_2)|
\;,\Eq(3.30c)$$
where $\xx_1$ and $\xx_2$ are the space-time points in $\t\bs
\t'$, where the two propagators of $l^*$ and $l_1$ are hooked on.
Note that $G^{(4)}_{\bar\o, \bar\o_1, \bar\o^*}(\xx, \yy, \xx_1,
\xx_2)$ has the same structure of a contribution to the effective
potential with four external fields with fixed isotropic sectors.
It follows that
$$[\ul v_0 \setminus v^*] \le \tilde G^{(4)}_{\o,
 \o_1, \o^*}\;,\Eq(3.30d)$$
where $\o, \o_1, \o_2$ are the anisotropic sectors containing
$\bar\o, \bar\o_1, \bar\o^*$, respectively and $\tilde
G^{(4)}_{\o, \o_1, \o^*}$ has the same structure of
$J^{(4)}_{h,n-m_4(v^*)}(\t\bs \t', \bP\bs \bP', T\bs T')$, see
\equ(2.76).

By using Lemma \lm(2.5), we get
$$\tilde G^{(4)}_{\o, \o_1, \o^*} \le (c|U|)^{n-m_4(v^*)}
\prod_{v\in V_c, v\not\in \t'} \g^{\d(|P_v|) (h_v-h_{v'})}\;,
\Eq(3.31a)$$

In order to complete the proof, we still have to bound the two
sums in \equ(3.30); the first one gives a factor $\prod_{v\in\ul
v_0}|P_v|\le \prod_{v\in V_c, v\not\in\t'}|P_v|$, the second a
factor $\g^{-h}$. Combining all these bounds, we find that also
the term with $L=1$ can be bounded by the r.h.s. of
\equ(3.6).\\

\0(A2) $s_{v_0}\ge 2$ and $T_{v_0}\= l_{\xx,\yy}$. In this case by
hypothesis all the vertices in $\ul v_0$ belong to $l_{\xx,\yy}$
and each of them has at least two free loop lines contracted into
the Gram determinant, see Fig. \graf(A2).

\insertplot{300pt}{100pt}{\ins{50pt}{40pt}{$v_\xx$}
\ins{245pt}{45pt}{$v_\yy$}
\ins{75pt}{45pt}{$l^*$}
}%
{A2}{\eqg(A2)}

\*

We consider the vertex $v_\xx$ and we note that, by construction,
$v_\xx$ has one external line coinciding with $f_\xx$ and the
other one belonging to the spanning tree $T_{v_0}$, let us call it
$f^*$. Calling $\tilde P_{\ul v_0}$ the set of loop lines
contracted into the Gram determinant and $\tilde P_{v_\xx}=\tilde
P_{\ul v_0}\cap P_{v_\xx}$, we again write $G^{h,T_{v_0}}$ in
blocks, as in \equ(3.9), with $A$ the block with both row and
column indices in $\tilde P_{v_\xx}$, $B$ the block with row
indices in $\tilde P_{v_\xx}$ and column indices in $\tilde P_{\ul
v_0}\setminus \tilde P_{v_\xx}$, $C$ the block with row indices in
$\tilde P_{\ul v_0}\setminus \tilde P_{v_\xx}$ and column indices
in $\tilde P_{v_\xx}$ and $D$ the block with both row and column
indices in $\tilde P_{\ul v_0} \setminus \tilde P_{v_\xx}$.

In this case, by a fourth order expansion of $\pmatrix{A&B\cr
C&D\cr}$ around $\pmatrix{A&0\cr 0&D\cr}$, we get the analogue of
\equ(3.12) and, since $|P_{v_0}|=2$, we get the bound analogue to
\equ(3.13):
$$\lis J^{(2)}_{h,n,\bar\o}(\t,\bP,T;0)\le\sum_{L=0,2,4}
\lis J^{(2)}_{h,n,\bar\o}(\t,\bP,T;0,L)\Eq(3.32)$$
where $\lis J^{(2)}_{h,n,\bar\o}(\t,\bP,T;0,L)$ with $L=0,2,4$ are
defined by an equation completely analogous to \equ(3.14), where,
if $L=2,4$, the sum $\sum^*_{\{i_qj_q,i_q'j_q'\}_{q=1}^L}$ has to
be interpreted as the sum over the indices
$\{i_qj_q,i_q'j_q'\}_{q=1}^L$ such that $f_{i_qj_q}\in P_{v_\xx}$
and $f_{i_q'j_q'}\in P_{\ul v_0}\setminus P_{v_\xx}$ or viceversa,
while, if $L=0$, the factor
$\fra1{L!}\sum^*_{\{i_qj_q,i'_qj'_q\}_{q=1}^L} \Big[\prod_{q=1}^L
G^{h,T_{v_0}}_{i_qj_q,i'_qj'_q}\Big]\cdot \det \wt
G^{h,T_{v_0}}\big(\{i_qj_q,i'_qj'_q\}_{q=1}^L;s_L,\tt_{v_0} \big)$
in \equ(3.14) must be interpreted as equal to $\det A\cdot \det
D$. In other words, the bound \equ(3.32) allows us to distinguish
between the contributions to $\lis
J^{(2)}_{h,n,\bar\o}(\t,\bP,T;0)$ such that: $v_\xx$ is connected
with $\ul v_0\setminus v_\xx$ only by a line of the spanning tree
($L=0$) or $v_\xx$ is connected with $\ul v_0\setminus v_\xx$ also
by exactly two loop lines ($L=2$) or $v_\xx$ is connected with
$\ul v_0\setminus v_\xx$ also by four or more loop lines ($L=4$).\\

We now want to show how to get the desired dimensional gain for
the three contributions $\lis J^{(2)}_{h,n,\bar\o}(\t,\bP,T;0,L)$
in \equ(3.32), $L=0,2,4$.\\
\\
(A2.1) $L=2$: there are exactly two loop lines connecting $\tilde
P_{v_\xx}$ with $\tilde P_{v_0}\setminus \tilde P_{v_x}$, call
them $l_1,l_2$ and $f_1,f_2$ their field labels, see Fig.
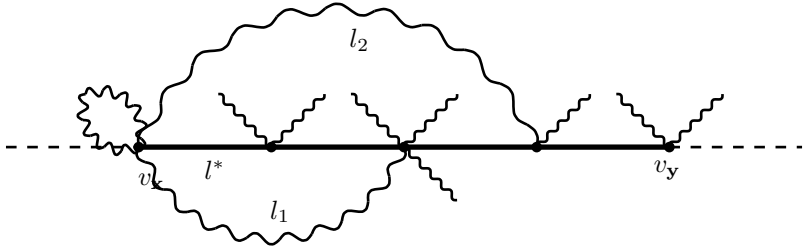
\graf(A21).

\insertplot{300pt}{120pt}{\ins{50pt}{40pt}{$v_\xx$}
\ins{245pt}{45pt}{$v_\yy$}
\ins{75pt}{45pt}{$l^*$}
\ins{100pt}{30pt}{$l_1$}
\ins{130pt}{95pt}{$l_2$}
}%
{A21}{\eqg(A21)}

\*

This case can be treated in a way similar to the case A1.1 above.
Let us consider the product
$\prod_{q=1}^2G^{h,T_{v_0}}_{i_qj_q,i'_qj'_q}=
\prod_{q=1}^2t_{i_q,i_q'}g_{\o(f_q)}^{(h)}(\xx_{l_q}-\yy_{l_q})
\d_{\o_{l_q}^+,\o_{l_q}^-}$ in \equ(3.14) and let us rewrite both
the extracted loop propagators $g_{\o(f_q)}^{(h)}$ with $q=1,2$
and the propagator of the spanning tree $g_{\o(f^*)}^{(h)}$ (we
recall that $f^*$ is the line of $T_{v_0}$ belonging to
$P_{v_\xx}$) as sums of isotropic propagators, to be denoted by
$\lis g_{\bar\o_q}^{(h)}$, $q=1,2$, and $\lis g_{\bar\o^*}^{(h)}$.
For any fixed configuration of internal anisotropic sectors, the
new sums on the isotropic sectors can be written as:
$$\sum_{\bar\o_q\prec \o(f_q)\atop q=1,2}\ \sum_{\bar\o^*\prec\o(f^*)}\;,
\Eq(3.33)$$
where the meaning of the symbol $\prec$ is the same as in
\equ(2.88). Note also that, by construction, the summand in the
r.h.s. of \equ(3.14) is $0$ unless
$\c(\{\lis\o,\lis\o_1,\lis\o_2,\lis\o^*\})=1$, so that we can
freely multiply the summand in the r.h.s. of \equ(3.14) by
$\c(\{\lis\o,\lis\o_1,\lis\o_2,\lis\o^*\})$. In this case it is
crucial the assumption that the external sector index
corresponding to $f_\xx$ is isotropic.

Having done this, we can bound the resulting expression by
replacing all the modified coupling functions, all the propagators
and all the Gram determinants with their absolute values. Now, we
can bound the integral over the absolute values of the modified
coupling functions as in \equ(2.82) (by using again \equ(2.71b) in
place of \equ(2.71a)), the integral over the absolute values of
the propagators of the spanning tree as in
\equ(2.81) and the product of the determinants times
$\prod_{q=1}^2|\lis g_{\bar\o(f_q)}^{(h)}(\xx_{l_q}-\yy_{l_q})|$
by the product over $v$ of the r.h.s. of \equ(2.80) {\it times a
dimensional gain} $\g^{h}$. After these bounds, the sum
$\sum^*_{\{i_qj_q,i'_qj'_q\}_{q=1}^2}$ can be bounded by
$\prod_{v\in\ul v_0}|P_v|^2\le \prod_{v\in V_c}|P_v|^5$ (the
latter is the same product appearing in the r.h.s. of \equ(3.6))
and we are still left with the sum over the sector indices:
$$\sum_{\cup_v \O_v\setminus\O_{v_0}}
\sum_{\bar\o_q\prec \o(f_q)\atop
\bar\o^*\prec\o(f^*)}\Big[\prod_{v>v_0\atop v\in V_c}\ \c
(\tilde\O_{v,\bar\o})\Big]\c(\{\lis\o,\lis\o_1,\lis\o_2,\lis\o^*\})
\Big[\prod_{l\in T}\d_{\o_l^+,\o_l^-}\Big]
\prod_{q=1}^2\d_{\o_{l_q}^+,\o_{l_q}^-}\;.\Eq(3.34)$$
where the product $\c(\{\lis\o,\lis\o_1,\lis\o_2,\lis\o^*\})
\prod_{q=1}^2\d_{\o_{l_q}^+,\o_{l_q}^-}$ takes into account the
new constraints coming from the extraction of the two loop
propagators.

We want to prove that \equ(3.34) can be bounded by the r.h.s. of
\equ(2.83) times a dimensional factor $|h|$. Combining this loss
with the gain $\g^{h}$ coming from the bound on the size of the
loop propagators and discussed above, we see that globally the
contribution under analysis has a dimensional gain $\g^h|h|$ with
respect to the bounds described in \sec(2.8), that is the desired
gain.

We proceed as in item A1.1 above. With the same notations as in
\equ(3.20), we can rewrite \equ(3.34) as
$$\eqalign{&\sum_{\O_{\ul{v}_0}\setminus\O_{v_0}}
\sum_{\bar\o_q\prec \o(f_q)\atop \bar\o^*\prec\o(f^*)}\Big[
\prod_{v\in\ul{\tilde v}_0} F_{v}(\tilde\O_{v,\bar\o})\Big]\cdot
\c(\{\lis\o,\lis\o_1,\lis\o_2,\lis\o^*\})\prod_{l\in
T_{v_0}}\d_{\o_l^+,\o_l^-} \prod_{q=1}^2\d_{\o_{l_q}^+,\o_{l_q}^-}
\;.\cr}\Eq(3.35)$$
Defining
$${\cal F}_{\ul{v}_0\setminus v_\xx}(\tilde\O_{\ul{v}_0
\setminus v_\xx})\defin \prod_{v\in\ul{v}_0\setminus v_\xx}
F_{v}(\tilde\O_{v,\bar\o}) \prod_{l\in T_{v_0}\setminus
l^*}\d_{\o_l^+,\o_l^-}\;,\Eq(3.36)$$
and calling $\O^{(3)}=\{\o(f_1),\o(f_2),\o(f^*)\}$, we can bound
\equ(3.35) by
$$\eqalign{&
\sup_{\O^{(3)}}\Big[\sum_{\tilde\O_{\ul{v}_0\setminus v_\xx}}^{*}
\ {\cal F}_{\ul{v}_0\setminus v_\xx}\big(\tilde\O_{\ul{v}_0
\setminus v^*}\big)\Big]\,\cdot\, \sup_{\bar\o,\O^{(3)}}\Big[
\sum_{\tilde\O_{v_\xx,\bar\o}}^*
F_{v^*}\big(\tilde\O_{v_\xx,\bar\o}\big)\Big]\,
\sum_{\bar\o_1,\bar\o_2,\bar\o^*}
\c(\{\lis\o,\lis\o_1,\lis\o_2,\lis\o^*\}) \;,\cr}\Eq(3.37)$$
where the $*$ on the sums means that we are not summing over the
indices in $\{\lis\o\}\cup\O^{(3)}$. Now, the first two sums can
be bounded using Lemma \lm(2.4) and, with respect to the cases in
which the two sums have just one fixed external sector (as it is
the case in the proof of \equ(2.83), see \equ(2.94)), they both
have a dimensional gain $\g^{h/2}$ (so combining the two, their
product has a gain of $\g^h$). By Lemma \lm(3.1) we see that the
contribution from the last sum can be bounded by $c\g^{-h}|h|$,
and we see that \equ(3.34) can be bounded by the r.h.s. of
\equ(2.83) times a dimensional factor $|h|$. As discussed after
\equ(3.34) this implies that $\lis
J^{(2)}_{h,n,\bar\o}(\t,\bP,T;0,L=2)$ with $|P_{v_0}|=2$
admits a dimensional bound given by the r.h.s. of \equ(3.6) with $j=0$.\\

\0(A2.2) L=4: there are four extracted loop lines connecting
$\tilde P_{v_\xx}$ with $\tilde P_{\ul v_0}\setminus \tilde
P_{v_\xx}$, call them $l_1,l_2,l_3,l_4$ and $f_1,f_2,f_3,f_4$
their field labels, see Fig. \graf(A22).

\insertplot{300pt}{120pt}{\ins{40pt}{45pt}{$v_\xx$}
\ins{250pt}{45pt}{$v_\yy$}
\ins{75pt}{45pt}{$l^*$}
\ins{120pt}{25pt}{$l_1$}
\ins{130pt}{95pt}{$l_2$}
\ins{85pt}{80pt}{$l_3$}
\ins{150pt}{20pt}{$l_4$}
}%
{A22}{\eqg(A22)}

\*

This case can be treated in a way similar to item A1.2 above: we
replace all the modified coupling functions, all the propagators
and all the Gram determinants with their absolute values. Having
done this, we bound the integral over the modified coupling
functions as in \equ(2.82), the integral over the propagators of
the spanning tree as in \equ(2.81) and we bound the product of the
determinants times
$\prod_{q=1}^4|g_{\o(f_q)}^{(h)}(\xx_{l_q}-\yy_{l_q})|$ by the
product over $v$ of the r.h.s. of \equ(2.80). After these bounds,
the sum $\sum^*_{\{i_qj_q,i'_qj'_q\}_{q=1}^4}$ can be bounded by
$\prod_{v\in\ul v_0}|P_v|^4\le \prod_{v\in V_c}|P_v|^5$ (the
latter is the same product appearing in the r.h.s. of \equ(3.6))
and we are still left with the sum over the sector indices:
$$\sum_{\cup_v\O_v\setminus\O_{v_0}}\Big[
\prod_{v>v_0\atop v\in V_c}\c(\tilde\O_{v,\bar\o})\Big]
\Big[\prod_{l\in T}\d_{\o_l^+,\o_l^-}\Big]
\prod_{q=1}^4\d_{\o_{l_q}^+,\o_{l_q}^-}\;,\Eq(3.38)$$
where the product $\prod_{q=1}^4\d_{\o_{l_q}^+,\o_{l_q}^-}$ takes
into account the new constraints coming from the extraction of the
three loop propagators.

We want to prove that \equ(3.38) can be bounded by the r.h.s. of
\equ(2.83) times a dimensional factor $\g^{h}$, that
is even more than the desired dimensional gain. With the same
notations as above, we rewrite \equ(3.38) in the form
$$\sum_{\O_{\ul{v}_0}\setminus\O_{v_0}}\Big[
\prod_{v\in\ul{\tilde v}_0}F_{v}(\tilde\O_{v,\bar\o})\Big]\cdot
\prod_{l\in T_{v_0}}\d_{\o_l^+,\o_l^-}
\prod_{q=1}^4\d_{\o_{l_q}^+,\o_{l_q}^-} \;.\Eq(3.39)$$
Defining ${\cal F}_{\ul{v}_0\setminus v_\xx}(\tilde\O_{\ul{v}_0
\setminus v_\xx})$ as in \equ(3.36) and calling $\O^{(5)}=
\{\o(f_q)\}_{q=1}^4,\cup\{\o(f^*)\}$, we can bound
\equ(3.39) by
$$
\sup_{\bar\o}\Big[\sum_{\tilde\O_{\ul{v}_0\setminus v_\xx} }^*
{\cal F}_{\ul{v}_0\setminus v_\xx}\big(\tilde\O_{\ul{v}_0
\setminus v_\xx}\big)\Big]\,\cdot\, \sup_{\bar\o,\O^{(5)}}\Big[
\sum_{\tilde\O_{v_\xx,\bar\o}\setminus\O^{(5)}}^*
F_{v_\xx}\big(\tilde\O_{v_\xx}\big)\Big]\;.\Eq(3.40)$$
where the $*$ on the sums means that we are not summing over the
sectors in $\O_{v_0}$. Now, the first sum can be bounded exactly
as in the proof of \equ(2.38), while the second one, using Lemma
\lm(2.4), has a dimensional gain $\g^h$, and this concludes the
proof in the present case.
\\

\0(A2.3)  L=0: there are no loop lines connecting $\tilde
P_{v_\xx}$ with $\tilde P_{v_0}\setminus \tilde P_{v_\xx}$, see
Fig. \graf(A23).

\insertplot{300pt}{120pt}{\ins{45pt}{45pt}{$v_\xx$}
\ins{245pt}{45pt}{$v_\yy$}
\ins{75pt}{45pt}{$l^*$}
}%
{A23}{\eqg(A23)}

\*

We consider the line $l^*$ in $T_{v_0}$ anchored to $v_\xx$ and we
rewrite the corresponding propagator $g_{\o(f^*)}^{(h)}$ as a sum
of isotropic propagators, to be denoted by $\lis
g_{\bar\o^*}^{(h)}$. For any fixed configuration of internal
anisotropic sectors, the new sum on the isotropic sectors can be
written as $\sum_{\bar\o^*\prec\o(f^*)}\,\openone
(|\lis\o-\lis\o^*|\le 1)$, where the characteristic function is
due to the remark that the value of the integrand in the r.h.s. of
\equ(3.14) is $0$ unless $|\lis\o-\lis\o^*|\le 1$. We recall that
if $L=0$ the factor $\fra1{L!}
\sum^*_{\{i_qj_q,i'_qj'_q\}_{q=1}^L} \Big[\prod_{q=1}^L
G^{h,T_{v_0}}_{i_qj_q,i'_qj'_q}\Big]\cdot \det \wt
G^{h,T_{v_0}}\big(\{i_qj_q,i'_qj'_q\}_{q=1}^L;s_L,\tt_{v_0} \big)$
in \equ(3.14) must be interpreted as equal to $\det A\cdot \det
D$. If we rename the matrix $A$ by $\wt
G^{h}_{v_\xx}(\tt_{v_\xx})$ and the matrix $D$ by $\wt
G^{h,T_{v_0}}_{\ul v_0\setminus v_\xx} (\tt_{v_0})$, we can bound
$\lis J^{(2)}_{h,n,\bar\o}(\t,\bP,T;0,0)$ in a way similar to
\equ(3.30):
$$\sum_{\bar\o^*} \openone(|\lis\o-\lis\o^*|\le 1) [v_\xx] \cdot
[\ul v_0\bs v_\xx] \int d\xx |\lis g_{\bar\o^*}^{(h)}(\xx)|
\;.\Eq(3.41)$$
Note that $[v_\xx]$ has a form similar to \equ(3.30a), so that it
can be bounded by using \equ(3.8), if $v_\xx$ is an endpoint, or
\equ(3.31), in the opposite case; this will produce a gain $\g^h|h|$
with respect to the dimensional estimate one would get simply
using \equ(2.71a) and Lemma \lm(2.2).

Let us now consider $[\ul v_0\bs v_\xx]$. It has the structure of
a contribution to the effective potential with two external fields
with fixed isotropic sectors, which we substitute in the bound
with the corresponding anisotropic ones, before using Theorem
\thm(2.1). Hence no gain or loss follows from this operation, as
well from the bound $\int d\xx |\lis g_{\bar\o^*}^{(h)}| \le
c\g^{-h}$. Since $\sum_{\bar\o^*} \openone(|\lis\o-\lis\o^*|\le 1)
\le c$, the r.h.s. of \equ(3.41) can be bounded by $(C|U|)^n
(|h|\g^{2h}) \g^h \g^{-h} = (C|U|)^n |h|\g^{2h}$, times the usual
factor $\prod_{v\in V_c, v> v_\xx}{1\over s_v!} |P_v|^5
\g^{\d(|P_v|)(h_v-h_{v'})}$.

\\
We now turn to the proof of the case $s_{v_0}=1$ and of the bound
\equ(3.31).
\\

\0(B) In this section we prove the dimensional gain for the
case $s_{v_0}=1$. The proof in this item {\it includes} the
unfinished proofs of items (A1.3) and (A2.3) above, that is the
proofs of the bounds \equ(3.31) and of the analogue bound for
$[v_\xx]$ in (A2.3).

If $s_{v_0}=1$, we call $v_m$ the first non trivial vertex
following $v_0$ on $\t$ and $v_1,\ldots,v_{m-1}$ the trivial
$c$--vertices preceding $v_m$ on $\t$, see Fig. \graf(treeB),
where the $c$--vertices are represented as dots of larger size
with respect to the others.\\
\\
\\

\insertplot{300pt}{120pt}{\ins{30pt}{85pt}{$r$}
\ins{55pt}{88pt}{$v_0$}
\ins{95pt}{88pt}{$v_1$}
\ins{135pt}{88pt}{$v_2$}
\ins{30pt}{-2pt}{$h$--$1$}
\ins{58pt}{-2pt}{$h$}
\ins{135pt}{-2pt}{$k$}
\ins{255pt}{-2pt}{$+1$}
}%
{treeB}{\eqg(treeB)}

\*\*

Let us call $k=h_{v_m}$ the scale of $v_m$ and $\ul v_m$ the set
of $c$--vertices immediately following $v_m$ on $\t$; we shall
also call {\it tadpole propagator} any propagator linking two
external fields of a single vertex. Very roughly, the strategy
will consist in getting a first dimensional gain $\g^{k}|k|$ on
scale $k=h_{v_m}$ by a loop extraction similar to that described
in \sec(3.1), by distinguishing again among several cases.
Moreover, in order to ``transfer the gain on scale $h=h_{v_0}$'',
we shall suitably ``extract from the determinant'' one tadpole
propagator of scale $h$ and expand it into a sum of isotropic
propagators, as described below.
\\
Calling $\det G^{h}_{v_0},\det G^{h_{v_1}}_{v_1},\ldots, \det
G^{h_{v_{m-1}}}_{v_{m-1}}$ the determinants of the tadpole
propagators contracted on scales $h,h_{v_1},\ldots,h_{v_{m-1}}$
respectively, we want to prove the following:
$$\eqalign{&
\sum_{\O\setminus \O_{v_0}}\Big[\prod_{v\in \t}
\c(\tilde\O_{v,\bar\o})\Big] \Big| \int d(\xx_{v_m}\bs
\xx^*){\lis{\FFF}}_{2,h,(\bar\o,\bar\o)} *\Big[
\big(\prod_{i=0}^{m-1}\det G^{h_{v_i}}_{v_i}\big)
W^{(mod)}_{\t',\bP',\O',T'}(\xx_{v_m})\Big]\Big|\le\cr
&\le(c|U|)^n \g^{2h}|h| \prod_{v\in V_c, v>v_0}{1\over
s_v!}|P_v|^5\g^{\d(|P_v|)(h_v-h_{v'})} \;,\cr}\Eq(3.43)$$
where the l.h.s. is equal to $\lis
J^{(2)}_{h,n,(\bar\o,\bar\o)}(\t,\bP,T;0)$ and $\t'$ is the
subtree of $\t$ such that $v_0'$, \ie the first vertex following
the root on $\t'$, coincides with $v_m$. Correspondingly,
$\bP',\O',T'$ are the field labels, the sector labels and the
spanning tree of $\t'$.

Comparing \equ(3.43) with the naive bound one would get by
repeating the proof of Theorem \thm(2.1) we see that, again, we
want to find a dimensional gain
of $\g^h|h|$ with respect to the estimates in Sect.2.\\
\\
{\bf Remark.} After a suitable (and obvious) identification of the
symbols in \equ(3.30a) with those in \equ(3.43), it becomes clear
that the bound \equ(3.31) and the analogue bound for $[v_\xx]$ in
item (A2.3) are essentially the same as \equ(3.43) (unless for the
specific values of the external fixed isotropic sectors). Then the
proof of \equ(3.43) described in the following applies unvaried to
the analysis of \equ(3.31) and of the analogue bound for $[v_\xx]$
so that the bound \equ(3.43) will also complete the proof of the
dimensional
gain in items (A1.3) and (A2.3) above.\\
\\
We look into the structure of $v_m$. Since, by construction,
$s_{v_m}\ge 2$, the spanning tree $T_{v_m}$ can be represented as
in Fig. \graf(A1) or in Fig. \graf(A2); the only difference is
that the loop fields may have different scales between $k$ and
$h$, with the important constraint that at least four fields have
scale $h$ (indeed two fields of scale $h$ would be sufficient).
Let us choose, as before, $v_\xx$ as the root of $T_{v_m}$  and
select one leaf of $T_{v_m}$, to be called again $v^*$. If it is
possible, as in the case of Fig. \graf(A1), we choose $v^*\neq
v_\yy$, otherwise (that is if $T_{v_m}\= l_{\xx,\yy}$, as in Fig.
\graf(A2)) it will be necessarily $v^*\=v_\yy$. By construction
$v^*$ has one external line belonging to the spanning tree
$T_{v_m}$, call it $f^*$ and call $\o^*$ its sector label. If
$v^*\neq v_\yy$, $P_{v^*}$ will have an odd number of fields
associated to loop lines contracted into the Gram determinant. If
$v^*\=v_\yy$, $P_{v^*}$ will have an even number of fields
associated to loop lines contracted into the Gram determinant and
one more field belonging to $P_{v_0}$ (associated to a fixed
isotropic sector index).

We consider the product of determinants
$$\big(\prod_{i=0}^{m-1}\det G^{h_{v_i}}_{v_i}\big)\cdot \det
G^{h_{v_m},T_{v_m}} \;,\Eq(3.48)$$
where $\det G^{h_{v_m},T_{v_m}}$ is the determinant of the loop
lines contracted on scale $h_{v_m}$, if this family of lines is
not empty, otherwise it is equal to $1$. The product in
\equ(3.48) can be rewritten as the determinant of a single ``big''
matrix $G_{\ul v_m}$ of all the loop lines contracted on scales
$h,h_{v_1},\ldots,h_{v_m}$ (all the loop lines in the examples
similar to those of Fig. \graf(A1) and \graf(A2)). Let us call
$\tilde P_{v^*}\subset P_{v^*}$ the set of loop lines of $P_{v^*}$
contracted into $G_{\ul v_m}$ and $\tilde P_{\ul v_m}$ the whole
set of loop lines in $\cup_{w\in\ul v_m}P_w$ contracted into
$G_{\ul v_m}$. In analogy with the procedure described in item
(A1), we rewrite $G_{\ul v_m}$ in blocks, as in the r.h.s. of
\equ(3.9), with the blocks $A,B,C,D$ defined in a way similar to
that introduced after
\equ(3.9): that is $A$ is the block with both row and column indices in
$\tilde P_{v^*}$, $B$ is the block with row indices in $\tilde
P_{v^*}$ and column indices in $\tilde P_{\ul v_m}\bs\tilde
P_{v^*}$, and so on. Then we Taylor expand the determinant
$\det\pmatrix{A&B\cr C&D\cr}$ around $\det\pmatrix{A&0\cr 0&D\cr}$
up to fourth or fifth order, depending whether $v^*$ is equal to
$v_\yy$ or not: in this way we can distinguish the contributions
to \equ(3.43) in which there are exactly $L=0,1,2,3$ loop lines
connecting $v^*$ with $\ul v_m\bs v^*$ from those with at least
$4$ or $5$ loop lines, depending on whether $v^*$ is equal to
$v_\yy$ or not. We do not write explicitly here the expressions
corresponding to these various contributions: they are similar to
the r.h.s. of \equ(3.14) with $\wt G^{h,T_{v_0}}$ replaced by a
suitable minor $\wt G_{\ul v_m}$ of the matrix $G_{\ul v_m}$. As
in Section \secc(3.1) above, in the contributions with
$L=0,1,2,3$, the minor $\wt G_{\ul v_m}$ can be rewritten in a
factorized form, analogue to \equ(3.29), allowing to distinguish a
contribution associated to the leaf $v^*$ from a rest associated
to $\ul v_m\bs v^*$.

We now turn to the detailed analysis of the cases $L=0,1,2,3,4,5$.\\
\\
(B.1a) $L=3$: there are exactly three loop lines connecting
$\tilde P_{v^*}$ with $\tilde P_{v_m}\setminus \tilde P_{v^*}$,
call them $l_1,l_2,l_3$ and denote by $f_1,f_2,f_3$, $h_1,h_2,h_3$
and $\o_1,\o_2,\o_3$ their field, scale and sector labels. With no
loss of generality, we can assume $h\le h_1\le h_2\le h_3\le k$.

In this case we consider the propagators of $f_1,f_2,f_3,f^*$ and
we expand them as sums of isotropic propagators, to be denoted by
$\lis g^{(h_i)}_{\bar\o_i}$, $i=1,2,3$ and $\lis
g^{(k)}_{\bar\o_{f^*}}$.  If $h_1=h$, we rewrite the propagator
$\lis g^{h_1}_{\bar\o_1}(\xx_{f_1^+}-\yy_{f_1^-})$ in the form
$$\lis g^{h_1}_{\bar\o_1}(\xx_{f_1^+}-\yy_{f_1^-})=
\lis g^{h_1}_{\bar\o_1}({\bf 0})+(\xx_{f_1^+}-\yy_{f_1^-})\int_0^1
ds \dpr_\xx\lis
g^{h_1}_{\bar\o_1}\big(s(\xx_{f_1^+}-\yy_{f_1^-})\big)
\;.\Eq(3.49)$$
If $h_1>h$, we expand the determinant of $\wt G_{\ul v_m}$
appearing in the analogue of \equ(3.14) along the row
corresponding to a field label $f_0^-\in P_{v_1}^-\setminus
P_{v_0}^-$ (that is a field of scale $h$) and we rewrite
$$\eqalign{&\det\wt G_{\ul v_m}=\sum_{f_0^+\in P_{v_1}^+\bs P_{v_0}^+}
\,(-1)^{\e_0}\cdot g_{\o_0}^{(h)}(\xx_{f_0^+}-\yy_{f_0^-})
\cdot\det\wt G_{\ul v_m}(\{f_0^+,f_0^-\})\cdot
\d_{\o(f_0^+),\o(f_0^-)}\;,\cr
&g_{\o_0}^{(h)}(\xx_{f_0^+}-\yy_{f_0^-})=g^{(h)}_{\o_0}({\bf 0})+
\sum_{\bar\o_0\prec\o_0}\big(\xx_{f_0^+}-\yy_{f_0^-}\big)\int_0^1ds\,
\dpr_\xx \lis
g^{(h)}_{\bar\o_0}\big(s(\xx_{f_0^+}-\yy_{f_0^-})\big)\;,\cr
}\Eq(3.50)$$
where $\wt G_{\ul v_m}(\{f_0^+,f_0^-\})$ is the minor of $\wt
G_{\ul v_m}$ corresponding to the entry $(f_0^+, f_0^-)$,
$(-1)^{\e_0}$ is the corresponding sign and we denoted by $\o_0$
the common value of $\o(f_0^\pm)$. Both in the case $h_1=h$ and
$h_1>h$, implementing the decompositions \equ(3.49) or \equ(3.50),
we see that the terms proportional to $\lis
g^{h_1}_{\bar\o_1}({\bf 0})$ or to $g^{(h)}_{\o_0}({\bf 0})$ have
a dimensional gain of $\g^h$, associated with the fact that the
bound on $\lis g^{(h)}_{\bar\o}({\bf 0})$ and on
$g^{(h)}_{\o}({\bf 0})$ are $\g^h$ smaller than the respective
bounds on $\lis g^{(h)}_{\bar\o}({\bf x})$ and on
$g^{(h)}_{\o}({\bf x})$, see Lemma \lm(2.2), Lemma \lm(2.2a),
Lemma \lm(2.3) and Lemma \lm(2.2b).

So, in both cases, let us focus on the contributions proportional
to the interpolated term in \equ(3.49) and \equ(3.50).

Such terms have essentially the same structure as the r.h.s. of
\equ(3.14) with $L=3$, unless for the following facts: the minor
$\wt G^{h,T_{v_0}}$ is replaced by a suitable minor of the
determinant $\det G_{\ul v_m}$ of all loop lines contracted on
scales $h=h_{v_0},h_{v_1},\ldots,h_{v_m}=k$; if $h_1>h$ the number
of extracted loop lines is 4 instead of 3 and their field labels
are denoted by $f_0,f_1,f_2,f_3$; one of the loop lines ($f_1$ if
$h_1=h$ or $f_0$ if $h_1>h$) is associated to an interpolated
propagator (the second term in\equ(3.49) or the second term in the
second line of \equ(3.50), respectively).

For any fixed configuration of anisotropic sectors, the new sums
on the isotropic sectors, including possibly the sum over
$\lis\o_0$, can be written as:
$$\prod_{i}^*\sum_{\bar\o_i\prec \o_{f_i}}\;.\Eq(3.50a)$$
where the * on the product means that the product runs over
$i=1,2,3$ if $h_1=h$ and over $i=0,1,2,3$ if $h_1>h$. Note also
that, by construction, the summand in the expression under
analysis (\ie the analogue of the summand in the r.h.s. of
\equ(3.14)) is identically $0$ unless
$\c(\{\lis\o_{1},\lis\o_{2},\lis\o_{3},\lis\o_{f^*}\})=1$. Then we
are free to multiply the summand in the expression under analysis
by $\c(\{\lis\o_{1},\lis\o_{2},\lis\o_{3},\lis\o_{f^*}\})$.
However we choose to multiply it by a slightly weaker constraint,
that is by
$\c(\{\lis\o_{1}^{(k)},\lis\o_{2}^{(k)},\lis\o_{3}^{(k)},\lis\o_{f^*}\})$,
where $\lis\o_{i}^{(k)}$ is the isotropic sector index on scale
$k$ such that $\lis\o_{i}\prec \lis\o_{i}^{(k)}$ (note that by
construction $\lis\o_{f^*}$ is already on scale $k$).

Having done this, we can bound the resulting expression by
replacing all the modified coupling functions, all the propagators
and all the Gram determinants with their absolute values. As
regarding the factor $(\xx_{f_1^+}-\yy_{f_1^-})$ or
$(\xx_{f_0^+}-\yy_{f_0^-})$ appearing in \equ(3.49) or \equ(3.50),
we can bound its absolute value by a sum over the lines $l\in T''$
of $|\xx_l-\yy_l|$, where $T''$ is a suitable subset of the
spanning tree $T'$. In particular we can associate each factor
$|\xx_l-\yy_l|$ to the corresponding propagator of the spanning
tree $T'$. Note that the number of terms in the sum $\sum_{l\in
T''}$ is of order $n$.

Now, we can bound the integral over the absolute values of the
modified coupling functions as in \equ(2.82) (by using \equ(2.71b)
in place of \equ(2.71a)), the integral over the absolute values of
the propagators of the spanning tree times the factor $\sum_{l\in
T''} |\xx_l-\yy_l|$ by the r.h.s. of \equ(2.81) {\it times a
dimensional loss} $cn\g^{-k}$, where we used that the propagators
associated to $l\in T''$ are of scale $j\ge k$ and that, by the
decay properties described in Lemma \lm(2.2), $\int d\xx|\xx|
|g^{(j)}_\o(\xx)|\le c\g^{-2 j}$. Moreover, the product of the
determinants times the product of the absolute values of the
extracted loop propagators can be bounded by the product over $v$
of the r.h.s. of \equ(2.80) {\it times a dimensional gain}
$\g^{h(1+\fra12\openonesix(h_1>h))} \g^{\fra12(h_1+h_2+h_3)}$: the
first factor $\g^h$ comes from the derivative $\dpr_\xx$ acting on
$\lis g^{(h)}_{\bar\o_1}$ or on $\lis g^{(h)}_{\bar\o_0}$, see
Lemma \lm(2.3), while the remaining factors come from the improved
bound on the size of the isotropic propagators with respect to the
anisotropic ones.

After these bounds, the sums over the choices of the extracted
loop propagators (including possibly the sum over $f_0^+$) can be
bounded by $|P_{v_1}|\prod_{w\in\ul v_m}|P_w|^3$.

We are still left with the sum over the sector indices:
$$\sum_{\O\setminus\O_{v_0}}
\Big[\prod_{i}^*\sum_{\bar\o_i\prec
\o_i}\d_{\o_{f_i^+},\o_{f_i^-}}\Big] \Big[\prod_{v>v_0\atop v\in
V_c}\ \c (\tilde\O_{v,\bar\o})\Big]\c(\{\lis\o_1^{(k)},
\lis\o_2^{(k)},\lis\o_3^{(k)},\lis\o_{f^*}\}) \Big[\prod_{l\in
T'}\d_{\o_l^+,\o_l^-}\Big]\;.\Eq(3.51)$$
We now want to bound \equ(3.51) and to compare the result we shall
find with the bound we would get by proceeding as in
Sect.\secc(2.8) above, that is with:
$$c^n\prod_{v\in V_c\atop v\ {\rm not}\ {e.p.}}\g^{\fra12(h_v-h_{v'})(|P_v|-3+
2\openonesix(|P_v|\ge 10))}\prod_{v\
{e.p.}}\g^{-\fra12h_{v}}\;.\Eq(3.52a)$$
Using a notation completly analogous to the notations of
\secc(2.8), see \equ(2.84) and following equations, we begin with
bounding the product of $\c$ functions in \equ(3.51) as
$$\prod_{v>v_0\atop v\in V_c}\ \c(\tilde\O_{v,\bar\o})\le
\prod_{v>v_m \atop v\in V_c} \c(\tilde\O_{v,\bar\o}^{(k)})\;,
\Eq(3.52)$$
that is we simply neglect the constraints associated to the
vertices $v_1,\ldots, v_m$ and we weaken the remaining constraints
by replacing sectors on scales $\le k$ with the corresponding
sectors on scale $k$.

Also, we rewrite the sums over the sector indices in the form
$$\eqalign{&\sum_{\O\setminus\O_{v_0}}
\Big[\prod_{i}^*\sum_{\bar\o_i\prec \o_i}
\d_{\o_{f_i^+},\o_{f_i^-}}\Big]=
\Big[\prod_{i}^*\sum_{\bar\o_i^{(k)}\in\bar O_k}\Big]\cdot
\Big[\prod_{i}^*\sum_{\bar\o_i\prec \bar\o_i^{(k)}}\Big]\cdot
\sum_{\O_{v_m}^{(k)}}^{**}\ \cdot\
\sum_{\O_{v_m}\prec\O_{v_m}^{(k)}}^{**} \ \cdot\ \sum_{\O\bs
\O_{v_m}}^{**} \cr}\Eq(3.53)$$
where the $**$ on the sums mean that we are not summing neither
over the sectors in $\O_{v_0}$ nor over the sectors
$\o_0,\o_1,\o_2,\o_3$. Let us denote by $P_L$ the set of field
labels of the half--lines corresponding to the extracted loop
propagators (equal to $\{f_1^\pm,f_2^\pm,f_3^\pm\}$ or to
$\{f_0^\pm,f_1^\pm,f_2^\pm,f_3^\pm\}$, depending whether $h_1=h$
or not) and by $n^L_i=|P_L\cap P_{v_i}|$ the number of field
labels in $P_L$ external to the cluster $v_i$. With this notation,
we can bound
$$\eqalign{&\Big[\prod_{i}^*\sum_{\bar\o_i\prec \bar\o_i^{(k)}}\Big]\le
c\g^{(k-h)\openonesix(h_1>h)}\prod_{i=1}^3\g^{(k-h_{i})} \cr
&\sum_{\O_{v_m}\prec\O_{v_m}^{(k)}}^{**}\le c^m
\prod_{i=1}^m\g^{\fra12(h_{v_i}-h_{v_{i-1}})(|P_{v_i}|-2-n^L_i)}=\cr
&\hskip1.4truecm=
c^m\g^{-(k-h)\openonesix(h_1>h)}\prod_{i=1}^3\g^{-(k-h_{i})}
\prod_{i=1}^m\g^{\fra12(h_{v_i}-h_{v_{i-1}})(|P_{v_i}|-2)}
\cr}\Eq(3.54)$$
Combining the two bounds in \equ(3.54), we see that the global
contribution from the sector sums corresponding to the vertices
$v_1,\ldots v_m$ is proportional to
$\prod_{i=1}^m\g^{\fra12(h_{v_i}-h_{v_{i-1}})(|P_{v_i}|-2)}$, to
be compared with the corresponding bound
$\prod_{i=1}^m\g^{\fra12(h_{v_i}-h_{v_{i-1}})(|P_{v_i}|-3+2\openonesix(|P_v|
\ge 10))}$ obtained with the procedure of Section \secc(2.8), see
\equ(3.52a). We see that, concerning to the sector sums
corresponding to the vertices $v_1,\ldots v_m$, with respect to
the corresponding bound in \equ(3.52a) the present bound has a
{\it dimensional loss} at most equal to $\g^{\fra12(k-h)}$.

After having bounded as above the sector sums corresponding to the
vertices $v_1,\ldots v_m$, we are still left with the sector sums
corresponding to the vertices $v> v_m$, that can be rewritten in
the form
$$\eqalign{&\Big[\prod_i^*\
\sum_{\bar\o_i^{(k)}\in\bar
O_k}\Big]\sum_{\cup_{v>v_m}\O_v^{(k)}}^{**}
\Big[\prod_{v>v_m}\c(\tilde\O_{v,\bar\o}^{(k)})\Big]\c(\{\lis\o_1^{(k)},
\lis\o_2^{(k)},\lis\o_3^{(k)},\lis\o_{f^*}\}) \Big[\prod_{l\in
T'}\d_{\o_l^+,\o_l^-}\Big]=\cr
&=\sum_{\O_{\ul{v}_m}^{(k)}\setminus\O_{v_0}} \Big[
\prod_i^*\sum_{\bar\o_{i}^{(k)}\prec
\o_i^{(k)}}\d_{\o_{f_i^+},\o_{f_i^-}}\Big] \prod_{l\in
T_{v_m}}\d_{\o_l^+,\o_l^-} \Big[ \prod_{v\in\ul v_m}
F_{v}(\tilde\O_{v,\bar\o}^{(k)})\Big] \c(\{\lis\o_1^{(k)},
\lis\o_2^{(k)},\lis\o_3^{(k)},\lis\o_{f^*}\}) \;.\cr}\Eq(3.55)$$
Now, if $h_1=h$ (\ie if the only extracted loop lines are
$f_1,f_2,f_3$), defining $\O^{(4)}=\{\o_1,\o_2,$
$\o_3,\o_{f^*}\}$, we see that we can soon bound \equ(3.55) by an
expression analogue to the r.h.s. of \equ(3.20), so that,
repeating the discussion after \equ(3.20), we find that \equ(3.55)
can be bounded by the factor in \equ(3.52a) corresponding to the
vertices $v>v_m$ times a dimensional loss of $\g^{-k/2}|k|$. If
$h_1>h$ (\ie if the extracted loop lines are $f_0,f_1,f_2,f_3$) we
bound the sum $\sum_{\bar\o_{0}^{(k)}\prec \o_{0}^{(k)}}$ by a
constant times $\g^{-k/2}$. After this, we are again left with an
expression analogue to the r.h.s. of \equ(3.20), so that we lose a
factor $\g^{-k/2}|k|$ more.

In conclusion we can say that \equ(3.55) can be bounded by a
quantity that, with respect to the factor in \equ(3.52a)
corresponding to the vertices $v>v_m$, has a loss of
$\g^{-\fra12 k(1+\openonesix(h_1>h))}|k|$.\\

Combining all the gain and losses discussed in this item we find
that the contributions under analysis, with respect to the bounds
obtained via the procedure in Section \secc(2.8), have a global
gain of:
$$\eqalign{&cn\g^{-k}\g^h\g^{\fra12 h\openonesix(h_1>h)}\g^{\fra12
(h_1+h_2+h_3)} |P_{v_1}|\Big[\prod_{w\in\ul
v_m}|P_{w}|^3\Big]\g^{{1\over 2}(k-h)} \g^{-\fra12
k(1+\openonesix(h_1>h))}|k|=\cr =&cn|P_{v_1}|\Big[\prod_{w\in\ul
v_m}|P_{w}|^3\Big]\g^h|k| \g^{{1\over 2}(h-k)
(\openonesix(h_1>h)-1)}\prod_{i=1}^3\g^{\fra12(h_i-k)} \le
c^n\g^h|h|\prod_{v\in V_c}|P_v|^5\cr}\Eq(3.56)$$
where in the last inequality we used the trivial bounds $cn\le
c^n$, $|k|\le |h|$ and $|P_{v_1}|\cdot $ $\cdot\Big[\prod_{w\in\ul
v_m}|P_{w}|^3\Big]$ $\le \prod_{v\in V_c}|P_v|^5$.
\\
\\(B.1b) $L=2$: there are exactly three loop lines connecting
$\tilde P_{v^*}$ with $\tilde P_{v_m}\setminus \tilde P_{v^*}$. We
call $f_1,f_2$ the field labels of the extracted loop lines,
$h_1,h_2$ their scale labels (assume $h\le h_1\le h_2\le k$) and
$\o_1,\o_2$ their sector labels. In this case we proceed in a way
very similar to item (B.1a) above and we only rapidly repeat the
proof of item B.1a adapted to the present case. The main
difference consists in the fact that one of the fields in
$P_{v^*}$, call it $f_3$, belongs to $\P_{v_0}$, so that its
sector label is fixed to be isotropic and equal to $\lis\o$.

As in item B.1a we expand the propagators of $f_1,f_2$ and $f^*$
as sums of isotropic propagators and, depending whether $h_1=h$ or
not, we also perform the expansions in \equ(3.49) and \equ(3.50).
Focusing on the contributions coming from the intepolated terms in
\equ(3.49),
\equ(3.50) (the others trivially admitting a gain of $\g^h$),
we note that, for any fixed configuration of isotropic sectors,
the new sums on isotropic sectors can be written as in
\equ(3.50a), where now the
* on the product now means that the product runs over $i=1,2$ if
$h_1=h$ and over $i=0,1,2$ if $h_1>h$. We again choose to multiply
the summand in the expression under analysis by
$\c(\{\lis\o_1^{(k)},
\lis\o_2^{(k)},\lis\o_3^{(k)},\lis\o_{f^*}\})$, where now
$\lis\o_3^{(k)}$ must be interpreted as the isotropic sector index
on scale $k$ such that $\lis\o\prec\lis\o_3^{(k)}$.

Then we bound the product of coupling functions, the integrals
over the propagators of the spanning tee (including the factor
$\sum_{l\in T''}|\xx_l-\yy_l|$, see item (B.1a)) and the product
of Gram determinants times the product of loop propagators as
explained in item B.1a above. With respect to the naive bounds
obtained by proceeding as in Sect. \secc(2.8), we lose a factor
$n\g^{-k}$, because of the presence of $\sum_{l\in
T''}|\xx_l-\yy_l|$, and we gain a factor
$\g^{h(1+\fra12\openone(h_1>h))}\g^{\fra12(h_1+h_2)}$ because of
the presence of the isotropic loop propagators and of the
derivative acting on one isotropic loop propagators. Then, the sum
over the choices of the extracted loop propagators gives a
combinatorial factor $|P_{v_1}|\prod_{w\in\ul v_m}|P_w|^2$.

After this we are left with the sum \equ(3.51) over the sector
indices, whose bound must be compared with the naive bound
\equ(3.52a). We proceed again through \equ(3.52) and \equ(3.53)
(where now the symbol $\prod_i^*$ must be interpreted as explained
above), then we bound
$\prod_i^*\sum_{\bar\o_i\prec\bar\o_i^{(k)}}$ and
$\sum^{**}_{\O_{v_m}\prec\O_{v_m}^{(k)}}$ exactly as in
\equ(3.54), unless for the fact that the product $\prod_{i=1}^3$
appearing in the r.h.s. of
\equ(3.54) must be replaced by $\prod_{i=1}^2$. So, exactly as in item B.1a
above, the sector sums corresponding to the vertices
$v_1,\ldots,v_m$ have a loss at most equal to $\g^{\fra12(k-h)}$
with respect to the correponding contribution in \equ(3.52a). We
are still left with the sum in \equ(3.55) and again, if $h_1=h$,
we can reduce the analysis of the expression in \equ(3.55) to the
analysis in item A2.1: then we find that such expression can be
bounded by the factor in \equ(3.52a) corresponding to the vertices
$v>v_m$ times a loss of $|k|$. If $h_1>h$ we bound the sum
$\sum_{\bar\o_0^{(k)}\prec\o_0^{(k)}}$ by $c\g^{-k/2}$ and then we
are left with an expression analogue to that studied in A2.1 and
we find a loss of a factor $|k|$ more.

Combining all gains and losses we find that the global gain has an
expression similar to \equ(3.56):
$$\eqalign{&cn\g^{-k}\g^h\g^{\fra12 h\openonesix(h_1>h)}\g^{\fra12
(h_1+h_2)} |P_{v_1}|\Big[\prod_{w\in\ul
v_m}|P_{w}|^2\Big]\g^{{1\over 2}(k-h)} \g^{-\fra12
k(\openonesix(h_1>h))}|k|=\cr =&cn|P_{v_1}|\Big[\prod_{w\in\ul
v_m}|P_{w}|^2\Big]\g^h|k| \g^{{1\over 2}(h-k)
(\openonesix(h_1>h)-1)}\prod_{i=1}^2\g^{\fra12(h_i-k)} \le
c^n\g^h|h|\prod_{v\in V_c}|P_v|^5\cr}\Eq(3.57)$$
\\
(B.2a) $L=5$: there are exactly five loop lines connecting $\tilde
P_{v^*}$ with $\tilde P_{v_m}\setminus \tilde P_{v^*}$, call them
$l_1,\ldots,l_5$ and denote by $f_1,\ldots,f_5$, by
$h_1,\ldots,h_5$ and by $\o_1\ldots,\o_5$ their field, scale and
sector labels. With no loss of generality, we can assume $h\le
h_1\le \cdots h_5\le k$. If $h_1=h$, then we rewrite the
propagator $g^{h_1}_{\o_1}(\xx_{f_1^+}-\yy_{f_1^-})$ in a form
similar to \equ(3.49):
$$g^{h_1}_{\o_1}(\xx_{f_1^+}-\yy_{f_1^-})=
g^{h_1}_{\o_1}({\bf 0})+\sum_{\bar\o_1\prec \o_1}
(\xx_{f_1^+}-\yy_{f_1^-})\int_0^1 ds \dpr_\xx\lis
g^{h_1}_{\bar\o_1}\big(s(\xx_{f_1^+}-\yy_{f_1^-})\big)
\;.\Eq(3.57aa)$$
If $h_1>h$, then as in item B.1a above we expand the determinant
of $\wt G_{\ul v_m}$ appearing in the analogue of \equ(3.14) along
the row corresponding to a field label $f_0^-\in
P_{v_1}^-\setminus P_{v_0}^-$ and we rewrite $\det\wt G_{\ul v_m}$
as in \equ(3.50).

Both in the case $h_1=h$ and $h_1>h$, implementing the rewritings
\equ(3.57) or \equ(3.50), we see that the terms proportional to
$g^{h_1}_{\o_1}({\bf 0})$ or to $g^{(h)}_{\o_0}({\bf 0})$ have a
dimensional gain of $\g^h$.

So, in both cases, let us focus on the contributions proportional
to the interpolated term in \equ(3.57) and \equ(3.50).

We replace all the modified coupling functions, all the
propagators and all the Gram determinants with their absolute
values. As regarding the factor $(\xx_{f_1^+}-\yy_{f_1^-})$ or
$(\xx_{f_0^+}-\yy_{f_0^-})$ appearing in \equ(3.57) or in
\equ(3.50), we can bound its absolute value by a sum over the
lines $l\in T''$ (here $T''$ is a suitable subset of the spanning
tree $T'$) of $|\xx_l-\yy_l|$. In particular we can associate each
factor $|\xx_l-\yy_l|$ to the corresponding propagator of the
spanning tree $T'$. Note that the number of terms in the sum
$\sum_{l\in T''}$ is bounded proportionally to $n$.

Now, we can bound the integral over the absolute values of the
modified coupling functions as in \equ(2.82) (by using \equ(2.71b)
in place of \equ(2.71a)), the integral over the absolute values of
the propagators of the spanning tree times the factor $\sum_{l\in
T''} |\xx_l-\yy_l|$ by the r.h.s. of \equ(2.81) {\it times a
dimensional loss} $cn\g^{-k}$, where we used that the propagators
associated to $l\in T''$ are of scale $j\ge k$ and that, by the
decay properties described in Lemma \lm(2.2), $\int d\xx|\xx|
|g^{(j)}_\o(\xx)|\le c\g^{-2 j}$. Moreover, the product of the
determinants times the product of the absolute values of the
extracted loop propagators can be bounded by the product over $v$
of the r.h.s. of \equ(2.80) {\it times a dimensional gain}
$\g^{\fra32 h}$, coming from the fact that the dimensional bound
of the derived isoptropic propagator on scale $h$ (the one
corresponding to $f_1$ or to $f_0$, depending whether $h_1=h$ or
not) is $\g^{\fra32 h}$ smaller than the corresponding bound for
an anisotropic popagator on scale $h$. After these bounds, the
sums over the choices of the loop propagators can be bounded by
$|P_{v_1}|\prod_{w\in\ul v_m}|P_w|^5$.

We are still left with the sum over the sector indices:
$$\sum_{\O\setminus\O_{v_0}}\sum_{\bar\o_{i^*}\prec \o_{i^*}}
\Big[\prod_{v>v_0\atop v\in V_c}\ \c (\tilde\O_{v,\bar\o})\Big]
\Big[\prod_{l\in T'}\d_{\o_l^+,\o_l^-}\Big]
\Big[\prod_{i}^*\d_{\o_i^+,\o_i^-}\Big]\;.\Eq(3.57a)$$
where the index $i^*$ attached to the sector indices in the second
sum is $i^*=1,0$, depending whether $h_1=h$ or not, and the * on
the last product means that the product ranges over $i=1,2,3,4,5$
or over $i=0,1,2,3,4,5$, depending whether $h_1=h$ or not.

We now want to bound \equ(3.51) and to compare the result we shall
find with \equ(3.52a). We again bound the product of $\c$
functions as in
\equ(3.51). We bound the sum $\sum_{\bar\o_{i^*}\prec \o_{i^*}}$ by a
constant times $\g^{-h/2}$ and, after this, we rewrite
$$\eqalign{&\sum_{\O\setminus\O_{v_0}}
\Big[\prod_i^*\d_{\o_i^+,\o_i^-}\Big]=
\Big[\prod_i^*\sum_{\o_i^{(k)}\in O_k}\Big]\cdot
\Big[\prod_i^*\sum_{\o_i\prec \o_i^{(k)}}\Big]\cdot
\sum_{\O_{v_m}^{(k)}}^{**}\ \cdot\
\sum_{\O_{v_m}\prec\O_{v_m}^{(k)}}^{**} \ \cdot\ \sum_{\O\bs
\O_{v_m}}^{**} \cr}\Eq(3.58)$$
where the $**$ on the sums mean that we are not summing neither
over the sectors in $\O_{v_0}$ nor over the sectors
$\o_0,\o_1,\ldots,\o_5$. If we denote by $P_L$ the set of field
labels of the half--lines corresponding to the extracted loop
propagators (equal to $\{f_1^\pm,\ldots,f_5^\pm,\}$ or to
$\{f_0^\pm,f_1^\pm,\ldots,f_5^\pm\}$, depending whether $h_1=h$ or
not) and by $n^L_i=|P_L\cap P_{v_i}|$ the number of field labels
in $P_L$ external to the cluster $v_i$, we can bound
$$\eqalign{&
\Big[\prod_i^*\sum_{\o_i\prec \o_i^{(k)}}\Big]\le
c\g^{\fra12(k-h)\openonesix(h_1>h)}\prod_{i=1}^5\g^{\fra12(k-h_i)}
\cr &\sum_{\O_{v_m}\prec\O_{v_m}^{(k)}}^{**}\le c^m
\prod_{i=1}^m\g^{\fra12(h_{v_i}-h_{v_{i-1}})(|P_{v_i}|-2-n^L_i)}=\cr
&\hskip1.4truecm=
c^m\g^{-(k-h)\openonesix(h_1>h)}\prod_{i=1}^5\g^{-(k-h_{i})}
\prod_{i=1}^m\g^{\fra12(h_{v_i}-h_{v_{i-1}})(|P_{v_i}|-2)}
\cr}\Eq(3.59)$$
Combining the two bounds in \equ(3.54), we see that the global
contribution from the sector sums corresponding to the vertices
$v_1,\ldots v_m$ is proportional to
$\prod_{i=1}^m\g^{\fra12(h_{v_i}-h_{v_{i-1}})(|P_{v_i}|-2)}$
$\g^{-\fra12(k-h)\openonesix(h_1>h)}\prod_{i=1}^5\g^{-\fra12(k-h_i)}$.
Comparing this with the corresponding bound in \equ(3.52a), we see
that the present bound has a {\it dimensional gain} at least equal
to $\g^{-\fra12(k-h)(\openonesix(h_1>h)-1)}$
$\prod_{i=1}^5\g^{-\fra12(k-h_i)}$.

After having bounded as above the sector sums corresponding to the
vertices $v_1,\ldots v_m$, we are still left with the sector sums
corresponding to the vertices $v> v_m$, that can be rewritten in
the form
$$\eqalign{&\Big[\prod_i^*\
\sum_{\o_i^{(k)}\in O_k}\Big]\sum_{\cup_{v>v_m}\O_v^{(k)}}^{**}
\Big[\prod_{v>v_m}\c(\tilde\O_{v,\bar\o}^{(k)})\Big]
\Big[\prod_{l\in T'}\d_{\o_l^+,\o_l^-}\Big]=\cr
&=\sum_{\O_{\ul{v}_m}^{(k)}\setminus\O_{v_0}}
\prod_i^*\d_{\o_{i}^+,\o_{i}^-} \prod_{l\in
T_{v_m}}\d_{\o_l^+,\o_l^-} \Big[ \prod_{v\in\ul v_m}
F_{v}(\tilde\O_{v,\bar\o}^{(k)})\Big] \;.\cr}\Eq(3.60)$$
If we now discard the constraint coming from
$\d_{\o_{f_0}^+,\o_{f_0}^-}$ (whenever present in \equ(3.60)) we
are left with an expression completely analogous to \equ(3.26) so
that, proceeding as in item (A1.2), we find that \equ(3.60) admits
a bound that, with respect to the corresponding bound in
\equ(3.52a), has a dimensional gain of $\g^k$.

Combining all the gain and losses discussed in this item we find
that the contributions under analysis, with respect to the bounds
obtained via the procedure in Section \secc(2.8), have a global
gain of:
$$\eqalign{&cn\g^{-k}\g^{\fra32 h}|P_{v_1}|\Big[\prod_{w\in\ul v_m}|P_w|^5\Big]
\g^{-\fra12 h}\g^{-\fra12(k-h)(\openonesix(h_1>h)-1)}
\Big[\prod_{i=1}^5\g^{-\fra12(k-h_i)}\Big]\g^k=\cr
&=cn|P_{v_1}|\Big[\prod_{w\in\ul v_m}|P_w|^5\Big]
\g^h\g^{-\fra12(k-h)(\openonesix(h_1>h)-1)}
\Big[\prod_{i=1}^5\g^{-\fra12(k-h_i)}\Big]\le c^n\g^h\prod_{v\in
V_c}|P_v|^5\;.\cr}\Eq(3.61)$$
\\(B.2b) $L=4$: there are exactly three loop lines connecting
$\tilde P_{v^*}$ with $\tilde P_{v_m}\setminus \tilde P_{v^*}$. We
call $f_1,f_2,f_3,f_4$ the field labels of the extracted loop
lines, $h_1,h_2,h_3,h_4$ their scale labels (assume $h\le h_1\le
\cdots\le h_4\le k$) and $\o_1,\o_2,\o_3,\o_4$ their sector
labels. In this case we proceed in a way very similar to item
(B.2a) above and we only rapidly repeat the proof of item B.2a
adapted to the present case. The main difference consists in the
fact that one of the fields in $P_{v^*}$, call it $f_5$, belongs
to $\P_{v_0}$, so that its sector label is fixed to be isotropic
and equal to $\lis\o$.

As in item B.2a, depending whether $h_1=h$ or not, we perform the
expansions in \equ(3.57aa) or in \equ(3.50). We again focus on the
contributions coming from the intepolated terms in \equ(3.57aa),
\equ(3.50) (the others trivially admitting a gain of $\g^h$).

We bound the product of coupling functions, the integrals over the
propagators of the spanning tee (including the factor $\sum_{l\in
T''}|\xx_l-\yy_l|$, see item (B.1a)) and the product of Gram
determinants times the product of loop propagators as explained in
item B.2a above. With respect to the naive bounds obtained by
proceeding as in Sect. \secc(2.8), we lose a factor $n\g^{-k}$,
because of the presence of $\sum_{l\in T''}|\xx_l-\yy_l|$, and we
gain a factor $\g^{\fra32h}$ because of the presence of one
derived isotropic loop propagators. The sum over the choices of
the extracted loop propagators gives a combinatorial factor
$|P_{v_1}|\prod_{w\in\ul v_m}|P_w|^4$.

After this we are left with the sum \equ(3.57a) over the sector
indices (where now the * on the last product means that the
product ranges over $i=1,2,3,4$ or over $i=0,1,2,3,4$, depending
whether $h_1=h$ or not), whose bound must be compared with the
naive bound \equ(3.52a).

We proceed again through \equ(3.52), we bound
$\sum_{\bar\o_{i^*}\prec\o_{i^*}}$ by a constant times $\g^{-h/2}$
and we consider the rewriting \equ(3.58) (where now the symbol
$\prod_i^*$ must be interpreted as explained above). We bound
$\prod_i^*\sum_{\o_i\prec\o_i^{(k)}}$ and
$\sum^{**}_{\O_{v_m}\prec\O_{v_m}^{(k)}}$ exactly as in
\equ(3.59), unless for the fact that the product $\prod_{i=1}^5$
appearing in the r.h.s. of
\equ(3.59) must be replaced by $\prod_{i=1}^4$. So, exactly as in item B.2a
above, the sector sums corresponding to the vertices
$v_1,\ldots,v_m$ have a gain at least equal to
$\g^{-\fra12(k-h)(\openonesix(h_1>h)-1)}$
$\prod_{i=1}^4\g^{-\fra12(k-h_i)}$ with respect to the
correponding contribution in \equ(3.52a). We are still left with
the sum in \equ(3.60) and again we discard the constraint coming
from $\d_{\o_0^+,\o_0^-}$ (whenever present in \equ(3.60)): in
this way we are reduced to an expression completely analogous to
the one considered in item (A2.2), see \equ(3.39), and we find
that such expression can be bounded by the factor in \equ(3.52a)
corresponding to the vertices $v>v_m$ times a gain of $\g^k$.

Combining all gains and losses we find that the global gain has an
expression similar to \equ(3.61):
$$\eqalign{&cn\g^{-k}\g^{\fra32h}|P_{v_1}|\Big[\prod_{w\in\ul v_m}|P_w|^4\Big]
\g^{-\fra12 h}\g^{-\fra12(k-h)(\openonesix(h_1>h)-1)}
\Big[\prod_{i=1}^4\g^{-\fra12(k-h_i)}\Big]\g^k=\cr
&=cn|P_{v_1}|\Big[\prod_{w\in\ul v_m}|P_w|^4\Big]
\g^h\g^{-\fra12(k-h)(\openonesix(h_1>h)-1)}
\Big[\prod_{i=1}^4\g^{-\fra12(k-h_i)}\Big]\le c^n\g^h\prod_{v\in
V_c}|P_v|^5\;.\cr}\Eq(3.62)$$
\\
(B.3) $L=0,1$: in these two cases either there is no loop line
connecting $\tilde P_{v^*}$ with $\tilde P_{v_m}\setminus \tilde
P_{v^*}$, or there is exactly one such line, depending whether
$v^*$ is equal to $v_\yy$ or not. If $L=1$, we call $l_1$ the
extracted loop line and $f_1$, $h_1$, $\o_1$ its field, scale and
sector labels. Let us rewrite the propagator of $f^*$ as a sum of
isotropic propagators, to be denoted by $\lis
g^{(k)}_{\bar\o_{f^*}}(\xx)$. If $L=1$, let us do the same with
the propagator of $f_1$, and let us denote by $\lis
g^{(h_1)}_{\bar\o_1}(\xx)$ the corresponding isotropic propagator.
Note that, as explained in item (A1.3) above, by momentum
conservation the summand in the expression under analysis can be
freely multiplied by $\openone(|\lis\o_{f^*}-\lis\o_1^{(k)}|\le
1)$, where $\lis\o_1^{(k)}$ is the isotropic sector of scale $k$
such that $\lis\o_1\prec \lis\o_1^{(k)}$.

Note also that, both in the case $L=0$ and $L=1$, the determinant
$\det \wt G_{\ul v_m}$ of the loop lines contracted on scales
$h,h_{v_1},\ldots,h_{v_m}$ (with the possible exception of $f_1$,
if $L=1$) can be factorized in a form analogous to \equ(3.29), to
be denoted by $\det\wt G_{v^*}\cdot \det\wt G_{\ul v_m\bs v^*}$.
Let us call $h\le \lis h_*\le k$ the smallest scale among those of
the propagators contracted into $\det\wt G_{v^*}$ and $h\le \lis
h_m\le k$ the smallest scale among those of the propagators
contracted into $\det\wt G_{\ul v_m\bs v^*}$.

Now, if $\lis h_*=h$, then we simply bound the expression under
analysis by the analogues of \equ(3.41) or of \equ(3.30). More
precisely, if $L=0$, using the same bound on $[\ul v_m\bs v^*]$
discussed in item (A2.3), we get
$$\eqalign{&\sum_{\bar\o_{f^*}}\openone(|\lis\o_{f^*}-\lis\o|\le 1)\cdot[v^*]
\cdot[\ul v_m\bs v^*]\int d\xx|\lis
g_{\bar\o_{f^*}}^{(k)}(\xx)|\le\cr &\le c \cdot[v^*]\cdot
\Big\{(c|U|)^{n-m_4(v^*)}\g^{\bar h_m} \prod_{v\in V_c, v\not\in
\t'} \g^{\d(|P_v|) (h_v-h_{v'})}\Big\}\cdot \g^{-k}\cr}\Eq(3.63)$$
where, again, $\t'$ is the subtree of $\t$ rooted in $v_m$ and
containing $v^*$, $m_4(v^*)$ is the number of endpoints following
$v^*$ on $\t$. If $v^*$ is an endpoint, $[v^*]$ is a contribution
of the form \equ(3.8), for which an improved dimensional bound
proportional to $\g^{2h}|h|$ is valid because of the inductive
assumption, and in this case the proof is complete. If $v^*$ is
not an endpoint, $[v^*]$ is given by the analogue of \equ(3.30a),
with $\det \wt G^{h}_{v^*}$ replaced by $\det\wt G_{\ul v_m}$ and
the scale of the first non trivial vertex following $v^*$ on $\t'$
{\it strictly larger than $k$} and we are still left with the
problem of bounding it. Note that a bound on $[v^*]$ of the form
\equ(3.31) would soon imply the desired improved bound.

If $L=1$ we get
$$\eqalign{&\sum_{l_1}^*\sum_{\bar\o_1,\bar\o_{f^*}}
\openone(|\lis\o_{f^*}-\lis\o_1^{(k)}|\le 1)\cdot[v^*] \cdot[\ul
v_m\bs v^*]\cdot\sup_{\xx}|\lis g^{(h_1)}_{\bar\o_1}(\xx)| \int
d\xx|\lis g^{(k)}_{\bar\o_{f^*}}(\xx)|\le\cr &\le c
\Big[\prod_{w\in\ul v_m}|P_w|\Big]\cdot \g^{(k-h_1)}\g^{-k}\cdot
[v^*] \cdot \Big\{(c|U|)^{n-m_4(v^*)} \prod_{v\in V_c, v\not\in
\t'} \g^{\d(|P_v|) (h_v-h_{v'})}\Big\} \cdot \g^{2 h_1}\cdot
\g^{-k}\;,\cr} \Eq(3.64)$$
where the notation and the interpretation of \equ(3.64) is the
same as
\equ(3.63). Again, if $v^*$ is an endpoint, the bound on $[v^*]$
follows from the inductive assumption \equ(3.8) and the proof is
complete. If $v^*$ is not an endpoint we are still left with the
problem of bounding $[v^*]$. As above, a bound on $[v^*]$ of the
form \equ(3.31) would soon imply the desired improved bound.

Before discussing how to bound $[v^*]$ let us briefly discuss the
case $\lis h_*>h$. In this case, if we proceed as above, we are
left with expressions like \equ(3.63) and \equ(3.64), with $[v^*]$
that can be bounded at best proportionally to $\g^{2\bar h_*}|\lis
h_*|$, as it is clear already in the case of $v^*$ an endpoint. So
in this case we must show how to gain a ``memory factor''
$\g^{h-\bar h_*}$ allowing to transfer the gain on scale $h$. Note
that if $\lis h_*>h$, then it must be either $h_1=h$ or $\lis
h_m=h$. Then, if $h_1=h$, we can expand $\lis
g_{\bar\o_1}^{(h)}(\xx)$ as in
\equ(3.49), while, if $\lis h_m=h$, we can extract from
$\det \wt G_{\ul v_m\bs v^*}$ a loop propagator on scale $h$ (by
expanding the determinant along a row labelled by an $f_0^-\in
P_{v_1}^-\bs P_{v_0}^-$) and such a propagator can be expanded as
in the second line of \equ(3.50). In both cases, the local
contributions $\lis g_{\bar\o_1}^{(h)}({\bf 0})$ or
$g_{\o_0}^{(h)}({\bf 0})$ obviously have a gain of $\g^h$ with
respect to the usual bound on isotropic or anisotropic
propagators. The interpolated contributions are associated to an
isotropic propagator, derived with respect to its argument,
multiplied by a factor $(\xx_{f^+}-\yy_{f^-})$, with $f$ equal to
$f_1$ or $f_0$. As already discussed in items (B.1a)--(B.2b)
above, the presence of the ``zero'' $(\xx_{f^+}-\yy_{f^-})$
worsens the bounds of a factor $c n\g^{-k}$, while the presence of
the derivative acting on the isotropic propagator improves the
bounds of a factor $\g^h$. Without entering in more details than
this (the bounds can be performed in a way similar to the one
discussed above for $\lis h_*=h$ and, as regarding the gains
associated to the interpolated terms, in a way similar to that
discussed in item (B.1a)), we find that the intepolated term is
associated to the gain of a factor $cn\g^{h-k}\le cn\g^{h-\bar
h_*}$, that is what we need in order to transfer
the improved bound on $[v^*]$ on scale $h$. \\

So, let us finally turn to the proof of \equ(3.31), where in
general the scale $h$ in the r.h.s. of \equ(3.31) must be replaced
by $\lis h_*$ (that is the lowest scale among those of the lines
contracted in the loop propagators).

In order to study \equ(3.30a) we simply repeat the analysis of
items (B.1a)--(B-2b) and of the present item. We again consider
the first non trivial vertex following $v^*$ on $\t'$, we call it
$v_m^*$ and we let it play the same role of $v_m$ above. Repeating
the procedure iteratively, up to the point where either one of the
cases (B.1a)--(B.2b) is realized or the analogue of $v^*$ is an
endpoint, we finally get the desired dimensional gain proportional
to $\g^{\bar h_*}$. We just have to be careful with the
combinatorial factors: a priori, if we were forced to repeat the
procedure $O(n)$ times, there could be the risk of a producing a
factorial more in the bounds. However, it is clear from the
discussion above, that, at each step of the iteration,
corresponding to a loop extraction on scale $j$ one loses at most
a combinatorial factor $\prod_{v\in V_c: h_v=j}c|P_v|^5$, bounding
the number of choices of possible loop propagators on that scale.
Then, repeating iteratively the procedure, we can at most lose a
combinatorial factor $c^n\prod_{v\in V_c}|P_v|^5$, that is the one
appearing in the r.h.s.
of \equ(3.6) and \equ(3.31).\\

This concludes the proof of \equ(3.6) with $j=0$ under the
assumptions
\equ(2.71a), \equ(2.71b) and \equ(3.8). The validity of \equ(3.6) with
$j\ge 1$ under the same assumptions is a trivial generalization of
the proof above. In fact, in order to take into account the factor
$|\xx-\yy|^j$ in the r.h.s. of \equ(3.5), we can bound it by
$(cn)^{j-1} \sum_{l\in T''}|\xx_l-\yy_l|^j$, where $T''\subset T$
is a path on $T$ such that $(\xx-\yy)=\sum_{l\in
T''}(\xx_l-\yy_l)$. Each term $|\xx_l-\yy_l|^j$ in this sum can be
associated to the corresponding propagator of $l\in T$: then, when
performing the integrations over the propagators of the spanning
tree, each of these terms contributes with a dimensional factor
$\le\g^{-hj}$, see Remarks following Lemma \lm(2.2) and Lemma
\lm(2.3). Then, for any integer $j\ge 0$, \equ(3.6) under the
assumptions
\equ(2.71a), \equ(2.71b) and \equ(3.8) follows. 
\qed\\

\sub(3.2) In this section we want to prove that if, given $h\le
0$, for any $j>h$ \equ(2.36) is satisfied, then \equ(2.71a) and
\equ(2.71b) are true for $h_i-1\ge h$ and \equ(3.8) is true for
$h\le k'\le k\le 0$. Combining this result with the results
discussed above in Sect.\secc(3.1) finally completes the inductive
proof of the validity of \equ(2.36) and of \equ(2.71a). 
Furthermore, this result, together with Theorem \thm(2.1), 
completes the proof of convergence 
of the expansion for the free energy and, together with the discussion 
in Section \secc(2.9), finally completes the proof of Theorem \thm(1.1).\\

Given a set $P_4=\{f_1,f_2,f_3,f_4\}$ of four field labels, a set
of sector indices $\O_4=\{\o_{f}\in O_j\,:\, f\in P_4\}$ and a set
of space--time points $\ul\xx=\{\xx_{f}\,:\,f\in P_4\}$ labelled
by $P_4$, let us begin with proving that, if $U_0|h_{\b}|=c_0$ is
small enough, then
$$\int d\xx_2 d\xx_3 d\xx_4|\wt\l_{j,\O_4}(\ul\xx)|\le
C|U|\;,\qquad j\ge h\Eq(3.65)$$
implying \equ(2.71a) for any tree $\t\in\TT_{h,n}$, $n\ge 1$. We
proceed by induction. Note that for $h=0$ the bound just follows
from the bounds on the kernels of the effective potential
$\VV^{(0)}$.

In order to prove \equ(3.65) for $j<0$, we substitute into the
definition of $\wt\l_{j,\O_4}$ the beta function equation in the
second line of \equ(2.34) (for notational convenience, in the
following we shall drop the dependence of $\b^4_j$ on
$(E_j,\l_j;\cdots;E_0,U)$). We find
$$\wt\l_{j,\O_4}(\ul\xx)=\FFF_{4,j,\O_4}*\l_0(\ul\xx)+
\sum_{j'=j+1}^0\FFF_{4,j,\O_4}*\b^4_{j'}(\ul\xx)\Eq(3.66)$$
where, repeating the same iterative construction leading to the
tree expansion for $\VV^{(h)}$, see \equ(2.68), and for $\b^2_h$,
see \equ(3.4), we can represent each term in the sum in the r.h.s.
of \equ(3.66) as:
$$\eqalign{&\b^4_{j',\O_4}(\ul\xx)\=\FFF_{4,j,\O_4}*\b^4_{j'}(\ul\xx)=\cr&
\quad=\sum_{n=2}^\io\sum_{\t\in\TT_{j',n}}\sum_{\bP\in\PP_\t \atop
|P_{v_0}|=4}\sum_{T\in\bT}\sum_{\O\setminus\O_{v_0}}
\Big[\prod_{v\in V_c}\c(\O_v)\Big]\cdot
\FFF_{4,j,\O_4}*\int d(\xx_{v_0}\setminus\ul\xx)
W^{(mod)}_{\t,\bP,\O,T}(\xx_{v_0}) \;.\cr} \Eq(3.67)$$
Note that the sum in the r.h.s. begins with the second order
($n=2$): by construction there is no trivial tree contributing to
$\b^4_{j'}$.

In analogy with \equ(2.98) and \equ(3.5), we can define $\lis
J^{(4)}_{j',n,\O_4}(\t,\bP,T)$ as the following quantity,
representing a bound for the terms appearing in the sum in the
r.h.s. of \equ(3.67):
$$\eqalign{&\lis J_{j',n,\O_4}^{(4)}(\t,\bP,T)= \cr
&\quad=\sum_{\O\setminus\O_{v_0}} \left[\prod_{v}\c(\O_v)
\right]\int d\xx_2d\xx_3d\xx_4\cdot \left|
\lis{\FFF}_{4,j,\O_4}*\int d(\xx_{v_0}\setminus\ul\xx)
W^{(mod)}_{\t,\bP,\O,T}(\xx_{v_0}) \right|\;.\cr}\Eq(3.68)$$
Assuming inductively the bound \equ(3.65) for $j'>j$ and repeating
the proof of Lemma \lm(2.5), leading to the first bound in
\equ(2.98), we find:
$$\lis J_{j',n,\O_4}^{(4)}(\t,\bP,T)\le (C|U|)^{n}\prod_{v\in V_c}
\fra1{s_v!}\g^{\d(|P_v|)(h_v-h_{v'})}\Eq(3.69)$$
Using \equ(3.69) into \equ(3.67), we find
$|\b^4_{j',\O_4}(\ul\xx)|\le c|U|^2$, so that, by \equ(3.66)
$$|\wt\l_{j,\O_4}(\ul\xx)|\le C_0|U|+c|j||U|^2\le C|U|\Eq(3.70)$$
where in the last passage we used that $|j||U|\le c_0$. This
completes the proof of \equ(2.71a). The proof of \equ(2.71b) is a
step
by step repetition of the proof above and we do not repeat it here.\\

Let us now turn to the proof of \equ(3.8). We begin with
considering $j=0$. We again proceed by induction. First, we want
to show the validity of
\equ(3.8) for $k=0$ and $h\le h'\le 0$, that is we want to prove
$$\sum_{\o\in O_{h'}}\int d(\xx_1-\xx_4) \,\left|
\int d\xx_2 d\xx_3\; g^{(h')}_{\o}(\xx_2-\xx_3)
\tilde\l_{0,\tilde\O_4}(\xx_1,\xx_2,\xx_3,\xx_4) \right| \le
C_0|U| \g^{2h'}\;, \Eq(3.71)$$
where $\tilde\O_4$ was defined after \equ(3.8) and
$\tilde\l_{0,\tilde\O_4}$ in \equ(3.8a). The strategy consists
again in rewriting $g^{(h')}_{\o}(\xx_2-\xx_3)$ as
$$g^{(h')}_{\o}({\bf 0})+(\xx_2-\xx_3)\sum_{\bar\o\prec\o}\int_0^1 ds\,
\dpr_\xx\lis
g^{(h')}_{\bar\o}\big(s(\xx_2-\xx_3)\big)\;.\Eq(3.72)$$
We substitute \equ(3.72) into \equ(3.71) and we bound the l.h.s.
of \equ(3.71) by the sum of two terms, corresponding to the two
terms in \equ(3.72). Now, the term proportional to
$g^{(h')}_{\o}({\bf 0})$ can soon be bounded as in the r.h.s. of
\equ(3.71), by Lemma \lm(2.2a) and by the estimate \equ(2.16) for
$l=2$. The term corresponding to the second addend in \equ(3.72)
can be bounded as
$$\sum_{\bar\o\in\bar O_{h'}}\sup_\xx|\dpr_\xx\lis g^{(h')}_{\bar\o}(\xx)|
\int
d\xx_2d\xx_3d\xx_4|\xx_2-\xx_3|\cdot|\tilde\l_{0,\tilde\O_4}(\xx_1,
\xx_2,\xx_3,\xx_4)|\le C_0'|U|^2\g^{-h'}\g^{3h'}\;,\Eq(3.73)$$
and \equ(3.71) follows.

Now, for any fixed $h\le h'\le 0$, we inductively suppose that
\equ(3.8) is valid for any $\lis k<k\le 0$, with $\lis k\ge h'$,
and we prove it for $k=\lis k$. We insert in the r.h.s. of
\equ(3.8) the beta function equation in the second line of
\equ(2.34) and we find that \equ(3.8) can be bounded by
$$\eqalign{&\sum_{\o\in O_{h'}}\int d(\xx_1-\xx_4) \, \left|
\int d\xx_2 d\xx_3\; g^{(h')}_{\o}(\xx_2-\xx_3)
\tilde\l_{0,\tilde\O_4}(\xx_1,\xx_2,\xx_3,\xx_4) \right|+\cr
&+\sum_{k=\bar k+1}^0 \sum_{\o\in O_{h'}}\int d(\xx_1-\xx_4) \,
\left| \int d\xx_2 d\xx_3\;
g^{(h')}_{\o}(\xx_2-\xx_3)\b^4_{k,\tilde\O_4}(\xx_1,
\xx_2,\xx_3,\xx_4)\right|\;,\cr}\Eq(3.74)$$
where $\b^4_{k,\tilde\O_4}(\ul\xx)$ admits the same representation
\equ(3.67), with $\O_4$ replaced by $\tilde\O_4$. The term in the
first row of \equ(3.74) can be bounded as in \equ(3.71). Repeating
the proof discussed in items (B.1a)--(B.3), it can be realized
that the $k$--th term in the sum in the second line of \equ(3.74)
can be bounded as:
$$\sum_{\o\in O_{h'}}\int d(\xx_1-\xx_4) \, \left|
\int d\xx_2 d\xx_3\;
g^{(h')}_{\o}(\xx_2-\xx_3)\b^4_{k,\tilde\O_4}(\xx_1,
\xx_2,\xx_3,\xx_4)\right|\le c|U|^2\g^{2h'}|k|\;.\Eq(3.75)$$
Inserting the bounds \equ(3.71) and \equ(3.75) into \equ(3.74) we
find that
\equ(3.8) with $k=\lis k$ can be bounded by
$$C_0|U|\g^{2h'}+c|U|^2\g^{2h'}\sum_{k=\bar k+1}^0|k|\le C|U|\g^{2h'}
|\lis k|\;, \Eq(3.76)$$
where in the last passage we used that $|U||\lis k|\le c_0$. So,
\equ(3.8) is proved in the case $j=0$.

As regarding the case $j\ge 1$, we can once more proceed by
induction. The term with $k=0$ and $h\le h'\le 0$ and $j\ge 1$ can
be bounded exactly as in \equ(3.71), with $C_0$ replaced by some
$C_{0,j}$. The term with $k=\lis k$, supposing inductively that
\equ(3.8) is valid for any $k>\lis k$, can be bounded by an
expression analogous to \equ(3.74), with a factor
$|\xx_1-\xx_4|^j$ more appearing under the integration. Via the
same strategy used to prove \equ(3.71), we can prove that the
analogue of the term in the first line of \equ(3.74) can be
bounded by $C_{0,j}|U|^2\g^{2h'}$: the reason for the factor
$|U|^2$ replacing the factor $|U|$ in \equ(3.71) is that the
function $\l_0(\ul\xx)$ is local (hence with vanishing derivative)
unless for terms of order $|U|^2$ coming from the ultraviolet
integration.

Similarly, the analogue of the $k$--th term in the sum in the
second line of \equ(3.74) can be bounded by the r.h.s. of
\equ(3.75) (with $c$ replaced by some new constant $c_j$) times a
dimensional factor $\g^{-jk}$ (obtained exactly as described at
the end of section \secc(3.2) above). Substituting such bounds in
the analogue of \equ(3.74) we find that \equ(3.8) with $k=\lis k$
and $j\ge 1$ can be bounded as
$$C_{0,j}|U|^2\g^{2h'}+c_j|U|^2\g^{2h'}\sum_{k=\bar k+1}^0|k|\g^{-jk}
\le C_j|U|^2|\lis k|\g^{2h'-j\bar k}\;.\Eq(3.77)$$
So the proof of \equ(3.8) is complete and, together with it, the
proof of convergence of the expansion for the free energy and for 
the two point Schwinger function is complete. In particular this concludes 
the proof of Theorem \thm(1.1).
\qed\\
\*

\appendix (A1, The ultraviolet integration)

In this Appendix we prove \equ(2.16). It is convenient to
introduce an ultraviolet cut-off $N$ by rewriting the kernels $W^{(0)}_{2n}$
as $\lim_{N\to\io}W^{[0,N]}_{2n}$, where $W^{[0,N]}_{2n}$ are 
the kernels of a cutoffed effective potential $\VV^{[0,N]}$
defined by an equation similar to \equ(2.13), with $\EE_1^T$ replaced by the
truncated expectation with propagator
$$g^{[1,N]}(\xx)=\sum_{n=0}^N g^{(1,n)}(\xx)\Eq(A1.1)$$
where
$$g^{(1,n)}(\xx)={1\over L^2\b}\sum_{\kk\in\DD_{\b,L}} f_1(\kk)
h_n(k_0)\fra{e^{-i\kk\xx}}{-ik_0+\e_0(\vec k)-\m}\Eq(A1.2)$$
with $h_0(k_0)=H_0(|k_0|)$ and $h_n(k_0)=H_0(\g^{-n+1}|k_0|)-
H_0(\g^{-n}|k_0|)$. 

Note that
$\lim_{N\to\io} g^{[1,N]}(\xx)=g^{(+1)}(\xx)$ and that,
for any integer $K\ge 0$, 
$g^{(1,n)}(\xx)$ satisfies the bound
$$|g^{(1,n)}(\xx)|\le {C_K\over 1+(\g^n|x_0|+|\vec x|)^K}\Eq(A1.3)$$
We associate to any propagator $g^{(1,n)}(\xx)$ a Grassmann field 
$\psi^{(1,n)}$ and a Gaussian integration $P(d\psi^{(1,n)})$ with 
propagator $g^{(1,n)}(\xx)$. Using repeatedly the addition principle 
\equ(2.12) we can rewrite $\VV^{(0)}$ as:
$$\VV^{(0)}(\phi)+L^2\b E_1=-\lim_{N\to\io}\log 
\int P(d\psi^{(1,0)})P(d\psi^{(1,1)})\cdots 
P(d\psi^{(1,N)}) e^{-V(\psi^{[1,N]}+\phi) }\Eq(A1.4)$$
We can integrate iteratively the fields on scale $N,N-1,\ldots,k+1$ and 
after each integration, using iteratively an identity like \equ(2.13),
we can rewrite the r.h.s. of 
\equ(A1.4) in terms of a new effective potential $\VV^{[1,k]}$:
$$\equ(A1.4)=\lim_{N\to\io}\Big\{
L^2\b\sum_{j=k+1}^N E_j-\log 
\int P(d\psi^{(1,0)})P(d\psi^{(1,1)})\cdots 
P(d\psi^{(1,k)}) e^{-\VV^{(k,N)}(\psi^{[1,k]}+\phi) }\Big\}\Eq(A1.5)$$
with $\VV^{(k,N)}(\psi^{[1,k]})$ admitting a representation similar to
\equ(2.15), with the fields $\psi^{(\le 0)}$ replaced by $\psi^{[1,k]}$
and the kernels $W^{(0)}_{2l}$ replaced by $W^{(k,N)}_{2l}$.
It is in fact well known that  
$\VV^{(k,N)}(\psi^{[1,k]})$ and the corresponding kernels 
can be written as sum over {\it trees} $\t$
of tree values 
$\VV^{(k,N)}(\t;\psi^{[1,k]})$ or $W^{(k,N)}_{2l}(\t;\xx_1,\s_1,\e_1;\ldots;
\xx_{2l},\s_{2l},\e_{2l})$, 
each of them computable as a product of 
truncated expectations $\EE^T_{(1,n)}$ (\ie of truncated expectations 
associated to the propagators $g^{(1,n)}$ of \equ(A1.2)). The definition of
tree is similar to the definition introduced in Section \secc(2.6) above
with the following modifications:
\item{1)} a tree $\t\in\TT_{(k,N),n}$, \ie a tree $\t$ belonging to the set
of trees $\TT_{(k,N),n}$ contributing to the $n$--th order in $U$ 
of $\VV^{(k,N)}$, has vertices $v$ associated to scale labels
$k\le h_v\le N+1$; the root $r$ has scale $k$ and all the endpoints
have scale $N+1$;
\item{2)} all the endpoints are ``of type $U$'', that is are associated
to a contribution $V(\psi)$, see \equ(2.6);
\item{3)} to any field variable with field label $f$ a spin label $\s(f)=\ud$
will be associated; the integer $m(f)$ (produced by the renolrmalization 
procedure) is identically $0$ (since no renormalization is needed to treat
the ultraviolet problem).
\item{4)} in analogy with definition 1 in \S(3.1) of [BGM], we call 
$c$--vertices the vertices $v$ of $\t$ such that the set of internal lines 
${\cal{I}}_v$ is not empty.\\

The values $W^{(k,N)}_{2l}(\t;\xx_1,\s_1,\e_1;\ldots;
\xx_{2l},\s_{2l},\e_{2l})$ can be computed via the iterative
rules described in \S(2.4) of [BGM] (with some obvious modifications 
needed to adapt the rules to the present case) and can be bounded via the same
strategy described in detail in \S(2.7) of [BGM]. 
Using the dimensional bound \equ(A1.3), it is easy to realize that
the contribution from a tree $\t\in\TT_{[1,k],n}$
associated to a kernel with $2l$ external legs can be bounded as:
$$\eqalign{&\fra{1}{L^2\b}\int d\xx_1\cdots d\xx_{2l}
|W^{(k,N)}_{2l}(\t;\xx_1,\s_1,\e_1;\ldots;
\xx_{2l},\s_{2l},\e_{2l})| \le\cr
&\hskip2.truecm \le C^n |U|^n \g^{-k(n-1+n^{tad})}
\prod_{v\ {\rm not}\ {\rm e.p.}}\g^{-(h_v-h_{v'})
(n_v-1+n^{tad}_v)}\;,\cr}\Eq(A1.6)$$ 
where $v'$ is the vertex immediately preceding $v$ on $\t$, 
$n_v$ is the number
of endpoints following $v$ on $\t$,
$n^{tad}$ is the total number of tadpoles on $\t$ (\ie 
the number of trivial $c$--vertices $v$ with $n_v=1$) and 
$n^{tad}_v$ is the number of tadpoles following $v$ on $\t$.\\
\\
{\bf Remark} In deriving \equ(A1.6) an explicit computation of the tadpole
contributions is needed. In particular one has to realize that 
a parity cancellation implies that the tadpole
contribution associated to the subtree rooted on 
a trivial $c$--vertex $v$ of scale $h_v$ with
$n_v=1$ can be bounded by $C\g^{-h_v}$ (instead of the naive
dimensional bound $C$).\\
\\
Note that the ``internal dimensions'' $(n_v-1+n^{tad}_v)$ are all $>1$,
so that the sum over $\t$ of \equ(A1.6) gives rise
to an exponentially convergent sum and the bound like \equ(2.16) follows.\\
\\
Finally, note that the possibility of rewriting the effective
potential on scale $0$ in the form \equ(2.15), with $W^{(0)}_{2l}$
independent of the spin labels, follows from the symmetries listed after
\equ(2.8) and the remark that $P(d\psi^{(+1)})$ itself is 
invariant under the same symmetries. \\

\appendix(A2, Proof of Lemma {\lm(3.1)})

Given $h\le 0$ and an isotropic sector index 
of scale $h$, $\lis\o\in \lis O_h$, let
us call $\lis S_{h,\lis\o}$ the corresponding isotropic s--sector
(here we choose a notation as close as possible to that introduced 
in [BGM] for the anisotropic sectors, see (2.72) of [BGM]).

Let assign $\lis\o_1\=\lis\o_{f_1}\in \lis O_h$ and let us call  
$A_h(\lis\o_1;\lis\o_2,\lis\o_3,\lis\o_4)$ the set of sequences 
$(\lis\o_2,\lis\o_3,\lis\o_4)$ in $\lis O_h\times\lis O_h\times\lis O_h$ 
such that\\

{\it there 
exists a sequence of vectors $(\vkk_1,\vkk_2,\vkk_3,\vkk_4)$ 
s.t. $\vkk_i\in \lis S_{h,\lis\o_i}$ and $\sum_{i=1}^4\vkk_i=
{\vec 0}$}.\\

We want to prove that 
$$|A_h(\lis\o_1;\lis\o_2,\lis\o_3,\lis\o_4)|\le C\g^{-h} |h|\;,\Eq(A2.0)$$ 
for some constant $C>0$. Let $\th_i$ be the center
of the $\th$--interval which the polar angle of $\vpp$ has to belong to, 
if $\vpp\in\lis S_{h,\lis\o}$. For any pair $(i,j)$, $i,j=1,2,3,4$, we define:
$$\phi_{i,j}=\min\{||\th_i-\th_j||, \p-||\th_i-\th_j||\}\Eq(A2.1)$$
where $||\cdot||$ is the distance on the torus. 
By a reordering of the sectors, we can always impose the condition:
$$\max\{\phi_{1,3},\phi_{1,4}\}\le \phi_{1,2}\Eq(A2.2)$$
In fact, calling $A_h^*(\lis\o_1;\lis\o_2,\lis\o_3,\lis\o_4)$ the subset
of $A_h(\lis\o_1;\lis\o_2,\lis\o_3,\lis\o_4)$ with $\lis\o_2,\lis\o_3,
\lis\o_4$ satisfying condition \equ(A2.2), it holds:
$$|A_h(\lis\o_1;\lis\o_2,\lis\o_3,\lis\o_4)|\le 3\,
|A_h^*(\lis\o_1;\lis\o_2,\lis\o_3,\lis\o_4)|\Eq(A2.3)$$
So, we describe how to bound $|A_h^*(\lis\o_1;\lis\o_2,\lis\o_3,\lis\o_4)|$.
With no loss of generality, we shall assume $\phi_{1,2}\le\p/2$
(the case $\phi_{1,2}\le\p/2$ can be reduced to $\phi_{1,2}\le\p/2$,
by the symmetry property (7.2) of [BGM]).

We first note that, given any positive constant $\k_0$, if we define
$$\AA_<(\k_0)=\{(\lis\o_2,\lis\o_3,\lis\o_4)\in 
A_h^*(\lis\o_1;\lis\o_2,\lis\o_3,\lis\o_4)\ :\ |\phi_{1,2}|\le \p \k_0 
\g^{h\over 2}\}\Eq(A2.4)$$
we have:
$$|\AA_<(\k_0)|\le 16 \k_0^2\g^{-h}\Eq(A2.5)$$
In fact, for any choice of $\phi_{1,2}=\p k\g^h$, $|k|=0,1,\ldots,
[\k_0\g^{-{h\over 2}}]\= N(\k_0)$, 
by condition \equ(A2.2) we have that
$\lis\o_3$ can be chosen at most in $|k|$ different ways. Finally,
once $\lis\o_2$ and $\lis\o_3$ are fixed, 
by momentum conservation $\lis\o_4$ is fixed in
a finite number of sectors (and it is easy to realize that such number is 
$\le 8$). Then $|\AA_<(\k_0)|$ is bounded by $8\sum_{k=-N(\k_0)}^{N(\k_0)}|k|=
8N(\k_0)[N(\k_0)+1]\le 16 \k_0^2 \g^{-h}$ and \equ(A2.5) follows. 

Then, in order to prove \equ(A2.0), it is sufficient to prove that a similar
bound is valid for the set $\AA_>(\k_0)
\=A_h^*(\lis\o_1;\lis\o_2,\lis\o_3,\lis\o_4)
\setminus \AA_<(\k_0)$. We start with computing the number $N$ of pairs of
angles $(\th_3,\th_4)$ compatible with a given choice of $\th_1,\th_2$ s.t.
$\phi_{1,2}\ge \p \k_0\g^{h/2}$. 
Rewriting $\th_3,\th_4$ in the form: $\th_i+\p+\p n_i\g^h$, $i=1,2$,
it is clear that $N$ can be bounded by the number of
pairs $\vec n =(n_1,n_2)$ of integers compatible with the condition
$$\sum_{i=1}^2\big[\vpp_F^{(h)}(\th_i+\p n_i\g^h)-\vpp_F^{(h)}(\th_i)\big]
=\g^h\vrr\qquad {\rm with}\qquad
|\vec r|\le R \Eq(A2.6)$$
for some $O(1)$ constant $R$. \equ(A2.6) can be 
rewritten in the form:
$$\vrr=\vec f(\vec n)\=\g^{-h}\sum_{i=1}^2\big[
\vpp_F^{(h)}(\th_i+\p n_i\g^h)-\vpp_F^{(h)}(\th_i)\big]\Eq(A2.7)$$
with $\vec f$ twice differentiable.
We now want to apply Dini's implicit function theorem in order to 
invert \equ(A2.7) in a neighborood of $\vec n=\vec r=\vec 0$
(that is an ``unperturbed'' solution to \equ(A2.7)). We define 
$A\defin (D \ul f(\ul 0))^{-1}$ and the norm $||A||$
as $||A||\defin\sum_{i,j}|A_{i,j}|$. We recall that 
Dini's Theorem claims that if $\vec n$ varies in a ball $B_\r(\vec 0)$
around $\vec 0$ of radius $\r$ so small that 
$$||D\vec f(\vec n)-D \vec f(\vec 0)||\le {1\over 4||A||}\;,\qquad\forall
\vec x\in B_\r(\vec 0)\Eq(A2.8)$$
then, if $r$ is chosen so that $r<\r/(2||A||)$, then for any
$\vec r\in B_r(\vec 0)$ we can invert \equ(5) as 
$$\vec n=\vec g(\vec r),\qquad \vec r\in B_r(\vec 0),\qquad r<{\r
\over 2||A||}\Eq(A2.9)$$
with $\vec g$ twice differentiable and such that 
the image of $B_r(\vec 0)$ through $\vec g$ is contained in $B_\r(\vec 0)$.
In order to have the condition $|\vec r|\le R$ in \equ(A2.6)
verified together with the condition $|\vec r|\le {\r
\over 2||A||}$ in \equ(A2.9), we can choose 
$$\r\defin 2||A|| R\Eq(A2.9a)$$
We now want to compute $||A||$ and check \equ(A2.8).
Using (7.3) and (7.6) of [BGM]
and calling $s'_i\=s'(\th_i+\p\g^h n_i)$, we see that 
the Jacobian of $\vec f(\vec n)$,
in the basis $\vec n_h(\th_1),\vec\t_h(\th_1)$, is:
$$\eqalign{&D \vec f(\vec n)
=\p\pmatrix{s_1'\sin\big(\a(\th_1)-\a(\th_1+\p\g^h n_1)\big)
& \ &s_2'
\sin\big(\a(\th_1)-\a(\th_2+\p\g^h n_2)\big)\cr \ & \ \cr
s_1'\cos\big(\a(\th_1)-\a(\th_1+\p\g^h n_1)\big)
& \ &s_2'
\cos\big(\a(\th_1)-\a(\th_2+\p\g^h n_2)\big)
}\cr}\Eq(A2.10)$$
so that $\det D\vec f(\vec 0)=\p^2 s'(\th_1)s'(\th_2)\sin\big(\a(\th_1)-
\a(\th_2)\big)$. Hence, recalling the definition of $A$ (see the lines
preceding \equ(A2.8)) and the property $c_1||\th_1-\th_2||\le
||\a(\th_1)-\a(\th_2)||\le c_2||\th_1-\th_2||$ (see (7.5) of [BGM]
and section 2 of the present paper), we have: 
$${C_1\over |\sin(\th_1-\th_2)|}\le ||A||\le {C_2\over |\sin(\th_1-\th_2)|}
\Eq(A2.11)$$
Using the bound \equ(A2.11), 
we now want to check that the definition of $\r$ in \equ(A2.9a) is compatible
with \equ(A2.8).
We first note that, since $s'(\th)$ is differentiable, the l.h.s.
of \equ(A2.8) is bounded by $c\g^h|\vec n|\le c\g^h\r=2cR||A||\g^h$, 
for some $O(1)$ constant $c>0$.
So, \equ(A2.8) holds if $8cR||A||^2\g^h\le 1$ and, by \equ(A2.11), this is 
surely true if 
$$ 8cRC_2^2{\g^h\over \sin^2(\th_1-\th_2)}\le 1\Eq(A2.12)$$
We now recall that, since we are assuming that $||\th_1-\th_2||\ge 
\p \k_0\g^{h/2}$, it holds: $|\sin(\th_1-\th_2)|\ge {2\over \p}||\th_1-\th_2||
\ge 2 \k_0\g^{h/2}$. This means that \equ(A2.12) is satisfied if $\k_0$
is big enough, that is if
$$\k_0^2\ge 2cR C_2^2\Eq(A2.13)$$
For instance we can choose $\k_0\=C_2\sqrt{2cR }$. 

With these choices, we finally have that \equ(A2.7) can be inverted into
\equ(A2.9) and, if $|\vec r|\le R$, then $|\vec n|\le 2||A|| R\le 
{2C_2R\over |\sin(\th_1-\th_2)|}$. 
This means in particular that, for any choice of $\th_2=\th_1+\p k\g^h$, 
$[\k_0\g^{-{h\over 2}}]\le |k|\le[{\g^{-h}\over 2}]$, $\lis\o_3$ can be 
chosen in at most ${4C_2R\over |\sin(\p k\g^h)|}$ different ways. Once
both $\lis\o_2$ and $\lis\o_3$ are chosen, by momentum conservation $\lis\o_4$
is essentially fixed (it can be chosen in finite number of ways, and one
can realize that this number is $\le 8$). Then the number of 
elements of $\AA_>$ can be bounded as:
$$|\AA_>|\le 2\sum_{k=[\k_0\g^{-{h\over 2}}]}^{[{\g^{-h}\over 2}]}
{16 C_2 R\over k\g^h}\le \big[16 C_2 R\log\g\big]\g^{-h}|h|\Eq(A2.14)$$
Combining \equ(A2.14) with \equ(A2.5) we get \equ(A2.0).\qed

\\

\appendix(A3, Improved proofs of geometric lemmas)

In this Appendix we want to prove the following Lemma (called {\it 
Sector Counting Lemma}).

\*\lemma(A3.1){\it Let $h',h,L$ be integers such 
that $h'\le h\le 0$ and $L\ge 4$. 
Given $\o_1^{(h')},\ldots$, $\o_L^{(h')}\in O_{h'}$
and $\o_2^{(h)},\ldots,\o_L^{(h)}\in O_{h}$, let us define
$\O_L^{(h')}=\{\o_i^{(h')}\}_{i=1}^L$,
$\O^{(h')}_{L-1}=\{\o_i^{(h')}\}_{i=2}^L$ and $\O^{(h)}_{L-1}=
\{\o_i^{(h)}\}_{i=2}^L$.
Then, for any choice of $\o_1^{(h')}$, it is
$$\sum_{\O_{L-1}^{(h')}\prec \O_{L-1}^{(h)}}
\c(\O_L^{(h')})\le c^L \g^{\fra12(h-h')(L-3)}\;,\Eq(A3.1)$$
uniformly in $\o_1^{(h')}$ and in $\O^{(h)}_{L-1}$. 
The symbol $\prec$ must be interpreted as explained after \equ(2.88).}\\

In [BGM] a slightly different version of Lemma \lm(A3.1) was proved, see 
Lemma (3.1) of [BGM]. The main differences between the Lemma \lm(A3.1)
and the one proved in [BGM] are the following: in [BGM] the case
of a fixed $C^\io$ Fermi surface was considered, while here we are studying
the case of a $C^2$ Fermi surface, changing step by step of 
$O(|U||h|\g^{2h})$. 

By a critical rereading of the proof of Lemma 3.1 of [BGM], see Section 7 
of [BGM], it can be realized that the proof in [BGM] can be easily adapted 
to the case of a $C^\io$ Fermi surface, changing step by step of 
$O(|U||h|\g^{2h})$. In order to see this it must be taken into account that, 
as already noted after \equ(2.69), given an s-sector $S_{h,\o}$ on scale $h$, 
there are exactly $\g^{\fra12(h-h')}$ s-sectors 
on scale $h'$ strictly contained into it, and their centers $\th_{h',\o'}$
are independent of the specific shape of the Fermi surface. Moreover
the fact that $\vpp_F^{(h')}(\th)$ is convex (uniformly in $h'$) must be used. 
Keeping these two remarks in mind, the proof of Lemma \lm(A3.1) above in the 
case of a $C^\io$ Fermi surface, changing step by step of 
$O(|U||h|\g^{2h})$, is easily obtained by replacing any $\vpp_F(\th)$ 
appearing in the proof in Section 7 of [BGM] by $\vpp_F^{(h')}(\th)$.

While the fact that the Fermi surface was fixed was not really used in 
the proof in Section 7 of [BGM], the fact that it was chosen as a $C^\io$ 
curve was used here and there and it is not so straightforward to adapt 
the proof of [BGM] to the present $C^2$ case. However
by a careful rereading of Section 7 of [BGM], it can be realized that the 
only places where the $C^2$ regularity of $\vpp_F(\th)$ was used 
were the proofs
of Lemma 7.1 and of Lemma 7.5, where some error terms were bounded by 
the third derivatives of $\vec p_F(\th)$. In this section we want to
reproduce the proofs of this two Lemmas in a more careful way and 
the result will be that the proofs also work in the case that the Fermi 
surface is $C^2$, and not more regular than this. This in particular 
implies the validity of Lemma \lm(A3.1) above.\\
 
In the following we refer for notation to Section 7 of [BGM]. 
All the quantities $\vpp_F(\th)$, $u(\th)$, $s(\th)$, $\vec n(\th)$, 
$\vec\t(\th)$ 
appearing below do depend on $h'$: $\vpp_F(\th)$ must be interpreted as 
equal to $\vpp_F^{(h')}(\th)$ and all the other quantities are obtained 
from $\vpp_F^{(h')}(\th)$ via the definitions in Section 7 of [BGM].
We chose to drop the dependence on $h'$ in order to unify the notation 
with that of Section 7 of [BGM].\\

\centerline{\bf Proof of Lemma 7.1}
\\
We want to prove that $c_1\le |\a'(\th)|\le c_2$, where
$$\a(\th)=\arcsin\Big\{{1\over s'(\th)}\big[u(\th)\sin\th-u'(\th)\cos\th
\big]\Big\}\Eq(1)$$
By explicitely performing the derivative of \equ(1), we find:
$${-{s''(\th)\over(s'(\th))^2}\big[u(\th)\sin\th-u'(\th)\cos\th\big]
+{1\over s'(\th)}\big[2u'(\th)\sin\th+(u(\th)-u''(\th))\cos\th\big]
\over
\sqrt{1-{1\over (s'(\th))^2}\big[u(\th)\sin\th-u'(\th)\cos\th\big]^2}
}\Eq(2)$$
Using that $s'(\th)\defin\sqrt{u^2(\th)+(u'(\th))^2}$, we see that 
the denominator in \equ(2) is equal to
$$\eqalign{&
{1\over s'(\th)}\sqrt{u^2(\th)+(u'(\th))^2-u^2(\th)\sin^2\th-(u'(\th))^2
\cos^2\th+2u(\th)u'(\th)\sin\th\cos\th}=\cr
&={1\over s'(\th)}\Big|
u(\th)\cos\th+u'(\th)\sin\th\Big|\cr}\Eq(3)$$
Using that $s''(\th)={u u'+u' u''\over\sqrt{u^2+(u')^2}}$, we can rewrite
the numerator in \equ(2) as:
$$\eqalign{&
{1\over s'(\th)}\Big[-{(u u'+u' u'')\over u^2+(u')^2}(u\sin\th-u'\cos\th)
+2u'\sin\th+(u-u'')\cos\th
\Big]=\cr
&={1\over s'(\th)[u^2+(u')^2]}\Big[
-(u u'+u' u'')(u\sin\th-u'\cos\th)+\cr
&\hskip3.truecm
+(2u'\sin\th+(u-u'')\cos\th)(u^2+(u')^2)\Big]=\cr
&={1\over s'(\th)[u^2+(u')^2]}\Big[
\sin\th\big(2(u')^3+u^2u'-uu'u''\big)
+\cos\th
\big(2u(u')^2+u^3-u^2u''\big)\Big]=\cr
&={1\over s'(\th)[u^2+(u')^2]}(u'\sin\th+u\cos\th)
\big(2(u')^2+u^2-u u''\big)=\cr
&={s'(\th)\over r(\th) [u^2+(u')^2]}(u'\sin\th+u\cos\th)
\cr}\Eq(4)$$
where in the last identity we used $[2(u')^2+u^2-u u'']=(s'(\th))^2/r(\th)$,
that is an easy consequence of (7.3) and (7.4) of [BGM].
Dividing \equ(4) by \equ(3), we find that $|\a'|=1/r(\th)$, that 
is bounded above and below, by hypothesis. \qed\\
\\
\centerline{\bf Proof of Lemma 7.5}\\
\\
With the definitions introduced in the proof of Lemma 7.5, we can rewrite 
(7.24) of [BGM] as:
$$\eqalign{\pmatrix{\wt r_1\cr\wt r_2}
&={1\over \h}\pmatrix{\phi^{-1}\Big[
[\vec p_F(\lis\th_1+\h x_1)-\vec p_F(\lis\th_1)]\cdot\vec n(\lis\th_1)
+[\vec p_F(\lis\th_2+\h x_2)-\vec p_F(\lis\th_2)]\cdot\vec n(\lis\th_1)
\Big]\cr
[\vec p_F(\lis\th_1+\h x_1)-\vec p_F(\lis\th_1)]\cdot\vec \t(\lis\th_1)
+[\vec p_F(\lis\th_2+\h x_2)-\vec p_F(\lis\th_2)]\cdot\vec \t(\lis\th_1)
}\=\cr
&\=\ul f(x_1,x_2)\cr}\Eq(5)$$
The equation $\ul{\wt r}=\ul f(\ul x)$ admits an unperturbed
solution $\ul 0=\ul f(\ul 0)$ and we want to apply Dini's
theorem to invert the equation as $\ul x=\ul g(\ul{\wt r})$
in a neighborood af $\ul{\wt r}=\ul 0$.

The Jacobian matrix of $\ul f$ is:
$$D \ul f(\ul x)=\pmatrix{\phi^{-1}\vec n(\lis\th_1)\dpr_\th\vec p_F
(\lis\th_1+\h x_1) & \ &\phi^{-1}\vec n(\lis\th_1)\dpr_\th\vec p_F
(\lis\th_2+\h x_2) \cr \ & \ \cr
\vec \t(\lis\th_1)\dpr_\th\vec p_F
(\lis\th_1+\h x_1) & \ &\vec \t(\lis\th_1)\dpr_\th\vec p_F
(\lis\th_2+\h x_2) }\Eq(6)$$
and, using (7.3) of [BGM] we can rewrite:
$$D \ul f(\ul x)=\pmatrix{s'(\lis\th_1+\h x_1)
\phi^{-1}\vec n(\lis\th_1)\vec\t
(\lis\th_1+\h x_1) & \ &s'(\lis\th_2+\h x_2)
\phi^{-1}\vec n(\lis\th_1)\vec\t
(\lis\th_2+\h x_2) \cr \ & \ \cr
s'(\lis\th_1+\h x_1)\vec \t(\lis\th_1)\vec\t
(\lis\th_1+\h x_1) & \ &s'(\lis\th_2+\h x_2)\vec \t(\lis\th_1)\vec\t
(\lis\th_2+\h x_2) }\Eq(7)$$
and, using (7.6) of [BGM] we can rewrite the latter as:
$$\eqalign{&D \ul f(\ul x)=\cr
&=\pmatrix{-{s'(\lis\th_1+\h x_1)\over
\phi}\sin\big(\a(\lis\th_1+\h x_1)-\a(\lis\th_1)\big)
& \ &-{s'(\lis\th_2+\h x_2)\over
\phi}\sin\big(\a(\lis\th_2+\h x_2)-\a(\lis\th_1)\big)\cr \ & \ \cr
s'(\lis\th_1+\h x_1)\cos\big(\a(\lis\th_1+\h x_1)-\a(\lis\th_1)\big)
& \ &s'(\lis\th_2+\h x_2)\cos\big(\a(\lis\th_2+\h x_2)-\a(\lis\th_1)\big)
}\cr}\Eq(8)$$
In particular, the Jacobian determinant at $\ul x=\ul 0$ is
$$\det D \ul f(\ul 0)=s'(\lis\th_1)s'(\lis\th_2)\phi^{-1}\sin(\a(\lis\th_2)-
\a(\lis\th_1))\;,\Eq(9)$$ 
that is bounded above and below by $O(1)$ constants. As a consequence,
if we define $A\defin (D \ul f(\ul 0))^{-1}$ and the norm $||A||$
as $||A||\defin\sum_{i,j}|A_{i,j}|$, we have that $||A||$ is bounded 
above and below by $O(1)$ constants. 

Now, Dini's Theorem claims that if $\xx$ varies in a ball $B_\r(\ul 0)$
around $\ul 0$ of radius $\r$ so small that 
$$||D\ul f(\ul x)-D \ul f(\ul 0)||\le {1\over 4||A||}\;,\qquad\forall\ul 
x\in B_\r(\ul 0)\Eq(10)$$
then, if $r$ is chosen so that $r<\r/(2||A||)$, then for any
$\ul{\wt r}\in B_r(\ul 0)$ we can invert \equ(5) as 
$$\ul x=\ul g(\ul{\wt r})\Eq(11)$$
and the image of $B_r(\ul 0)$ through $\ul g$ is contained in 
$B_\r(\ul 0)$. So, let us compute $||D\ul f(\ul x)-D \ul f(\ul 0)||$
in our case. We have that $D\ul f(\ul x)-D \ul f(\ul 0)$ is given by:
$$\pmatrix{-{s'(\lis\th_1+\h x_1)\over
\phi}\sin\big(\a(\lis\th_1+\h x_1)-\a(\lis\th_1)\big)
& \ &-{s'(\lis\th_2+\h x_2)\over
\phi}\sin\big(\a(\lis\th_2+\h x_2)-\a(\lis\th_1)\big)+
\cr \ & \ & {s'(\lis\th_2)\over
\phi}\sin\big(\a(\lis\th_2)-\a(\lis\th_1)\big) \cr\ &\ & \cr
s'(\lis\th_1+\h x_1)\cos\big(\a(\lis\th_1+\h x_1)-\a(\lis\th_1)\big)
-
& \ &s'(\lis\th_2+\h x_2)\cos\big(\a(\lis\th_2+\h x_2)-\a(\lis\th_1)\big)
-\cr s'(\lis\th_1) 
&\ & s'(\lis\th_2)\cos\big(\a(\lis\th_2)-\a(\lis\th_1)\big)}\Eq(12)$$
and its norm is bounded by $C(c_2+\h)|\ul x|$, where $c_2$ was defined
in lemma 7.5 of [BGM] and $C$ is constant depending only on the 
second derivative of $s(\th)$ and the first derivative of $\a(\th)$, that
is at most on second derivatives of $u(\th)$. Now, choosing 
$c_2$ sufficiently small, we can say that for any $|\ul x|\le 1$ \equ(10)
is satisfied and \equ(11) holds for any $|\ul{\wt r}|\le c$, where $c$
is a suitable $O(1)$ constant. \qed

\\

\appendix(A4, Proof of Lemma {\lm(2.4)})

Consider the sector sum \equ(2.97),
where the sum runs over the sector indices in $\O\bs\O_{ext}^{(F)}$ and, 
given $f_1,\ldots,f_F\in P_{v_0}$, we defined $\O_{ext}^{(F)}=\{\o_{f_i}
\}_{i=1}^5$. We want to prove that, if $F=3$ or $F=5$, the sum in \equ(2.97)
can be bounded by the expression in the r.h.s. of \equ(2.83), \ie by the
bound for the sector sum with $F=1$, times a dimensional gain of $\g^{h/2}$
or $\g^h$, depending whether $F=3$ or $F=5$.

First of all, let us note that, if $\t\in\TT_{h,1}$, \ie if $\t$ is trivial,
the bound we want to prove is obvious. In fact, if $\o_1,\ldots,\o_4\in O_h$
and $\O_{4}=\{\o_1,\ldots,\o_4\}$, it is
$$\sum_{\o_1\in O_h}\c(\O_4)\le c\;,\Eq(A4.1)$$ 
simply by momentum conservation. Another simple consequence of momentum 
conservation is the following.
\*\lemma(A4.1) {\it Let $h',h,L,F$ be integers such that $h'\le h\le
0$, $L\ge 4$ and $1\le F\le L-1$. Given 
$\o_1^{(h')},\ldots,\o_L^{(h')}\in O_{h'}$
and $\o_1^{(h)},\ldots,\o_F^{(h)}\in O_{h}$, let us define
$\O_L^{(h')}=\{\o_i^{(h')}\}_{i=1}^L$,
$\O_{L-F}^{(h')}=\{\o_i^{(h')}\}_{i=F+1}^L$ and $\O_{L-F}^{(h)}=
\{\o_i^{(h)}\}_{i=F+1}^L$.
Then, for any choice of $\o_{1}^{(h')},\ldots,\o_F^{(h')}\in O_{h'}$ it is
$$\sum_{\O_{L-F}^{(h')}\prec \O_{L-F}^{(h)}}
\c(\O_L^{(h')})\le c^L \g^{\fra12(h-h')(L-F-1)}.\Eq(A4.2)$$}
\0\proof If $F=L-1$ the statement 
of the Lemma is a simple consequence of momentum conservation.
If $F<L-1$, then we can bound \equ(A4.2) by 
$$\sum_{\o_{F+2}^{(h')}\prec\o_{F+2}^{(h)}}\cdots\sum_{\o_L^{(h')}\prec
\o_L^{(h)}}\cdot
\sup_{\o_i^{(h')}\in O_{h'}\atop i=2,\ldots,L}\Big[
\sum_{\o_{F+1}^{(h')}\prec \o_{F+1}^{(h)}}\c(\O_{L}^{h'})\Big]\;.\Eq(A4.3)$$
Now, the last sum can be bounded as $\sum_{\o_1^{(h')}\prec \o_1^{(h)}}
\c(\O_{L}^{h'})\le c^L$, simply by momentum conservation. After this bound, 
we see that the remaining sums contribute with a factor 
$\g^{\fra12(h-h')(L-F-1)}$ and the Lemma is proved. \qed\\

\sub(A4.1) We start with the analysis of the case $F=3$. We assume 
that the number of endpoints $n$ of the tree is $n\ge 2$, the case $n=1$ 
being already treated, see \equ(A4.1).

As in Sect.\secc(2.8), we denote by $\tilde v_0$ the first $c$--vertex
following the root, we call $h_0$ its scale. Then, we bound the product of 
$\c$--functions in \equ(2.97) as in \equ(2.87) and, using the latter estimate, 
we bound \equ(2.97) by the r.h.s. of \equ(2.88), where the * on the sums 
must be interpreted as meaning that all the sectors index in $\O^{(3)}_{ext}$
must not be summed over. With this interpretation of the * on the sum and
using Lemma \lm(A4.1) for $4\le L\le 8$,
we perform the bound analogue to \equ(2.89):
$$\sum^*_{\O_{\tilde v_0}^{(h)}\prec\O_{\tilde v_0}^{(h_0)}}\c
(\O_{\tilde v_0}^{(h)})\le c \g^{\fra12(h_{0}-h)\left[|P_{\tilde v_0}|-4+
\openonesix(|P_{\tilde v_0}|>10)\right]}\;,\Eq(A4.4)$$
and we see that, with respect to \equ(2.89), we gain at least a factor 
$\g^{\fra12(h-h_0)}$. Now, proceeding through the analogues of \equ(2.90)
and \equ(2.91), we rewrite the left over expression in the same form as the 
r.h.s. of \equ(2.91):
$$\sum_{\O_{\ul{\tilde v}_0}^{(h_0)}\bs\O^{(3)}_{ext}}\prod_{v\in
\ul{\tilde v}_0}F_v(\O_v^{(h_0)})\prod_{l\in T_{\tilde v_0}}
\d_{\o_l^+,\o_l^-}\;.\Eq(A4.5)$$
We now distinguish three possible cases.\\
\\
(1) If the lines with field labels $f_1,f_2,f_3$ are all incident to the same
vertex $v_\xx\in\ul{\tilde v}_0$, we choose $v_\xx$ as the root of $T_{\ul{
\tilde v}_0}$ and, ``pruning $T_{\ul{\tilde v}_0}$ of its leaves'', one after 
the other, via the same procedure described after \equ(2.91), we can bound
\equ(A4.5) by the analogue of \equ(2.95) that takes the form:
$$\sup_{\O^{(3)}_{ext}}\sum_{\O_{v_\xx}^{(h_0)}\bs\O^{(3)}_{ext}}
F_{v_\xx}(\O_{v_\xx}^{(h_0)})\prod_{v\in \ul{\tilde v}_0\bs v_\xx}
\sup_{\o_v^*}\sum_{\O_v^{(h_0)}\bs\o_{v}^*}F_{v}(\O_v^{(h_0)})\Eq(A4.6)$$
where $\o_v^*$ is defined as after \equ(2.95). Now, each factor in the product
over $v\in \ul{\tilde v}_0\bs v_\xx$ can be bounded exactly as explained after
\equ(2.95), that is as in \equ(2.96) if 
$v$ is an endpoint or, if it is not, 
by the r.h.s. of \equ(2.89), with $(h_0-h)$ replaced by
$(h_v-h_0)$ and $|P_{\tilde v_0}|$ replaced by $|P_v|$. 
As regarding the first factor in \equ(A4.6), if $v_\xx$ is an endpoint,
it can be bounded by constant, that is it gives a contribution 
$\g^{h_0/2}$ smaller than the corresponding \equ(2.96); in this 
case, combining this gain with the gain $\g^{\fra12(h-h_0)}$ found above, see 
\equ(A4.4) and the following comment, the desired dimensional gain is found. 
If $v_\xx$ is not an endpoint, we are 
still left with $\sup_{\O^{(3)}_{ext}}\sum_{\O_{v_\xx}^{(h_0)}\bs\O^{(3)}_{
ext}}F_{v_\xx}(\O_{v_\xx}^{(h_0)})$ and we still have to establish whether 
this factor admits a dimensional gain with respect to the corresponding factor 
in \equ(2.95). We shall discuss this below, after item (3).\\
\\
(2) If two of the lines with field labels in $\{f_1,f_2,f_3\}$, say $f_1,f_2$, 
are incident to the same vertex $v_\xx$, while $f_3$ is incident to a distinct
vertex $v_\yy\in \ul{\tilde v}_0$, we identify the path $l_{\xx,\yy}$
on $T_{\tilde v_0}$ connecting $v_\xx$ and $v_\yy$, we denote by $T_\xx$
the subtree of $T_{\tilde v_0}$ rooted on $v_\xx$ and with no lines in common
with $l_{\xx,\yy}$ and by $\ul{\tilde v}_0^\xx\subset \ul{\tilde v}_0$ 
the set of vertices in $T_\xx$ (note that possibly $T_\xx$ is trivial, 
that is $\ul{\tilde v}_0^\xx=\{v_\xx\}$). In analogy with \equ(2.92), 
we define 
$$\eqalign{&{\cal F}_{\ul{\tilde v}_0\bs \ul{\tilde v}_0^\xx}
(\O^{(h_0)}_{\ul{\tilde v}_0\bs \ul{\tilde v}_0^\xx})
\defin \prod_{v\in\ul{\tilde v}_0\bs \ul{\tilde v}_0^\xx}
F_{v}(\O_{v}^{(h_{0})}) \prod_{l\in T_{\tilde v_0}\bs \{T_\xx\cup l^*_\xx\}}
\d_{\o_l^+,\o_l^-}\;,\cr
&{\cal F}_{\ul{\tilde v}_0^\xx}(\O^{(h_0)}_{\ul{\tilde v}_0^\xx})
\defin \prod_{v\in\ul{\tilde v}_0^\xx}F_{v}(\O_{v}^{(h_{0})}) 
\prod_{l\in T_\xx}\d_{\o_l^+,\o_l^-}\;,\cr}\Eq(A4.7)$$
and, denoting by $l_\xx^*$ the line of $T_{\tilde v_0}\bs T_\xx$
incident to $v_\xx$ (let $f^*_\xx$ and $\o^*_\xx$ be its field ans sector 
indices), we rewrite \equ(A4.5) as
$$\eqalign{&\sum_{\o^{*,+}_\xx,\o^{*,-}_\xx}\d_{\o^{*,+}_\xx,\o^{*,-}_\xx}
\sum_{\O_{\ul{\tilde v}_0^\xx}\bs \{\o_1,\o_2,\o^*_\xx\}}
{\cal F}_{\ul{\tilde v}_0^\xx}(\O^{(h_0)}_{\ul{\tilde v}_0^\xx})
\sum_{\O^{(h_0)}_{\ul{\tilde v}_0\bs \ul{\tilde v}_0^\xx}\bs\{\o^*_\xx,\o_3\}}
{\cal F}_{\ul{\tilde v}_0\bs \ul{\tilde v}_0^\xx}
(\O^{(h_0)}_{\ul{\tilde v}_0\bs \ul{\tilde v}_0^\xx})\;,\le\cr
&\le \sup_{\o_1,\o_2,\o^*_\xx}\Big[
\sum_{\O_{\ul{\tilde v}_0^\xx}\bs \{\o_1,\o_2,\o^*_\xx\}}
{\cal F}_{\ul{\tilde v}_0^\xx}(\O^{(h_0)}_{\ul{\tilde v}_0^\xx})\Big]
\cdot \sup_{\o_3}\Big[
\sum_{\O^{(h_0)}_{\ul{\tilde v}_0\bs \ul{\tilde v}_0^\xx}\bs\{\o_3\}}
{\cal F}_{\ul{\tilde v}_0\bs \ul{\tilde v}_0^\xx}
(\O^{(h_0)}_{\ul{\tilde v}_0\bs \ul{\tilde v}_0^\xx})\Big]\;.\cr}\Eq(A4.8)$$
Now, the second factor in \equ(A4.8) can be bounded following the same 
procedure
described in Section \secc(2.8) and we do not find neither a gain nor a 
loss with respect to the estimates in Section \secc(2.8). The first factor 
can be studied as in item (1) above and, if $v_\xx$ is an endpoint, we gain 
$\g^{h_0/2}$ with respect to the estimates in Section \secc(2.8); in this 
case, combining this gain with the gain $\g^{\fra12(h-h_0)}$ found above,
the desired dimensional gain is found. If $v_\xx$ is not an endpoint, we are 
still left with $\sup_{\O^{(3)}_{ext}}\sum_{\O_{v_\xx}^{(h_0)}\bs\O^{(3)}_{
ext}}F_{v_\xx}(\O_{v_\xx}^{(h_0)})$ and we still have to establish whether 
this factor admits a dimensional gain with respect to the corresponding factor 
in \equ(2.95). We shall discuss this below, after item (3).\\
\\
(3) If the lines with field labels $f_1,f_2,f_3$ are all incident to different
vertices in $\ul{\tilde v}_0$, call them $v_\xx,v_\yy,v_\zz$, 
we again identify the path $l_{\xx,\yy}$
on $T_{\tilde v_0}$ connecting $v_\xx$ and $v_\yy$ and we define $T_\xx$ and 
$\ul{\tilde v}_0^\xx$ as in item (2). With no loss of generality we can assume 
that $v_\zz\in \ul{\tilde v}_0^\xx$: it is in fact easy to realize that, 
if $v_\zz\not\in\ul{\tilde v}_0^\xx$, we can always permute the names 
$v_\xx,v_\yy,v_\zz$ in such a way that, after the permutaion, 
$v_\zz\in \ul{\tilde v}_0^\xx$. Repeating the discussion in item (2) we can 
again bound the expression under analysis by the r.h.s. of \equ(A4.8), 
in which now the first factor has the same structure as the term studied 
in item (2) (it corresponds to a sector sum in which two of the lines
corresponding to external fixed sectors are incident to $v_\xx$, while the 
third one is incident to a different vertex -- $v_\zz$). So, by the proof
in item (2), we find that, if $v_\xx$ is an endpoint, we soon find the desired
dimensional gain, otherwise we are still
left with $\sup_{\O^{(3)}_{ext}}\sum_{\O_{v_\xx}^{(h_0)}\bs\O^{(3)}_{ext}}
F_{v_\xx}(\O_{v_\xx}^{(h_0)})$ and we still have to establish whether this 
factor admits a dimensional gain with respect to the corresponding factor in 
\equ(2.95).\\

In all the three cases discussed above, either $v_\xx$ is an endpoint, in 
which case the desired dimensional bound is found, or it is not, 
and we are left with an expression completely analogous to the initial one, 
with the external scale $h$ replaced by $h_0$. Then it is clear that 
we can iteratively enter the structure of $v_\xx$ following the same
procedure described in items (1)--(3) above and, proceeding by induction, 
we find the desired gain.\\

\sub(A4.2) Let us now consider $F=5$. Using the same notations as in Section 
\secc(A4.1) and following the same procedure, we bound the product of 
$\c$--functions in \equ(2.97) as in \equ(2.87) and, using the latter estimate, 
we bound \equ(2.97) by the r.h.s. of \equ(2.88), where the * on the sums 
must be interpreted as meaning that all the sectors index in $\O^{(5)}_{ext}$
must not be summed over. With this interpretation of the * on the sum,
we perform the bound analogue to \equ(2.89) and \equ(A4.4):
$$\sum^*_{\O_{\tilde v_0}^{(h)}\prec\O_{\tilde v_0}^{(h_0)}}\c
(\O_{\tilde v_0}^{(h)})\le c \g^{\fra12(h_{0}-h)\left[|P_{\tilde v_0}|-6+
\openonesix(|P_{\tilde v_0}|>10)\right]}\;,\Eq(A4.9)$$
and we see that, with respect to \equ(2.89), we gain at least a factor 
$\g^{(h-h_0)}$. We now rewrite the left over expression in the a form
similar to \equ(A4.5):
$$\sum_{\O_{\ul{\tilde v}_0}^{(h_0)}\bs\O^{(5)}_{ext}}\prod_{v\in
\ul{\tilde v}_0}F_v(\O_v^{(h_0)})\prod_{l\in T_{\tilde v_0}}
\d_{\o_l^+,\o_l^-}\;.\Eq(A4.10)$$
We now distinguish three possible cases.\\
\\
(1) If the lines with field labels $f_1,f_2,f_3,f_4,f_5$ are all incident to 
the same vertex $v_\xx\in\ul{\tilde v}_0$, we choose $v_\xx$ as the root 
of $T_{\ul{\tilde v}_0}$ and, proceeding as in item (1) of Section \secc(A4.1),
we are left with the analogue of \equ(A4.6):
$$\sup_{\O^{(5)}_{ext}}\sum_{\O_{v_\xx}^{(h_0)}\bs\O^{(5)}_{ext}}
F_{v_\xx}(\O_{v_\xx}^{(h_0)})\prod_{v\in \ul{\tilde v}_0\bs v_\xx}
\sup_{\o_v^*}\sum_{\O_v^{(h_0)}\bs\o_{v}^*}F_{v}(\O_v^{(h_0)})\Eq(A4.11)$$
As discussed in item (1) of Section \secc(A4.1),
each factor in the product over $v\in \ul{\tilde v}_0\bs v_\xx$ is not 
associated to any gain or loss (it can be bounded as in Section \secc(2.8)).
Note that now $v_\xx$ cannot be an endpoint (it has at least 6 external lines),
so we are still left with $\sup_{\O^{(3)}_{ext}}\sum_{\O_{v_\xx}^{(h_0)}
\bs\O^{(3)}_{
ext}}F_{v_\xx}(\O_{v_\xx}^{(h_0)})$ and we still have to establish whether 
this factor admits a dimensional gain with respect to the corresponding factor 
in \equ(2.95). We shall discuss this below, after item (3).\\
\\
(2) Calling $v_{\xx_1},\ldots, v_{\xx_5}$ the vertices to which the lines
$l_1,\ldots,l_5$ labelled by $f_1,\ldots f_5$ are incident to,
if $v_{\xx_1},\ldots, v_{\xx_5}$ are not all coinciding, we can assume 
without loss of generality that $v_{\xx_1}\neq v_{\xx_5}$ and we can consider 
the path 
$l_{\xx_1,\xx_5}$ on $T_{\tilde v_0}$ connecting $v_{\xx_1}$ and $v_{\xx_5}$.
Given $l^*\in l_{\xx_1,\xx_5}$, we call $T_1$ and $T_2$ the two disjoint 
subtrees of $T_{\tilde v_0}$ obtained by disconnecting from 
$T_{\tilde v_0}$ the line $l^*$: we assume that $T_1$ is anchored to 
$v_{\xx_1}$ and $T_2$ is anchored to $v_{\xx_5}$. 
We call $\ul{\tilde v}_0^1, \ul{\tilde v}_0^2$ the two disjoint subsets of 
$\ul{\tilde v}_0$ connected by the lines of $T_1,T_2$ respectively. 

We now further distinguish two more subcases.\\
(2.a) If there is a way of choosing $l^*\in l_{\xx_1,\xx_5}$ so that 
three among the lines in $\{l_1,\ldots,l_5\}$ are incident to $T_1$ and 
two of them are incident to $T_2$ (or viceversa), we can assume without 
loss of generality that $f_1,f_2,f_3$ are incident to $T_1$ and $f_4,f_5$
are incident to $T_2$. In this case, in analogy with the definition \equ(A4.7),
we define
$$\eqalign{&{\cal F}_{\ul{\tilde v}_0^1}
(\O^{(h_0)}_{\ul{\tilde v}_0^1})
\defin \prod_{v\in\ul{\tilde v}_0^1}
F_{v}(\O_{v}^{(h_{0})}) \prod_{l\in T_1}\d_{\o_l^+,\o_l^-}\;,\cr
&{\cal F}_{\ul{\tilde v}_0^2}(\O^{(h_0)}_{\ul{\tilde v}_0^2})
\defin \prod_{v\in\ul{\tilde v}_0^2}F_{v}(\O_{v}^{(h_{0})}) 
\prod_{l\in T_2}\d_{\o_l^+,\o_l^-}\;,\cr}\Eq(A4.12)$$
and, calling $\o_*$ the sector index of $l^*$, we rewrite \equ(A4.10) as:
$$\eqalign{&\sum_{\o_*^+,\o^{-}_*}\d_{\o_*^+,\o^{-}_*}
\sum_{\O_{\ul{\tilde v}_0^1}\bs \{\o_1,\o_2,\o_3,\o_*\}}
{\cal F}_{\ul{\tilde v}_0^1}(\O^{(h_0)}_{\ul{\tilde v}_0^1})
\sum_{\O^{(h_0)}_{\ul{\tilde v}_0^2}\bs\{\o_*,\o_4,\o_5\}}
{\cal F}_{\ul{\tilde v}_0^2}
(\O^{(h_0)}_{\ul{\tilde v}_0^2})\;,\le\cr
&\le \sup_{\o_1,\o_2,\o_3,}\Big[
\sum_{\O_{\ul{\tilde v}_0^1}\bs \{\o_1,\o_2,\o_3\}}
{\cal F}_{\ul{\tilde v}_0^1}(\O^{(h_0)}_{\ul{\tilde v}_0^1})\Big]
\cdot \sup_{\o_*,\o_4,\o_5}\Big[
\sum_{\O^{(h_0)}_{\ul{\tilde v}_0^2}\bs\{\o_*,\o_4,\o_5\}}
{\cal F}_{\ul{\tilde v}_0^2}(\O^{(h_0)}_{\ul{\tilde v}_0^2})\Big]
\;.\cr}\Eq(A4.13)$$
Now both factors can be bounded as in Section \secc(A4.1) and we get 
a gain $\g^{h_0/2}$ from both, and in this case the desired dimensional
gain is found.\\
(2.b) If any choice of $l^*\in l_{\xx_1,\xx_5}$ is such that 
four among the lines in $\{l_1,\ldots,l_5\}$ are incident to $T_1$ and 
only one of them is incident to $T_2$ (or viceversa), we can assume without 
loss of generality that $f_1,f_2,f_3,f_4$ are incident to $T_1$ and $f_5$
is incident to $T_2$. In this case, with the same definitions \equ(A4.12)
and calling $\o_*$ the sector index of $l^*$, we can rewrite \equ(A4.10) as:
$$\eqalign{&\sum_{\o_*^+,\o^{-}_*}\d_{\o_*^+,\o^{-}_*}
\sum_{\O_{\ul{\tilde v}_0^1}\bs \{\o_1,\o_2,\o_3,\o_4,\o_*\}}
{\cal F}_{\ul{\tilde v}_0^1}(\O^{(h_0)}_{\ul{\tilde v}_0^1})
\sum_{\O^{(h_0)}_{\ul{\tilde v}_0^2}\bs\{\o_*,\o_5\}}
{\cal F}_{\ul{\tilde v}_0^2}
(\O^{(h_0)}_{\ul{\tilde v}_0^2})\;,\le\cr
&\le \sup_{\o_1,\o_2,\o_3,\o_4,\o_*}\Big[
\sum_{\O_{\ul{\tilde v}_0^1}\bs \{\o_1,\o_2,\o_3,\o_4,\o_*\}}
{\cal F}_{\ul{\tilde v}_0^1}(\O^{(h_0)}_{\ul{\tilde v}_0^1})\Big]
\cdot \sup_{\o_5}\Big[
\sum_{\O^{(h_0)}_{\ul{\tilde v}_0^2}\bs\{\o_5\}}
{\cal F}_{\ul{\tilde v}_0^2}(\O^{(h_0)}_{\ul{\tilde v}_0^2})\Big]
\;.\cr}\Eq(A4.14)$$
Now the second factor can be bounded exactly as in Section \secc(2.8) and then
it is not associated to any gain or loss. The first factor has again the same
structure as \equ(A4.10) and we can again bound it following the same
procedure described in items (1), (2.a) and (2.b) above. 
Following iteratively the procedure, either we get the desired dimensional 
gain (if we recover the case of (2.a) above) or we are left with an expression
analogue to \equ(A4.11).

But the first factor in \equ(A4.11) has exactly the same structure as 
\equ(A4.10), with th external scale $h$ replaced by $h_0$. Then it is clear 
that we can iteratively enter the structure of $v_\xx$ following the same
procedure described in items (1)--(2.b) above and, proceeding by induction, 
we find the desired gain.
This concludes the proof of Lemma \lm(2.4). \qed\\
\\
{\bf Acknowledgments.} Part of this work was done in Vienna at the 
Erwin Schr\"oendiger Institute for Mathematical Sciences during 
the program ``Many body quantum theory''. We thank the organizers 
M. Salmhofer and J. Yngvason for their nice invitation. AG was partially 
supported by a Junior Research Fellowship from ESI, which is gratefully
acknowledged. AG wants also to thank M. Salmhofer for some enlightening 
discussions.
\baselineskip=12pt
\vskip1cm

\centerline{\titolo References}
\*
\halign{\hbox to 1.2truecm {[#]\hss} &%
\vtop{\advance\hsize by-1.25truecm\0#}\cr
A& {P.W. Anderson. The theory of superconductivity on high $T_c$ cuprates, 
Princeton University Press, Princeton 
(1997)}\cr
AMR& {S. Afchain, J. Magnen, V. Rivasseau 
{\it Ann. Henri Poincare'} 6,3, 399--488 and 449--483 (2005)}\cr
BGM& {G.Benfatto, A.Giuliani, V.Mastropietro {\it Ann. Henri Poincare'} 
1,3, 137--193 (2003)}\cr
BM& {G. Benfatto, V.Mastropietro.
{\it Rev. Math. Phys.} 13 (2001), 11, 1323--143}\cr
DR& {G.Disertori, V. Rivasseau, {\it Comm. Math. Phys}. 215, 2, 291--341 
and 251--390 (2000)}\cr 
FKT& {J.Feldman, H. Knoerrer, E. Trubowitz {\it Comm. Math. Phys.} 247, 
1--320 (2004)}\cr
FMRT& {J.Feldman, J.Magnen, V.Rivasseau, E. Trubowitz. {\it Helv. Phys. 
Acta} 65, 679--721 (1992)}\cr
FST& {J.Feldman, M.Salmhofer, E. Trubowitz.  
{\it J. Statist. Phys.}  84  (1996),  no. 5-6, 1209--1336.
{\it Comm. Pure Appl. Math.}  51  (1998),  no. 9-10, 1133--1246.
{\it Comm. Pure Appl. Math.}  52  (1999),  no. 3, 273--324.
{\it Comm. Pure Appl. Math.}  53  (2000),  no. 11, 1350--1384.
}\cr
G& {P. Goldbaum. {\it Comm. Math. Phys.} 258,2, 317--337 (2005). }\cr
GLM& {G. Gallavotti, J. L. Lebowitz, V. Mastropietro. {\it 
J. Statist. Phys.}  108  (2002),  no. 5-6, 831--861.}\cr
GS& {V.M. Galitski, S. Das Sarma, {\it Phys. Rev. B} 70,  35111 (2004)}\cr
HM& {C.J. Halboth and W. Metzner
{\it Z. Phys. B} 102, 501-504 (1997).}\cr
KL& {W. Kohn, J. M. Luttinger {\it Phys. Rev. Lett.} 15, 524--526 (1965)}\cr
L& {E.H.Lieb in {\it The Hubbard model; its physics and mathematical physics}, 
Edited by D. Baeriswvy, D. Campbell et al, Nato asi series, 343.}\cr
LW& {E.H.Lieb, F.Y.Wu. {\it Phys. Rev. Lett.} 20, 1445--1449 (1968).}\cr
M& {Mastropietro, V. to appear in {\it J. Stat. Phys.}.}\cr
Me& {W. Metzner {\it Int. J. Modern Phys. A} 16, 11, 1889--1898 (2001).}\cr
NO& {J. W. Negele, H. Orland. Quantum Many-Particle Systems, Addison Wesley,
New York (1998).}\cr
PS& {S.Pedra, M.Salmhofer, Proceeding of ICMP 2003}\cr
R& {V.Rivasseau, {\it J. Stat. Phys.} 106, 3-4, 693--722 (2002).}\cr
S& {J. Solyom. {\it Adv. Phys.} 28, 201--303 (1979).}\cr
Sa1& {Salmhofer, M. {\it  Comm. Math. Phys.}  194,  no. 2, 249--295 (1998).}\cr
Sa2& {Salmhofer, M. {\it Rev. Math. Phys.}  10, 4, 553--578 (1998).}\cr
VR& {A.Viroszteck, J.Ruvals, {\it Phys. Rev. B} 42, 4064 (1990).}\cr
}

\bye

\bye